\documentclass[10pt, a4paper]{article}
\usepackage{graphicx}
\usepackage{bbm}
\usepackage[usenames]{color}
\usepackage{stmaryrd}
\usepackage{amssymb,amsmath,latexsym,amscd,amsfonts}

\begin{document}

\newcommand{\B}[1]{\mathbf{#1}}
\newcommand{\Bv}{\B{v}}
\newcommand{\Bu}{\B{u}}
\newcommand{\BQP}{\sf{BQP}}
\newcommand{\Un}[1]{\underline{#1}}
\newcommand{\Eq}[1]{Eq.~(\ref{#1})} 
\newcommand{\Fig}[1]{Fig.~\ref{#1}}
\newcommand{\Sec}[1]{Sec.~\ref{#1}}
\newcommand{\EqDef}{\stackrel{\mathrm{def}}{=}}
\newcommand{\bra}[1]{\left< #1\right|}
\newcommand{\ket}[1]{\left| #1\right>} 
\newcommand{\la}{\langle}
\newcommand{\ra}{\rangle}

\newcommand{\enote}[1]{}
\newcommand{\znote}[1]{}
\newcommand{\inote}[1]{}
\newcommand{\dnote}[1]{}
\newcommand{\ir}[1]{\textcolor{Red}{#1}}
\newcommand{\ia}[1]{\textcolor{Green}{#1}}
\newcommand{\ignore}[1]{} 

\newcommand{\tr}{\mbox{Tr}}
\newcommand{\mat}[4]{\left(\begin{array}{cc} #1 & #2\\ #3 & #4\end{array}\right)}
\newcommand{\Id}{\mathbbm{1}}

\newcommand{\mcI}{\mathcal{I}}
\newcommand{\mcT}{\mathcal{T}}
\newcommand{\mcS}{\mathcal{S}}
\newcommand{\mcA}{\mathcal{A}} 
\newcommand{\mcB}{\mathcal{B}} 
\newcommand{\mcC}{\mathcal{C}} 
\newcommand{\mcO}{\mathcal{O}} 

\newcommand{\mfI}{\mathfrak{I}}
\newcommand{\mfT}{\mathfrak{T}}
\newcommand{\mfS}{\mathfrak{S}}
\newcommand{\mfA}{\mathfrak{A}} 
\newcommand{\mfB}{\mathfrak{B}} 
\newcommand{\mfC}{\mathfrak{C}} 
\newcommand{\mfO}{\mathfrak{O}}

\newtheorem{thm}{Theorem}[section]
\newtheorem{proposition}{Proposition}[section]
\newtheorem{deff}{Definition}[section] 
\newtheorem{algorithm}{Algorithm}[section] 
\newtheorem{claim}{Claim}[section] 
\newtheorem{lem}{Lemma}[section] 
\newtheorem{conjecture}{Conjecture}[section] 
\newtheorem{hypothesis}{Hypothesis}[section] 
\newtheorem{speculation}{Speculation}[section]
\newtheorem{notation}{Notation}[section] 
\newtheorem{corol}{Corollary}[section] 
\newtheorem{fact}{Fact}[section] 
\newtheorem{comment}{Comment}[section] 
\newtheorem{remark}{Remark}[section]

\newenvironment{proof}{\noindent\textit{Proof: }}{$\Box $}
\newenvironment{proofof}[1]{\noindent{\textit Proof}
of $\bf{#1}$:\hspace*{1em}}{$\Box $}

\title{Polynomial Quantum algorithms for additive
approximations of the Potts model and other points of the 
Tutte plane \\ {\it Preliminary Version}}
\author{ 
 Dorit Aharonov 
 \thanks{School of Computer Science and Engineering, 
 The Hebrew University, Jerusalem, Israel. 
 doria@cs.huji.ac.il.} 
 \and Itai Arad 
 \thanks{School of Computer Science and Engineering, 
 The Hebrew University, Jerusalem, Israel. 
 itaia@cs.huji.ac.il.} 
\and 
Elad Eban 
 \thanks{School of Computer Science and Engineering, 
 The Hebrew University, Jerusalem, Israel. 
 elade@cs.huji.ac.il.} 
\and 
 Zeph Landau 
\thanks{Department of Mathematics, The City College of  New York, NY}}

\date{Version of \today}

\maketitle

\begin{abstract}

In the first part of this paper, we provide polynomial quantum
algorithms for additive approximations of the Tutte polynomial, at
any point in the Tutte plane, for any planar graph.  This includes
an additive approximation of the partition function of the Potts
model for any weighted planer graph at any temperature, as well as 
approximations to many other combinatorial graph properties
described by the (multivariate or not) Tutte polynomial. 

To achieve these algorithms, we generalize the Temperley Lieb
algebra representations, used in \cite{ref:Aha05}, to apply for any
graph (not necessarily coming from a braid).  Moreover, our
representations are non-unitary, as are all representations of the
Temperley Lieb algebra not corresponding to Jones polynomial related
parameters. It might seem at first sight that this makes it
impossible to apply them by a quantum circuit.  We show how to do
this nevertheless. The approximation window size turns out to be
inverse polynomial in $|G|$ times the product of the norms of the
operators we apply. 

Additive approximations are tricky; the range of the possible
outcomes, might be smaller than the size of the approximation
window, in which case the outcome is meaningless. Unfortunately, 
ruling out this possibility is difficult: If we want to argue that
our algorithms are meaningful, we have to provide an estimate of the
scale of the problem, which is difficult here exactly because no
efficient algorithm for the problem exists! 

In the second part of the paper we provide an indirect but very
convincing proof that our approximation is meaningful for a large
range of parameters, by showing that in those cases, the problems
our algorithms solve are complete for quantum polynomial time.  We
thus get a large class of new $\BQP$-complete problems.  This result
is particularly interesting in the case where the relevant
representations are non-unitary, since in this case even the notion
of universality seems counter intuitive. Indeed, the universality
proof is extremely involved technically, and requires many new
innovative ideas. 

The case of the Potts model parameters deserves special attention. 
Unfortunately, despite being able to handle non-unitary
representations, our methods of proving universality seem to be
non-applicable for the \emph{physical} Potts model parameters.  We
can provide only weak evidence that our algorithms are non-trivial
in this case, by analyzing their performance for instances for which
classical efficient algorithms exist. The characterization of the
quality of the algorithm for the Potts parameters is thus left as an
important open problem.  

To summarize, the main progress in this work is in our ability to
handle non-unitary representations; by doing so, we provide many new
quantum complete problems corresponding to the approximation of the Tutte
polynomial at various points.  We believe this work, and in 
particular, the progress we make in handling both algorithmically,
as well as from the universality point of view, the non-unitary 
representations, is an important step towards understanding the 
quantum complexity of the Potts model problem.  This is the first
non-trivial progress on this problem since it was posed, more than a
decade ago, as a challenge to quantum algorithms.  Many open
problems are raised by this work, other than the clarification of
the complexity of the Potts parameters.  A particularly interesting
one is finding other scenarios where (non-unitary) representations
of algebras can be used to derive efficient quantum algorithms for
combinatorial problems. 

\end{abstract}

\section{Introduction} 

In the search of the past decade for fast quantum algorithms for
problems for which no efficient classical algorithm is known,
several problems have been marked as good candidates.  The most
commonly known one is the graph isomorphism problem, but much effort
was devoted also to lattice problems, such as the shortest and
closest vector in a lattice problem, and to other problems.
Unfortunately, no progress was made in any of these questions on the
algorithmic front (but see interesting complexity theoretic results
\cite{ref:lattices, ref:gi}.) 

One of the problems that attracted the scientific community effort 
was that of the approximation of the Potts model partition 
function. The Potts model (also known as the $q$-state Potts model)
is a famous model originating from statistical physics
\cite{ref:Wu82}, which is a generalization of the Ising model
\cite{ref:Isi25} to more than two components. It was proposed by
C.~Domb and his then research student R.~B.~Potts almost five
decades ago \cite{ref:Pot52}, and has since become a rich and
active area of statistical physics.

One considers an edge-weighted graph (in this paper we consider only
planar graphs), the nodes of which can be colored in one out of $q$
colors. Each color configuration of the nodes, $\sigma$, is given an
{\it energy} $H(\sigma)$ which depends on how many edges in the
graph are monochromatic when colored by $\sigma$, and on the weights
of the edges. The probability for the system to be in a given
configuration is determined by the Boltzmann-Gibbs distribution,
$e^{-H(\sigma)/(k_BT)}$ where $T$ is the temperature and $k_B$ is the
Boltzmann constant.  This model captures the essentials of many
physical systems related to solid states, and physicists and
mathematicians are therefore very interested in the properties of
the model .

It turns out that almost all important properties of the Potts
model, as well as other statistical physical systems, can be derived
from a certain quantity called the {\it partition function} of the
system.  This is given by a sum running over all possible colorings
of the graph, of the Boltzmann-Gibbs distribution (which is
unnormalized).  The problem of evaluating, or even approximating,
the Potts model partition function, has become a very important
problem in mathematical physics. 

The Potts model partition function is well known to be a special
case of the \emph{Multivariate Tutte Polynomial} of a graph
\cite{ref:Sok05}.  This is a polynomial that can be defined for
every edge-weighted graph $G=(V,E)$ and an additional variable $q$
that is related to the number of colors in $q$ of the Potts model.
It is conventionally denoted by $Z_G(q,\Bv)$ with $\Bv=\{v_e | e\in
E\}$ being the set of weights of all edges. The multivariate Tutte
polynomial is a generalization to more than two variables of the
well-known Tutte polynomial \cite{ref:Tut47}. The latter captures an
extremely wide range of interesting combinatorial properties of
graphs. It also generalizes the Jones polynomial of alternating
links, the reliability of a network, the number of spanning trees,
and more. As in this paper we will mainly work with the multivariate
Tutte polynomial, we will occasionally refer to it simply as the
``Tutte polynomial''. The original Tutte polynomial will be referred
as the ``standard'' Tutte polynomial.

The exact evaluation of the standard Tutte polynomial for planar
graphs turns out to be $\#P$ hard at all but several trivial points
\cite{ref:Jae90}.  But what about approximations?  Researchers have
devoted much effort to the attempts of providing approximating
algorithms for the Tutte polynomial at various points, and in
particular, for the Potts model partition function.  One is
ultimately interested in providing an FPRAS (fully polynomial
randomized approximation scheme), which gives as good an
approximation as one desires with polynomial overhead.  The most
common approach for this matter is that of the celebrated Markov
chain Monte Carlo method \cite{ref:Har05,ref:Wan01}.  For high
temperatures, it is known that this approach works for the 
ferromagnetic case (where the edge weights are positive), and so an
FPRAS exists \inote{we must give a ref here to support this claim.
It is strange that such ref is not found in the \cite{ref:Gol06}
paper}.  On the other hand, it is known that an FPRAS is NP hard to
achieve for the anti-ferromagnetic case \cite{ref:Gol06}, and in
fact, this hardness result holds for about three quarters of the
Tutte plane, so to speak.  The question of achieving an
approximation for the ferromagnetic case at various points, as well
as for other points in the Tutte plane for which an FPRAS might
still be possible, is an extremely important question. We are not
aware of any complexity theoretic restrictions that would imply such
an approximation unlikely; see Ref.~\cite{ref:Gol06}.

\subsection{Results} 

Our first result is an \emph{additive}, rather than multiplicative,
quantum approximation algorithms for the standard Tutte plane at any
point, including all points corresponding to the Potts model at any
number of colors and any temperature, and any set of weights on the
edges. Roughly speaking, an FPRAS algorithm approximates the
quantity $X$ by a number within the range $[X-X/poly(n),
X+X/poly(n)]$, with $n$ being the order parameter of the problem. On
the other hand, our additive approximation scheme provides an
approximation within the range $[X-\Delta/poly(n),
X+\Delta/poly(n)]$, with $\Delta$ being some parameter that can be
easily calculated from the input. We call $\Delta$ the
\emph{approximation scale} of the problem. Obviously, if $\Delta =
\mathcal{O}(X)$ then our scheme is equivalent to FPRAS - but this is
usually unknown.

\begin{thm}[Quantum Algorithm, rough version]
\label{thm:alg:rough}
  There exist an efficient quantum mechanical algorithm for the
  following problem. The input is a planar graph, with (complex) 
  weights on the edges, and a (complex) number $q$. The output is an
  additive approximation of the (multivariate) Tutte polynomial of
  the graph with those weights. 
\end{thm} 

The approximation scale of the algorithm is not specified in this
rough version of the theorem; it requires some prior definitions
before we can state it, and so it will be given later on. 

The size of the approximation scale is crucial when one considers 
additive approximations. This is because one has to be convinced 
that the size of the window is not exponentially larger than the
scale of the problem - otherwise the problem the algorithm solves is
trivial. Unfortunately, it is quite difficult to give bounds on the
scale of the problem given the hardness to approximate it. 

To show that in many cases our algorithms do not fall
into the trivial category, we would like to show that the 
problems the algorithms solve are hard. 
We start by proving: 

\begin{thm} [$\BQP$ hardness, rough version]
\label{thm:hard:rough} 

  There exists a wide range of complex weights and complex values of
  $q$, for which the additive approximation of the multivariate
  Tutte polynomial to within a certain scale, is $\BQP$-hard.
\end{thm}  
As in the previous theorem, the exact size of the approximation
window and the precise definition of the complex weights and $q$ for which 
the theorem holds, will be given later on. 
Examples for such parameters are
\begin{itemize}
  \item $q=3$, and the weights 
    $\left\{ 3\left(e^{\pi i/3}-1\right)^{-1}, \ 
             3\left(1-e^{-\pi i/3}\right)^{-1},\ 
             e^{\pi i/3}-1,\  1-e^{-\pi i/3} \right\}$ \ ,

  \item $q=2i$, and the weights 
    $\left\{ 100, -2i-100, 1, -1/2 \right\}$ \ ,

  \item $q=3$, and the weights 
    $\left\{ 100, -103, 1, -1/2 \right\}$ \ .
\end{itemize}
We shall see later that these three examples actually fall into
three different classes of weights that we call \emph{unitary
weights}, \emph{non-unitary complex weights} and \emph{non-unitary
real weights}. Their names imply the type of operators that they
define. 

The proof of universality of the parameters which correspond to
unitary operators, includes as a special case the universality proof
of the approximation of the Jones polynomial
\cite{ref:Aha05,ref:Fre02,ref:Aha06}. Indeed, the proof of the
unitary case follows closely the proof of \cite{ref:Aha06}. The
result, however, applies to many more parameters than the Jones
related corresponding result \cite{ref:Aha06}.  Essentially, the
generalization comes from the following fact.  An edge weight in our
construction corresponds to two different matrices, depending on the
orientation of the edge.  In the Jones polynomial case, those two
matrices need to be unitary.  When we consider the multivariate
Tutte polynomial, we can use two different edge weights, and
therefore we can relax this restriction and require that only one of
the matrices corresponding to each of these weights is unitary. This
gives many more parameters for which the result holds. This will be
explained later on. 

Proving $\BQP$ hardness for the complex non-unitary case requires
much more extra work, compared to the unitary case.  In fact, even
the very notion of universality when the generators are non unitary
seems counter intuitive, and it is not clear how to even start.  The
proof follows very roughly the same outline of \cite{ref:Aha06}, but we
need to develop many new tools that enable us to deal with the
non-unitarity of the operators; we will elaborate on that later on.
The proof for the case of real non-unitary matrices requires yet
another separate treatment since now universality is interpreted as
density and efficiency in the orthogonal group rather than the
unitary group, which poses more obstacles. 

Comparing Theorem \ref{thm:alg:rough} and Theorem
\ref{thm:hard:rough}, we see that we get complementary results: an
algorithm (for unrestricted parameters) and a hardness result (for a
restricted set of parameters).  We would like to deduce that the
problems for which both results hold, are $\BQP$ complete.  However,
for this to hold, we need the approximation windows in both results
to match. It turns out that for the parameters of the unitary case,
this happens without any extra effort. Therefore the unitary set of
weights provides us with a wealth of new $\BQP$-complete problems. 

For the non-unitary cases, this is not automatically true; due to
reasons that we will see later on, the approximation scale of the
algorithm is larger than what is required for universality, and
thus, the algorithm solves an easier problem.  To match the two
problems, we modify the definition of the problem in a somewhat
artificial way, but that does not seem to damage the main point of
the result. In the newly defined problem, the input is a pair of a
planar graph, together with a \emph{partition} of its edges into
groups. The approximation scale is then defined using this
partition, and it is for this problem that we are able to show
completeness. 

\begin{thm}[Completeness, rough version]
\label{thm:comp:rough}
  The problem of approximating the Tutte polynomial of a given
  planar graph with a given partition of its edges, to within an
  additive approximation scale which is defined by the input graph
  and partition, is $\BQP$ complete. 
\end{thm}

Consequently, we get a wide range of new $\BQP$-complete problems,
corresponding to unitary and non-unitary parameters of the Tutte
polynomial. 

We remark that unfortunately, our universality proof does not hold
for the Potts model parameters, (regardless of the size of the
approximation window.) This raises the question, of whether our
algorithms perform any non-trivial task in this range of parameters.
We provide very weak evidence for non-triviality, by analyzing the
performance of the algorithm in the case of a family of graphs for
which we can easily calculate the Potts model: the line, with
various edge weights. We see that the algorithm's approximation 
window is such that it distinguishes, with high probability, 
between the Potts partition functions for the graphs with different
weights.  Of course, for those
graphs the exact Potts model can be calculated exactly efficiently,
and so this cannot be regarded as a proof of non-triviality.  The
very interesting open question of characterizing the quantum
complexity of the physical Potts problem is left open for future
research.  

We proceed to outline the main new ideas introduced in this work.

\subsection{Main Ideas in the Algorithm}

The work is divided into two unequal parts: the algorithm and the
universality proof. We start with the main ideas underlying the
algorithm. 

In the work of Aharonov, Jones and Landau regarding the Jones
polynomial \cite{ref:Aha05}, a special case of our current problem
was solved.  One can view the solution there as follows: One
considers a braid, and represents it in an algebra spanned by
pictures similar to braids but with no crossings, like those in
Section \ref{sec:kauffman:tl}.  This algebra is called the
\emph{Temperley Lieb algebra} (see Ref~\cite{ref:Tem71} and \S 12.4
in \cite{ref:Bax82}), and is denoted by $TL_n(d)$, where $n$ is the
number of strands in the braid.  It turns out that the Jones
polynomial is in fact equal to a certain function called the {\it
Markov trace} of the Temperley Lieb element that corresponds to the
given braid, and this trace function satisfies a certain property
called the {\it Markov property}.  To compute this trace, the idea
is to use the following fact: If a matrix representation of the
$TL_n(d)$ algebra can be assigned a weighted trace which satisfies
the Markov property, then it will be equal to the Markov trace of
the corresponding Temperley Lieb elements. It thus suffices to
compute this weighted trace of the matrices by a quantum computer.
Fortunately, there are known unitary representations of braids,
induced by the so called path representation of the Temperley Lieb
algebra.  Thus, the computation can be performed quite easily by a
quantum computer: for each crossing in the braid, the algorithm
applies the corresponding unitary matrix.  The weighted trace of the
overall unitary matrix is easy to estimate using standard techniques
in quantum computation. 

We will build on the above method. We start by generalizing the
Temperley Lieb algebra that is used in \cite{ref:Aha05} to an
infinite algebra, which we denote by $GTL(d)$, where the number of
strands is not fixed; in the physics language, we allow creation and
annihilation operators.  This allows us to handle any graph, and not
just graphs originating from braids. What more, it allows us to
relax the requirement of the Markov property, and use \emph{any}
representation! This is because miraculously, once we deal with
representations of the more general algebra, it is no longer the
Markov trace of the Temperley Lieb element that corresponds to the
Tutte polynomial, but simply the overall factor multiplying the
identity element of the $GTL(d)$ algebra.

We are thus looking for ways to approximate this norm efficiently
using a quantum computer.  The next step is to apply the matrices
from the representation by a quantum computer.  The most important
apparent obstacle here is that it seems that one is restricted to
use unitary representations of the Temperley Lieb algebra if one is
to apply it by a quantum computer.  It turns out, however, that this
is not at all a necessary restriction.  We find a way to apply the
representation we use even when it is non-unitary. Our only
restriction is that the norm of the operators we apply needs to be
$1$; if it is not, we divide the whole operator by its norm, and
book keep the extra factor.  We lose in the approximation scale a
factor which is exactly this norm.  

This gives an efficient quantum algorithm that approximates the 
Tutte polynomial to within a given additive approximation, where the
scale of the approximation is an inverse polynomial times the
product of the norms of the operators it applies. 

We can perhaps understand better now the notion of additive
approximation in this context.  The norm of the overall product of
operators is what we are looking for: it is equal to the value of
the Tutte polynomial.  Clearly, this norm, the norm of the product
of operators, is smaller than the product of the norms. It is this
latter bigger quantity, the product of norms, that we view as the
\emph{scale} of the approximation. Our algorithm approximates the
overall norm, to within inverse polynomial times the scale.

\subsection{Main ideas in the proof of universality} 

We next show that finding an additive approximation of the Tutte
polynomial, to within a certain approximation scale is as hard as
quantum computation, for many weights and $q$'s. Note that we first
show hardness of approximation to within a smaller scale than our
algorithms provide; that is, we show universality of a slightly more
difficult problem, with a smaller scale. This leaves a gap between
the performance of the algorithm and the hardness result for most
cases; We resolve this issue later on. 
  
As a first step, we would like to show universality for a set of 
parameters which correspond to the case in which the operators 
being applied are unitary (we include in this case also operators
which are a scalar times a unitary).  The universality proof itself
follows quite closely the proof of universality of \cite{ref:Aha06},
except for issues related to the book keeping of overall factors.

The proof is significantly harder when we move to the non-unitary
case, where we have to overcome a true difficulty: The matrices we
are supposed to use in order to express any quantum circuit, are non
unitary!  In universality proofs, one usually shows density in the
unitary group, and then uses the Solovay-Kitaev theorem that shows
that density implies efficiency. In our case, both seemed to be
impossible; working with general linear operators, which apply
stretching and shrinking, how can one show density in the unitary
group? And moreover, even if density applies, how can one use the
Solovay Kitaev theorem? To explain how we overcome these
difficulties, let us first recall the main ideas in the proof of
\cite{ref:Aha06}. 

The main idea is to encode the $n$ qubit Hilbert space into the
space of paths on finite graphs the shape of a line. Then using the
path representation, elements of the $TL_n(d)$ algebras are mapped
to operators on that space. The paths are encoded by strings in
which $0$ means a step to the left and $1$ means a step to the
right.  This is not a tensor product space, but one can easily
encode a tensor product space into it; A path of length $4$ of the
form $0101$ corresponds to the state $|0\ra$, and a path of length
$4$ of the form $1100$ corresponds to the state $|1\ra$.  Thus,
paths of $4n$ steps include the encoding of $n$ qubits.  Two qubits
gates are encoded as transformations that work on $8$ steps paths.
We note that the $8$ steps paths include many paths that are not 
one of the $4$ legitimate $8$ steps paths that correspond to encoded
two qubits. What more, as we will see, those four legitimate paths
are not an invariant subspace, and so density will have to be proven
in the smallest invariant subspace that contains them - which is the
subspace of all $8$ steps paths that start and end at the first
vertex of the graph. There are $14$ such paths.  To approximate two
qubit gates, we will thus have to approximate a matrix in $SU(14)$.

In \cite{ref:Aha06}, the first building block is to show density in
$SU(2)$, on two of those $14$ paths. This is called {\it the seed}; 
It is pretty easy to establish.  Then one can use various tools to
build up the dimensionality of the dense subgroup, to get to the
$14$ relevant dimensions, in which all two qubit gates are encoded;
one main tool is the Bridge lemma \cite{ref:Aha06} which states that
given matrices which generate dense subgroups on two orthogonal
subspaces, and a matrix mixing those subspaces up (the ``bridge''),
the resulting group is dense on the direct sum of the two spaces.
Another tool is the Decoupling lemma: if we have density on two
unitary groups on two orthogonal subspaces of different dimensions,
then even though these two generated groups might in principle be
coupled or correlated, they are in fact decoupled and we can
continue as if we have universality on each one of them separately. 
 
Let us now tentatively explain how one modifies this proof to handle
non-unitary parameters. We start with the first building block,
namely, constructing density on a two dimensional subspace.  Unlike
in the unitary case, here we are led to prove density in the special
linear group $SL(2,\mathbbm{R})$ or $SL(2,\mathbbm{C})$ instead of
$SU(2)$.  In the quantum literature, as far as we know, density of
non-unitary operators was not dealt with before; in fact, the theory
here is very different than that in the unitary case.  Fortunately
one can use a wealth of results from theory of complex M\"obius
transformations for that matter. In particular, we use Jorgensen's
inequality \cite{ref:Jor76} to find sets of parameters for which
density holds. We then reprove the Bridge lemma and the Decoupling
lemma for the non-unitary case; once again there are some technical
issues which one has to deal with when considering non-unitary
matrices. Finally, we need to reprove the Solovay Kitaev theorem. 
In this theorem, much technical effort is given to the issue of
accumulated errors.  Now we have new errors to deal with since our
matrices are only close to, but not inside, the unitary group. For
that matter, we generalize the theorem for the non-unitary groups
$SL_n(\mathbbm{C})$ and $SL_n(\mathbbm{R})$.

We get that any two-qubit gate can be replaced by
polylogarithmically many $GTL(d)$ elements, which are mapped under
the path representation to an operator that acts on 8-steps paths.

The final result is a mapping of the quantum circuit to a planar
graph, such that the result of the circuit can be read from an
appropriate approximation of the Tutte polynomial of the graph at
the relevant parameters.

\subsection{Proof of $\BQP$ Completeness} 

In the hardness result, the approximation scale turns out to be the
same as the algorithmic scale only in the unitary parameters case.
In this case, completeness is already proved. 

To match the approximation scales of the algorithm and the
universality results also in the non-unitary cases, we note that our
algorithm in the non-unitary case, was quite wasteful in its
approximation quality. Recall that the size of the approximation
scale is the product of the norms of the matrices being applied.
However, note that the norm of the product of operators might be
significantly smaller than the product of the norms, i.e., the scale
of the problem.  It is thus beneficial to apply the operators in
groups, and not one by one. This can be done easily if the operators
we would like to apply in one group, all operate on a small
dimensional space in a sequence. In this case, before applying them,
we calculate their product by a classical computer on the side, and
apply the resulting product by the quantum computer. 

In the context of the graph described in before, corresponding to
the quantum circuit, there is a very natural grouping: Each gate in
the circuit was replaced by polylogarithmically many operators,
which we would like to view as one group.  If we could group all
operators corresponding to one quantum gate together, the resulting
operator, which is approximately a unitary operator over 8-steps
paths, has norm approximately $1$, and thus does not increase the
approximation scale at all.  We would have liked to do this, rather
than to apply the approximating sequence of gates one by one, which
may contribute a lot to the approximation window. 

Unfortunately, to be able to achieve this improved approximation
scale, the quantum algorithm needs to first recognize all operators 
coming from the same gate as belonging to one group. More generally,
the algorithm would need to find a good way to partition the 
operators. So far we have not been able to find an efficient
grouping algorithm that would achieve this goal.  We are thus led to
a less elegant solution: we modify the problem slightly. Our input
is now a pair of a graph together with a partition of its edges to
groups.  The output is the Tutte polynomial of the graph, to within
the approximation window defined by the given grouping.  When this
is done this way, the two approximation scales match, and we get
both $\BQP$ hardness as well as an algorithm.

The proof that this works is in fact quite involved, due to a subtle
point in the above argument, which causes a lot of trouble in the
proof. Note that the idea above relied heavily on the fact that the
norm of the product of all operators in one group is $1$ (or very
close to $1$). However, the operators we consider operate on the
direct sum of many subspaces (denoted $A_{k,l}$), whereas only the
first subspace, $A_{1,1}$, is relevant for the quantum computation.
It could be that the approximation is unitary in the relevant
subspace, whereas in the other subspaces the resulting product of
operators has a much larger norm; the quantum algorithm only
guarantees approximation to within this (possibly too large) norm. 
To really match the sizes of the approximation windows, we have to
somehow address this issue. 

Our solution is as follows. We make sure that the norm of each group
of operators is indeed approximately $1$, even when the entire space
is considered, by proving density and efficiency in the unitary
group simultaneously on all those subspaces. This requires some work
since we need to deal with the additional subspaces one by one.
Moreover, we need to generalize the Solovay Kitaev to hold also for
direct sums of unitary groups, and we can only show this for a
finite sum. Unfortunately, all the above manipulations imply that we
have to limit the parameters with which we work, and so the set of
points for which we get completeness seems smaller than that for
which hardness holds.

The above seems like a very complicated solution to a rather minor
and technical issue; we speculate that a better solution to this
technical point exists, which would enable to handle it in a more
elegant way. 

The final result is that for a large class of parameters, we can
show that the problems with the same approximation scale is both
doable in quantum polynomial time and also universal - namely, it is
complete.

\subsection{Remarks on the Complexity of the Potts model case}

Unfortunately, our methods are not applicable for showing that 
values of the graph weights and $q$ that correspond to physical
Potts model parameters, are $\BQP$ hard, (for any additive
approximation).

The reason is essentially this. The partition function of physical
parameters is, by definition, a function that corresponds to some
Gibbs distribution, and is thus, always positive. Our methods,
however, translate a unitary matrix to a graph, such that, say, the
first entry of the unitary matrix is equal to the partition
function. Proving universality would mean that any unitary matrix
can be presented this way, including unitary matrices whose first entry is
negative. But there is no graph whose Potts model partition function 
would be negative, as it is a positive number!

The exact argument is slightly more complicated, and will be given
in a later version of the paper. The conclusion is that at least
with our methods, $\BQP$ completeness seems impossible to prove.
Perhaps, $\BQP$ completeness does not hold.

As we mentioned, we provide weak evidence for non-triviality in a 
different way. We consider two families of weighted graphs for which 
the evaluation of the Potts model can be done efficiently classically. 
The two families have the same underlying graph: the line of $n$ sites. 
However, we consider this graph with different weights on the edges. 
It is quite easy to
show that the quantum algorithm distinguishes between the Partition 
functions in the different cases, namely, the window in which one answer might 
lie does not intersect the window in which the other answer lies. 
It is unclear how well the algorithm performs 
for more difficult graphs, in which no efficient classical algorithm is 
known. 

Proving anything about the complexity of our algorithm for the Potts
model, remains a very
important open problem. It is still possible that this case of 
the Tutte polynomial, with our additive approximation window, 
can be solved by an efficient classical algorithm.

\subsection{Conclusions and Open Questions} 

In this work we provide new quantum algorithms for many new
problems, all casted in the framework of approximations of the Tutte
polynomial.  Our quantum algorithms are distinctly different from
the well trodden path of quantum algorithms for the Hidden Subgroup
problem.  They take the approach of \cite{ref:Aha05} in which local
structures of the problem are encoded into linear operators which
are then applied; however, we take a significant step forward, by
leaving the restriction of unitarity behind, which allows us to
apply our algorithms and the hardness results to many new points. 

Our methods rely on representations of a generalized version of the
Temperley Lieb algebras, which we define.  We believe that the main
achievement here is that we demonstrate how to handle non-unitary
representations, and in particular, we are able to prove
universality using non-unitary matrices.  This might open up the way
to applying non-unitary representations in completely different
contexts, and perhaps will serve as an important advancement in the
understanding of the physical Potts model case, which remains as the
main open question.

In our proofs of completeness, we have only given examples in which 
$\BQP$ completeness can be proved. In fact, many more points can
also be shown to be $\BQP$ complete; In this work we have only taken
the effort of showing a few examples, but the characterization of
which points exactly can be proven to be quantum hard or complete is
left for a further study.

A far reaching hope is that progress in the direction of the 
hardness of the Potts model parameters, would lead to insights 
regarding open questions in the field of statistical physics, and in
particular, understanding the conditions for rapid mixing in the
Potts model, and its relation to the phase transition of the Potts
model.  

Another very interesting question is whether it is possible to find
an algorithmic way to calculate the optimal grouping, that would
derive the best approximation window, or even some approximation of
it.  This would allow us to define the problem in a more natural
way, without artificially providing the partition in the input. 

One more interesting question is whether these results can be
generalizes to non-planar graphs.

Last but not least, it would be extremely interesting to see our
methods applied to other approximation algorithms of $\#P$ complete
problems, which are not described by the Tutte polynomial. 

{~}

\noindent\textbf{Organization of paper:}\\

\indent In \Sec{sec:background:tutte} we start with some background
on the Tutte polynomial and its connection to the Potts model and to
the Jones polynomial. We proceed in \Sec{sec:tutte:kauffman} to
establish a well-known mapping of a planar graph to its \emph{medial
graph}, which is a 4-regular graph. On the latter, we define the
\emph{Kauffman bracket}, which is equivalent to the multivariate
Tutte polynomial of the original graph, and which we will mainly
use. Section~{sec:kauffman:tl} defines our generalization of the
Temperley Lieb algebras, and provides the way to write the Kauffman
bracket in terms of a Temperley Lieb algebra scalar element. Section
\ref{sec:pathmodel} gives the exact representations we will use,
called the path model representations. Section~\ref{sec:algorithm}
first states the exact version of Theorem~\ref{thm:alg:rough},
together with the definition of the size of the approximation
window, and then provides the algorithm and proves that it works.

In the second part of the paper, we prove the results concerning the 
hardness and completeness. 
Section~\ref{sec:universality} proves the universality proof for two
cases: one is the case where the parameters are such that the
matrices are almost unitary, namely, they are a scalar times a
unitary.  This already provides many new points in the Tutte plane,
but is essentially the same proof as in the Jones case
\cite{ref:Aha06}. Then we provide the universality proof for the
non-unitary cases (both complex and real). It is here that we need
to use the novel ideas of how to apply the Solovay-Kitaev theorem
despite not being in the safe zone of unitarity. Finally, in
\Sec{sec:complete}, we define the version of the problem in which
the grouping of the graph is also provided, and prove the
completeness of this version.

\section{Background: The Tutte polynomial}\label{sec:background:tutte} 

\subsection{The Tutte polynomial}

The multivariate Tutte polynomial (which will often be referred here
simply as the ``Tutte polynomial'') is defined for finite
graphs with weighted edges and a scalar $q$. Below we give
a short definition and description of this important polynomial. Our
treatment and notation closely follow Ref~\cite{ref:Sok05}, which we
strongly recommend as an introductory text for this subject.
 
Given a graph $G=(V,E)$ and a subset of edges $A\subseteq E$, we
define $k(A)$ to be the number of connected components in the
subgraph $(V,A)$. Notice that an isolated vertex is considered a
connected component, hence when $A$ is empty, $k(A)=|V|$. The
multivariate Tutte polynomial is then defined by

\begin{deff}[The multivariate Tutte polynomial]
\label{def:MTP}
  Let $G=(V,E)$ be a finite graph with variables $\Bv=\{v_e\}$
  assigned to its edges $e\in E$. Then the multivariate Tutte
  polynomial of $G$ is a polynomial in $\Bv$ and an extra variable
  $q$, defined by
  \begin{equation}
    \label{def:Z}
         Z_G(q;\Bv) = \sum_{A\subseteq E} q^{k(A)}\prod_{e\in A}v_e \ .
  \end{equation}
  When considering a particular set of values of the variables
  $\Bv=\{v_e\}$, it is customary to call them weights, and $G$ a
  weighted graph. In addition, it is common to substitute the
  variables $\Bv=\{v_e\}$ with a single variable $v$ and obtain a
  two-variables polynomial
  \begin{equation}
    Z_G(q,v) = \sum_{A\subseteq E} q^{k(A)}v^{|A|} \ .
  \end{equation}

\end{deff}

As an example, consider the graph $G=$ \includegraphics[bb=0 20 80 80,
scale=0.3]{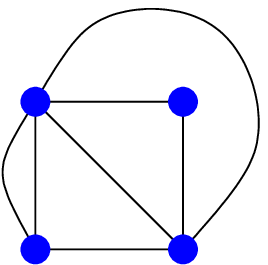} with all edges set as $v_e=v$. Its expansion
contains terms like these:
\begin{itemize}
  \item \includegraphics[bb=0 20 80 80, scale=0.5]{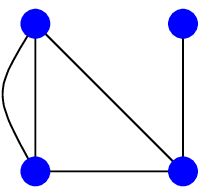}
    $k(A) = 1, |A|=5 \Longrightarrow \mbox{Term} = qv^5$
  \item \includegraphics[bb=0 20 80 80, scale=0.5]{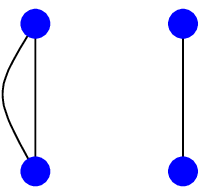}
    $k(A) = 2, |A|=3\Longrightarrow \mbox{Term} = q^2v^3$
  \item \includegraphics[bb=0 20 80 80, scale=0.5]{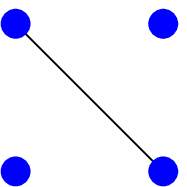}
    $k(A) = 3, |A|=1 \Longrightarrow \mbox{Term} = q^3v$
\end{itemize}

\ignore{
Let us also mention an alternative formula for the MTP. If
$c(A)$ is the number of independent cycles in the graph $(V,A)$ then
it is easy to see that $|V| + c(A) = |A| + k(A)$. Therefore we can
replace $k(A)$ in \Eq{def:MTP} and get
\begin{equation}
  Z_G(q;\Bv) = \sum_{A\subseteq E} q^{|V|+c(A)-|A|}\prod_{e\in A}v_e 
  = q^{|V|}\sum_{A\subseteq E} q^{c(A)}\prod_{e\in A}\frac{v_e}{q} \ .
\end{equation}
}

\subsection{Relation with the standard Tutte polynomial} 

When the variables $\Bv=\{v_e\}$ are all equal to the variable $v$,
it is easy to relate the resulting polynomial $Z_G(q,v)$ to the
standard Tutte polynomial $T_G(x,y)$ \cite[pp.~45]{ref:Wel93} which is defined by
\begin{equation}
\label{def:standard-Tutte}
  T_G(x,y) \EqDef \sum_{A\subseteq E}
      (x-1)^{k(A)-k(E)}(y-1)^{|A|+k(A)-|V|} \ .
\end{equation}
Indeed, a simple algebra yields
\begin{equation}
  T_G(x,y) = (x-1)^{-k(E)}(y-1)^{-|V|}Z_G\Big( (x-1)(y-1), y-1\Big) \ .
\end{equation}
In other words, $T_G(x,y)$ and $Z_G(q,v)$ are essentially equivalent
under the change of variables 
\begin{eqnarray}
  x = 1+q/v \ , \quad y=1+v \ , \\
  q=(x-1)(y-1) \ , \quad v=y-1 \ .
\end{eqnarray}

\subsection{Relation with the Potts model} 

In this section we elaborate on the connection between the Potts
model and the Tutte polynomial. A broad review of this model, in a
more physics-theoretical context, can be found in
Ref~\cite{ref:Wu82}.

The Potts model, which is also known as the $q$-state Potts model,
is physical model that is defined on a graph $G=(V,E)$ for some
$q\in \mathbbm{Z}_+$. We call the vertices of the graph sites. They
hold physical objects like ``atoms'' or ``spins'', which can be in
one of $q$ possible states (colors, or spin states) $\{1,\ldots,
q\}$. A coloring of all sites is called a \emph{configuration} of
the system and is denoted by a map $\sigma:V\to \{1, \ldots, q\}$.

The energy of the system is defined by assigning each edge $e\in E$,
a coupling constant $J_e$. Then if $e_1$ and $e_2$ denote the
adjacent sites of $e$, and $\sigma(e_1),\sigma(e_2)$ denote their
colors respectively, the energy of the edge is
\begin{equation}
  \epsilon_e = -J_e\delta_{\sigma(e_1), \sigma(e_2)} \ ,
\end{equation}
where $\delta_{a,b}$ is the usual Kronecker delta. Therefore the
energy of an edge is $-J_e$ if its sites share the same color, and
is zero otherwise. The total energy of the configuration $\sigma$ is
now
\begin{equation}
  H(\sigma) = \sum_{e\in E} \epsilon_e  
    = -\sum_{e\in E} J_e\delta_{\sigma(e_1), \sigma(e_2)} \ .
\end{equation}

In statistical physics we assume that the system constantly changes
its microscopic configuration. The actual state of the system is
therefore of little importance, and instead we try to estimate the
probability $P(\sigma)$ of the system to be in a configuration
$\sigma$, and use this information to understand the global
properties of the system. In this context, it is common to consider
the scenario in which the system is attached to another, much bigger
system (a thermal bath), with a constant temperature $T$, and the
two systems are allowed to exchange energy (heat). Then the system
is described by the so-called \emph{canonical ensemble}.  In this
case, the probability distribution of the system is the
Boltzmann-Gibbs distribution,
\begin{equation}
  \label{eq:dist}
  P(\sigma) = \frac{1}{Z^{\mbox{\tiny Potts}}} 
     e^{-\beta H(\sigma)} \ .
\end{equation}
Here $\beta=1/(k_B T)$ is the inverse temperature with $k_B$ being
the Boltzmann constant.  $Z^{\mbox{\tiny Potts}}$ is the
normalization factor which is called the \emph{partition function}
of the system, and is given by
\begin{equation}
  Z^{\mbox{\tiny Potts}} \EqDef \sum_{\sigma} e^{-\beta H(\sigma)} \ .
\end{equation}
The partition function $Z^{\mbox{\tiny Potts}}$ has a central role
in statistical physics. Knowing it allows us to calculate important
global properties of the system such as its average energy, entropy,
heat capacity etc. 

For a positive temperature $T>0$ (equivalently $\beta>0$) and real
couplings $J_e$, the total energy $H(\sigma)$ is a real number,
hence \Eq{eq:dist} is a valid probability distribution. In such
case, configurations with low energies are preferable. An edge with
$J_e>0$ is called ``ferromagnetic''. Its adjoint sites will prefer
to be mono-color. On the other hand, an edge with $J_e<0$, is called
``anti-ferromagnetic'', and its adjoint sites will tend to have
non-identical colors. An interesting case is when all sites are
anti-ferromagnetic and $T\to 0$ (equivalently $\beta\to+\infty$). In
such case, if there are configurations of the graph where adjacent
sites have different colors (i.e., ``legal coloring''), then only
those configurations will have a non-vanishing probability. In
particular, we find 
\begin{equation}
  \lim_{T\to 0} Z^{\mbox{\tiny Potts}} = \mbox{\# of $q$-colorings} \ .
\end{equation}
Deciding whether a graph has one or more $q$-colorings for $q>2$ is
a well-known NP-complete problem. Hence the above example
suggests that calculating, or even approximating the partition
function, can be, at least in some cases, a highly non-trivial task.
Currently, most common approximations to this quantity use
Monte-Carlo type algorithms (See Refs.~\cite{ref:Har05, ref:Wan01},
for example).

If we now define
\begin{equation}
\label{eq:v-J}
  v_e = e^{\beta J_e} - 1 \ ,
\end{equation}
then a simple algebra shows that the partition function can be
written as 
\begin{equation}
  Z^{\mbox{\tiny Potts}}(q,\Bv) = \sum_\sigma \prod_{e\in E}
    \left[1+v_e\delta_{\sigma(e_1),\sigma(e_2)} \right] \ .
\end{equation}
It is far from obvious that $Z^{\mbox{\tiny Potts}}(q,\Bv)$, which
is defined separately for each positive integer $q$, is in fact a
polynomial in $q$, but as noted by Fortuin and Kastelyn in the late
1960's \cite{ref:For72, ref:Kas69}, this is indeed the case:
\begin{thm}[Fortuin-Kasteleyn representation of the Potts model]
\label{thm:For-Kas}
  For integer $q>1$, and for all $\Bv$, 
  \begin{equation}
    Z^{\mbox{\tiny Potts}}(q,\Bv) = Z_G(q,\Bv).
  \end{equation}
\end{thm}
A proof of this theorem can be found in Ref.~\cite{ref:Sok05}.

Approximating the partition function of the $q$-state Potts model is
therefore equivalent to approximating the multivariate Tutte
polynomial of the underlying graph with weights given by
\Eq{eq:v-J}. The complexity of the two problems is identical.

Theorem~\ref{thm:For-Kas} enables us to assign a probability to
every coloring of a graph when integer $q>0$ and $v_e>-1$ (which is
equivalent to $J_e$ being real for real temperatures). We can
actually extend this probabilistic interpretation for non-integer
$q$'s by using the original definition~(\ref{def:MTP}). The idea is
to view the elements of sum in \Eq{def:Z} as the new configuration
space. In other words, a configuration is defined by a subset of the
edges $A\subseteq E$, and is assigned a probability which is
proportional to $q^{k(A)}\prod_{e\in A}v_e$.
Theorem~(\ref{thm:For-Kas}) guarantees that these two definitions
coincide in the appropriate region.  We are thus led to define
\begin{deff}[Potts parameters]
  We say that ${\Bv},q$ are Potts parameters (or physical
  parameters) when
  \begin{itemize}
    \item $q$ is an integer larger than $0$, and for all edges
      $v_e>-1$; or
    \item $q>0$  and for all edges $v_e>0$.
  \end{itemize}
\end{deff}
Notice that the conditions of the second case guarantee that
$q^{k(A)}\prod_{e\in A}v_e$ is real and positive. Also note that in
the first case, $J_e >0$ (ferromagnetic case) corresponds to
$v_e>0$, and $J_e<0$ (anti-ferromagnetic) corresponds to $-1<v_e<0$.
In both cases, we are assured that $Z_G(q,\Bv)$ is a real positive
number.

\subsection{Relation with the Jones polynomial} 

Another interesting special case of the multivariate Tutte
polynomial of planar graphs is the Jones polynomial. By constructing
the so-called ``medial graph'', which will be defined in the next
section, one can translate a planar graph to a knot in the three
dimensional space. Furthermore, for a particular choice of weights
and $q$, there is a simple connection between the Tutte polynomial
of the original graph and the Jones polynomial of the knot. We will
elaborate on this connection in a later version of this paper.

\section{Background: From the Tutte polynomial to the Kauffman brackets}
\label{sec:tutte:kauffman}

The Tutte polynomial of a planar graph $G$ can be recast in terms
of \emph{Kauffman brackets} of the \emph{medial graph} $L_G$. The
medial graph is constructed from the original graph in some
well-defined way. This definition of the Tutte polynomial turns out
to be more convenient for our purposes. 

We first define the medial graph.  For each planar graph $G$ we can
build 4-regular graph $L_G$ which is called the \emph{medial graph}
of $G$. It is obtained from $G$ by first encircling the facets of
$G$ with lines, and then crossing the lines that surround each edge
by putting a 4-regular vertex in its middle. 

\begin{figure}
  \center \includegraphics[scale=0.5]{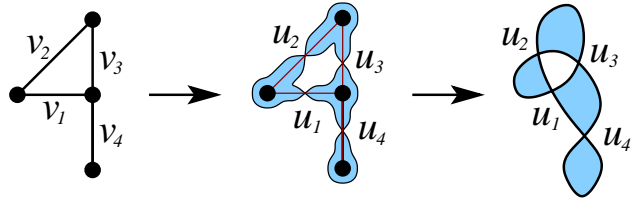} \caption{From a
  planar graph to a medial graph. In the planar graphs, the weights
  $v_i$ are assigned to the edges. These weights are then
  transformed to the \emph{crossing} weights $u_i$ in the medial
  graph, which are related by $v_i = du_i$. \label{fig:medial}}
\end{figure}

The regions of the medial graph can be colored in a consistent way
by black and white, such that the regions on two sides of an edge
are colored by opposite colors.  Its black and white coloring is
uniquely determined by setting its outmost area to be white.  It is
easy to convince oneself that this coloring is unique and well
defined. An example for such construction is given in
\Fig{fig:medial}. 

The vertices of the medial graph, which we also call
\emph{crossings}, are assigned weights, denoted $\Bu=\{u_e\}$, where
$u_e$ is the weight of the crossing corresponding to the edge $e$ in
$G$. If $G$ is a weighted graph, then one can define a natural,
one-to-one mapping between its weights $\Bv=\{v_e\}$ to the weights
$\Bu=\{u_e\}$. The nature of this mapping will be clear shortly, when
we define the Kauffman bracket of a medial graph.

To define the \emph{Kauffman brackets} of $L_G$, we first define the
notion of a \emph{state}.  A state $\sigma=\{\sigma_e | e\in E\}$ is
an assignment of $0$ or $1$ to every crossing in $G$. $\sigma$ is
understood as a prescription of how to open every crossing
\includegraphics[bb=-10 20 70 50, scale=0.3]{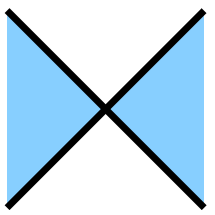} in $L_G$ in
one of two ways: \includegraphics[bb=-10 20 70 60,
scale=0.3]{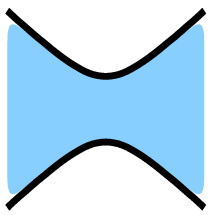} or \includegraphics[bb=-10 20 70 60,
scale=0.3]{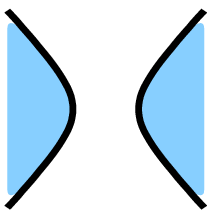}. When $\sigma_e=1$ the shaded areas are
connected (and consequently the original edge $e$ of $G$ is included
in the resulting shaded area), while when $\sigma_e=0$ the shaded
areas are disconnected (and the edge is not included in the shaded
area). Once all crossings have been opened, the resulting diagram
contains only loops. We let $|\sigma|$ denote their total number. 

The Kauffman brackets of $L_G$ is defined as the following sum
\begin{equation} 
  \langle L_G \rangle = \sum_{\sigma} \langle L_G |\sigma \rangle \ .
\end{equation}

where $\langle L_G |\sigma \rangle$ is defined by
\begin{equation}
  \langle L_G |\sigma \rangle = d^{|\sigma|}\prod u_e^{\sigma_e} \ .
\end{equation}
In other words, we take the product of $d^{|\sigma|}$ with the $u_e$
of all crossings that were opened in such a way that their edge was
left alive. We note that our definition of the Kauffman bracket is
different from the standard definition of the Kauffman brackets,
which is originated from the theory of knots. The main difference is
that in our definition we allow a different weight $u_e$ for each
crossing, whereas in the standard definition only one weight is
used. It might therefore be more appropriate to call this object the
``weighted Kauffman bracket''. However, for the sake of brevity and
at the risk of a slight confusion, we will continue to call it
simply the Kauffman bracket. A definition of the standard Kauffman
brackets and their relation to Knots theory, the Jones polynomial,
the Potts model, and many other topics in mathematical physics can
be found in Refs~\cite{ref:Kau87, ref:Wel93, ref:Kau01}.

We will now show the connection between the Kauffman bracket of $L_G$ 
and the multivariate Tutte
polynomial of $G$.
\begin{claim} 
  \label{cl:connection}
  For any scalar $d$ and a set of weights $\Bu=\{u_e\}$, the
  following equation holds:
  \begin{equation}
  \label{eq:connection}
    \langle L_G \rangle(d,\Bu) = d^{-|V|} Z_G(d^2, d\Bu) \ .
  \end{equation}
\end{claim}

\begin{proof} 
The proof is an adaptation of well-known ideas due to Kauffman from
the subject of the Jones polynomial in Knots theory
\cite{ref:Kau87}. Each state $\sigma$ of the medial graph $L_g$
uniquely defines a subgraph of $G$ whose edges are those with
$\sigma_e=1$. Denote this set of edges by $A\subseteq E$. At the
same time $\sigma$ corresponds to a particular opening of the
crossings. The opening is such that each facet of the subgraph
$(V,A)$ is encircled by exactly one loop. Therefore the number of
loops, $|\sigma|$, is exactly the number of facets in $(V,A)$. Now
the number of facets in $(V,A)$ is simply $k(A) + c(A)$, where
$c(A)$ is the number of circles in the subgraph. Using the identity
$|V| + c(A) = k(A) + |A|$ (see, e.g, \cite{ref:Sok05} Sec.~2)
\znote{Can't say I really followed but I think what must be invoked
is Euler's formula which says for a planar graph V-E+F= \# of
connected components-- this seems to imply that the $k(A) + c(A)$
should have just been $c(A)$ above. . . does this resonate with
anyone who really I understands the argument?}we have
\begin{equation}
  |\sigma| = k(A) + c(A) = 2k(A) + |A| - |V| \ .
\end{equation}
Hence,
\begin{eqnarray}
  \langle L_G \rangle &=& \sum_{A\subseteq E} d^{2k(A) + |A| - |V|}
  \prod_{e\in A} u_e 
  = d^{-|V|}\sum_{A\subseteq E} (d^2)^{k(A)}
  \prod_{e\in A} (du_e) \\
  &=& d^{-|V|} Z_G(d^2, d\Bu) \ .
\end{eqnarray}
\end{proof} 

Therefore the Kauffman bracket of $L_G$ is proportional to the Tutte
polynomial of $G$ with $q=d^2$ and $\Bv=d\Bu$.

\section{The Kauffman brackets in terms of the Generalized Temperley
Lieb algebras}\label{sec:kauffman:tl} 

Our first goal is to design an algorithm that calculates Kauffman
brackets of a given medial graph. How can the Kauffman bracket be
evaluated?  The main mathematical tool is the Temperley Lieb
algebras. It enables us to break the medial graph into small pieces,
which are then translated to local operators. Here, we use a
generalization of the well-known Temperley Lieb algebras to an
algebra of pictures with an arbitrary number of strands, so that we
can talk about product of elements with non-matching numbers of in-
and out-going strands. We call it the $GTL(d)$ algebra. It is this
definition that enables us to deal with all planar graphs, and not
just graphs that are related to braids - which have a fixed number
of strands.

\subsection{The Generalized Temperley-Lieb algebra $GTL(d)$} 
\label{sec:GTL}

\begin{figure}
    \center \includegraphics[scale=0.4]{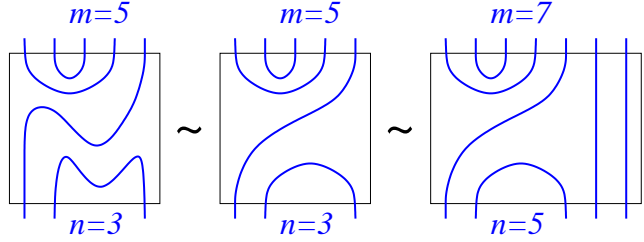} \caption{Three
      different diagrams that represent the \emph{same} $GTL(d)$
      basis element. The first two have 3 in-pegs and 5 out-pegs and
      are isotopic to each other. The third one has 5 in-pegs and 7
      out-pegs and is a result of adding two straight strands to the
      right of the second diagram.} \label{fig:tangle}
\end{figure}

The $GTL(d)$ algebra is an algebra that is generated by diagrams of
the form that is shown in \Fig{fig:tangle}. Its precise definition
is as follows:

\begin{deff}[The $GTL(d)$ algebra]
\label{def:GTL}
  The $GTL(d)$ algebra is an infinite dimensional algebra that is
  defined for every scalar $d$. Its basis elements are represented
  classes of diagrams that are made from a finite number of strands
  that connect $n$ lower pegs to $m$ upper pegs. The diagrams must
  not contain any crossings and loops, but they may contain local
  minima/maxima. Two diagrams, belong to the same class if they are
  topologically equivalent, or if one diagram can be obtained from
  the other by adding a number of straight strands to its right. See
  \Fig{fig:tangle}. We will denote algebra elements by capital calligraphic
  letters such as $\mcT, \mcA, \mcB, \ldots$, and diagrams by Gothic letters
  $\mfT, \mfA, \mfB, \ldots$. Notice that every $GTL(d)$ basis element is
  represented by an infinite number of diagrams, but every diagram
  represents only one basis element.

  \begin{description}  
    \item [Product Rule:] For every two basis elements $\mcT_1,\mcT_2$, the
      product $\mcT_3=\mcT_2\cdot \mcT_1$ is defined by the diagram
      that we obtain by placing a diagram that represents $\mcT_2$
      \emph{on top} of a diagram that represents $\mcT_1$ and
      matching their pegs \emph{starting from the left}. If the
      number of lower pegs in $\mcT_2$ is different than the number
      of upper pegs in $\mcT_1$ we pad one of them with straight
      strands until they match.
      
    \item [Identity Element:]
      From the above definition it is easy to see that the identity
      element of the algebra, $\mcI$, corresponds to diagrams with
      only straight strands, in which the $i$'th bottom peg is
      connected to the $i$'th upper peg.

    \item [Relations:] {~}
      \begin{itemize} 
        \item Two isotopic elements in the $GTL(d)$ algebra are 
              equal.
        \item Two elements in the $GTL(d)$ algebra are
              equal if a diagram of one element can be obtained from
              a diagram of the other element padded with straight
              strands at its right side.
        \item A single loop is equal to the scalar $d$ times the
              identity $\mcI$. 
              The parameter $d$ is a (complex) number which is a
              fixed parameter of the algebra, and is called the \emph{loop value}.
              
        \item An element in the algebra which contains (but is not isotopic to) 
              an empty loop, is equal to the same element, except that the
              loop is removed, and the result is multiplied by the loop value $d$.
      \end{itemize}  
      
  \end{description}

\end{deff}

We extend the $GTL(d)$ algebra to linear combinations of basis
elements by extending linearly the product rule.

\subsection{Tangles and elementary tangles}

It will be convenient to talk also about \emph{tangles}.  These are
diagrams, which are similar to the diagrams that represent basis
elements of the $GTL(d)$ algebra, except that they may also contain
loops and weighted crossings.  We use them to denote certain linear
combinations of basis elements in $GTL(d)$. 

Following the product rule in Definition~\ref{def:GTL}, a diagram
with $\ell$ loops denotes the $GTL(d)$ basis element of the corresponding
diagram without the loops, times a factor of $d^\ell$.

We use a tangle with exactly one crossing and a complex weight attached to it,
to denote a different linear combination of $GTL(d)$ elements. This
is done as follows: let $\mfT$ be such a tangle, with one crossing weighted 
with the 
weight $u$. We perform a black-and-white coloring of the tangle, 
starting from its left-most area, which is always painted in white.
We then look at the two possible ways of opening the crossing:
connecting or disconnecting the two shaded areas of the crossing.
Let $\mathcal{X}$ be the $GTL(d)$ element that corresponds to the
diagram in which the shaded areas are connected, and let
$\mathcal{Y}\in GTL(d)$ correspond to the diagram in which the
shaded areas are disconnected. Then we use the diagram $\mfT$ to
denote the following $GTL(d)$ element:
\begin{equation}
\label{def:crossing}
  \mcT = u\mathcal{X} + \mathcal{Y} \ .
\end{equation}
Of course, if the tangle contains more than one crossing, we can apply the 
above procedure on all crossings, recursively, 
thereby associating $\mcT$ with a linear combination of basis elements
of $GTL(d)$ and loops. It is easy to convince oneself that this 
definition
is independent of the order in which we open each
crossing, and so the mapping from tangles with weighted crossings 
to $GTL(d)$ is well defined. 

\begin{figure}
  \center
  \includegraphics[scale=0.4]{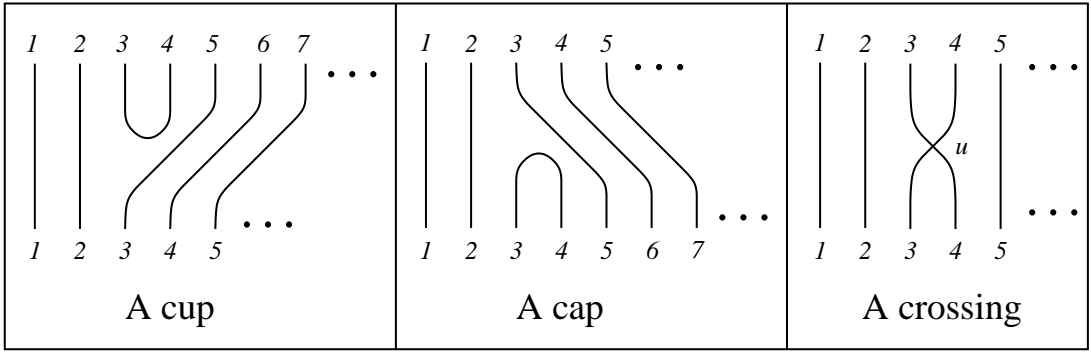}
  \caption{An illustration of the elementary tangles (from left to
  right) $\mcA_3$, $\mcB_3$ and $\mcC_3(u)$.}
 \label{fig:cupcapcross}
\end{figure}

Next, we define the notion of \emph{elementary tangles}, which are
simple tangles that can generate all tangles.
\begin{deff}[Elementary tangles]
\label{def:element-t}
  A tangle is called an elementary tangle if it is represented by
  one of the following diagrams (up to isotopy and/or addition of
  straight strands at the right)
  \begin{description}
    \item [1. Cup $\mcA_i$] The cup
      diagram $\mcA_i$ is the diagram which is identical to the
      identity $\mcI$ diagram except for the upper $i$,$i+1$ pegs,
      which are connected to each other. The diagram thus contains a
      minima after the $i-1$ strand.

    \item [2. Cap $\mcB_i$] Similar to the cup diagram, $\mcB_i$ is identical to the
      identity diagram except for having a maxima which is created
      by connecting the lower $i$,$i+1$ pegs.

    \item [3. Crossing $\mcC_i(u)$] The crossing diagram $\mcC_i(u)$
      contains a crossing with weight $u$ between the $i$, $i+1$
      strands, while all the other strands are trivial.
  \end{description}
  An illustration of these three diagrams is given in
  \Fig{fig:cupcapcross}.
\end{deff}
Notice that the three elementary tangles are not independent.
Indeed, it is easy to see that for odd $i$, $\mcC_i(u) = u\mcI +
A_i\cdot B_i$, whereas for even $i$, $\mcC_i(u) = \mcI + uA_i\cdot
B_i$. It turns out that it is only the cup and cap tangles that are
needed to generate the full $GTL(d)$ algebra. Nevertheless, we will
find the crossing operator extremely useful to describe the
algorithm and the universality result.

For a tangle that corresponds to a basis element we can define the number
of \emph{in-pegs} to be 
 the number of lower pegs in the diagram, and the number of \emph{out-pegs} 
to be the number of
upper pegs in the diagram (see \Fig{fig:tangle}). This definition
can be extended to general tangles, since by opening a crossing we obtain
two diagrams with the same number of in-pegs and the same number of
out-pegs. We note that these numbers are not well defined for  
$GTL(d)$ elements, but just for tangles, 
since the definition of the $GTL(d)$ elements allows 
adding a number of straight strands to the right without changing the element.

\subsection{Connection between the Kauffman brackets 
 and the $GTL(d)$ algebra.}
\label{sec:Kauffman-medial}

Our goal is to calculate the Kauffman bracket $\la L_G \ra (d,{\Bu})$
of a medial graph using the algebraic structure of the $GTL(d)$
algebra.  We do this by defining a map $\Psi_d$ which associates a
medial graph with an element in $GTL(d)$. This map was essentially
defined in the previous section. Indeed, we may simply consider an
embedding of $L_G$ in $\mathbbm{R}^2$ as a tangle. 
Then, according to the previous section, $\Psi_d(L_G)$ is a 
linear combination of basis elements of $GTL(d)$, and hence is an
element of $GTL(d)$. 

Moreover, as noted in the previous section,
tangles have a well-defined number of in-pegs and out-pegs. It is easy 
to see that
$\Psi_d(L_G)$ has zero in-pegs and zero out-pegs, and is actually equal to
a scalar times $\mcI$. This is because as we open up its crossings,
we end up with tangles that contain only loops; these are then
proportional to $\mcI$ by the loop-value relation. It turns out that
the coefficient in front $\mcI$ element is exactly the scalar we are
looking for!  
\begin{proposition}
\label{prop:id}
  $\Psi_d(L_G)= \langle L_G \rangle (d,{\Bu})\mcI$.
\end{proposition}
\begin{proof}
  Opening all the crossings of the $L_G$ element in accordance with
  \Eq{def:crossing}, we are left with a sum of $2^{|E|}$ terms. Each
  term corresponds to a possible opening of all the crossings, i.e.,
  to a state $\sigma$ of the Kauffman bracket. It is a diagram with
  only loops and must therefore be proportional to $\mcI$. The
  proportionality factor is exactly $\langle L_G|\sigma\rangle$.
  Indeed, assume that the opened diagram contains $\ell$ loops and
  corresponds to a state $\sigma$. Then the proportionality factor
  is $d^\ell$ times $u_e$ for every crossing with $\sigma_e=1$. This
  is because such opening joins the shaded areas and is therefore
  multiplied by a $u_e$ factor according to \Eq{def:crossing}.
  Summing up all terms we obtain the desired result. 
\end{proof}

This shows that if we want to approximate the Tutte polynomial, or
rather, $\la L_G \ra (d,{\Bu})$, it suffices to approximate the
coefficient in front  of $\mcI$ in $\Psi_d(L_G)$.

\subsection{Connection with Representations of $GTL(d)$}

To approximate the above coefficient, we consider representations of
the $GTL(d)$ algebra. We will define
the representation \emph{for diagrams} rather than for abstract
$GTL(d)$ elements, as it would depend on the particular in-pegs and
out-pegs of the diagrams.
\begin{deff}
\label{def:GTLrep} Let $\{H_n\}$ be a series of Hilbert spaces
  (which do not have to be different from each other). A map $\rho$
  that maps diagrams of tangles to linear operators over the Hilbert
  spaces is said to be a representation of $GTL(d)$ if it maps every
  tangle diagram $\mfT$ with $n$ in-pegs and $m$ out-pegs to a
  linear operator $\rho(\mfT):H_n\to H_m$ such that:
  \begin{itemize}
    \item $\rho(\mfT)$ only depends on the isotopy class of $\mfT$
    \item For diagram $\mfT$ with $n$ in-pegs and $m$ out-pegs,
      $\rho(\mfT)$ is a linear transformation from $H_n$ to $H_m$.
      
    \item \textbf{Preserving the multiplicative structure:}
      If $\mfT_1$ has $n$ in-pegs and $m$ out-pegs and $\mfT_2$
      has $m$ in-pegs and $\ell$ out-pegs then
      $\rho(\mfT_2\cdot\mfT_1) = \rho(\mfT_2)\cdot\rho(\mfT_1)$.
      
    \item \textbf{Preserving the additive structure:} 
    If $\mfT_1, \mfT_2$ represent the $GTL(d)$ elements
    $\mcT_1, \mcT_2$ and $\mfT_3$ represents the $GTL(d)$ element
    $c_1\mcT_1+c_2\mcT_2$ then $\rho(\mfT_3) = c_1\rho(\mfT_1) +
    c_2\rho(\mfT_2)$.
    \inote{Not sure if we should also add linearity req here.}
  \end{itemize}
\end{deff} 

We will later provide a definition of such a representation of
$GTL(d)$, called the \emph{path-model representation}. Nevertheless, any
representation will do to evaluate the desired coefficient: 
 
\begin{claim} 
  If $\rho$ is a representation of $GTL(d)$, then 
  $\rho\big(\Psi_d(L_G)\big)=\la L_G \ra (d,{\Bu})\Id$.
\end{claim} 
\begin{proof} 
  This is true because any representation will take the identity
  $\mcI$ to the identity operator $\Id$.  
\end{proof}  

We can thus perform our desired approximations on the image of a
representation of the $GTL(d)$ algebras. 

{~}

\noindent\textbf{Remark:} We note here an important difference
between the connection that is made here between the representation
of the $GTL(d)$ algebra, and the Kauffman bracket, versus a similar
connection that was used in \cite{ref:Aha05}.  In \cite{ref:Aha05},
a similar result was true for $TL_n(d)$, a restricted version of
$GTL(d)$, in which the number of strands was finite, and fixed.  The
representation however, was required to exhibit a property called
the Markov property, for the connection to the Kauffman bracket to
hold.  Here, due to the fact that a single algebra element is
associated with a class of diagrams, which have different in- and
out-pegs numbers, no such property is needed, and any representation
will do.  This is an advantage of working with the $GTL(d)$ algebra.

We now proceed to describe the representations we will work with.


\section{Background: The Path Model Representation of $GTL(d)$} 
\label{sec:pathmodel} 

We use the path model representations. They are essentially the
representations used in Ref.~\cite{ref:Aha05}, except here we allow
also non-Hermitian representations, and also consider the infinite
algebra rather than finite versions of it. In addition, in
accordance with Definition~\ref{def:GTLrep}, the representation that
we describe is defined for diagrams of tangles, rather than for the
tangles themselves, since it depends on the number of in-pegs and
out-pegs of the diagram (it does not depend, however, on the
particular geometry of the diagram - only on its topology).
Specifically, for each diagram we define a linear transformation
with \emph{different} domain and target spaces, the dimensions of
which, depend on the number of in-pegs and out-pegs of the diagram.
We begin by motivating the definition, and then move to the exact
definitions.

\dnote{add ref} 
\znote{I'm not positive, but I don't think they
are due exclusively to Jones, if at all.  Perhaps a safe first pass
would be to reference the path model described in "Coxeter Graphs
and Towers of Algebras" by Jones, Goodman and de la Harpe. Maybe in
"Exactly solved models in statistical mechanics" by Baxter, chapter
12.} 

\subsection{Diagrams as operators over paths on an auxiliary graph} 

Let $\mcT\in GTL(d)$ be a basis element, which is represented by
a diagram $\mfT$ with $m$ in-pegs and $n$ out-pegs.

\begin{figure}
  \center 
  \includegraphics[scale=0.4]{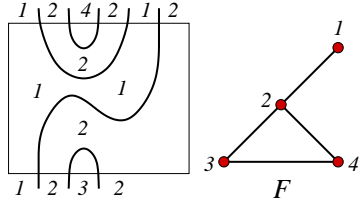}
  \caption{Labeling of a diagram according to the vertices of an
    auxiliary graph $F$.}
  \label{fig:coloring}
\end{figure}

We would like to associate the diagram with an operator.  How can
this be done? We draw $\mfT$ and confine it to a rectangular
box, thereby dividing the box into a set of disjoint regions. Then
the $n$ lower-pegs of the diagram, which are connected to the lower
edge of the box, divide it into $n+1$ gaps, while the upper pegs of
the diagram divide the upper edge of the box into $m+1$ gaps. We
would like to treat $\mfT$ as an operator acting on a sequence of labels
of the upper gaps, taking it to a sequence of labels of the lower
gaps.  For this, we use the notion of an \emph{auxiliary graph}
$F$, and allow the labels of the different regions of the boxed
diagram to be vertices of $F$. Further more, we restrict the
labeling of the diagram to be such that two adjacent regions are
labeled by adjacent vertices in $F$.  We observe that under those
conditions, the sequence of labels of the upper gaps, as well as
that of the lower gaps, is in fact a \emph{path} in the auxiliary
graph $F$, and that the diagram can be seen as a transformation the
takes the path of the lower gaps to the path of the upper gaps. As
an example, see \Fig{fig:coloring}. 

We will thus view $\mfT$ (and through it $\mcT$) as an operator on
paths on $F$. We can now define the Hilbert space of paths formally.

\subsection{The Hilbert Space of Paths on $F$}

For simplicity, denote the vertices of the auxiliary connected graph
$F$ by $\{ 1, 2, 3, \ldots, \}$. We define a \emph{family} of
Hilbert spaces $\{H_n\}$, corresponding to paths of length $n$ of
$F$, as follows:
\begin{deff}[The spaces $H_n$]
  For any integer $n\ge 0$, the $n$-steps paths space $H_n$ of an
  auxiliary graph $F$ is a Hilbert space that is defined as
  follows: let $p=\{p_1,p_2, \ldots, p_{n+1}\}$ denote an $n$-steps
  path on $F$ which starts at the vertex $p_1$ and ends at the vertex
  $p_{n+1}$. We demand that $p_1=1$ and that $p_i$ is adjacent to
  $p_{i+1}$ in $F$ (denoted as $p_i\sim p_{i+1}$). Then we associate
  every path $p$ with a vector $\ket{p}$, and consider the set
  $\{\ket{p}\}$ of all $n$-steps paths to be an \emph{orthonormal}
  basis for the space $H_n$.
  
  Note that $H_0$ is a one dimensional space which is spanned by the
  vector $\ket{1}$.
\end{deff}

\subsection{Compatible paths} 

Let $\mfT$ be a diagram with $m$ in-pegs and $n$ out-pegs that
represents a basis element of $GTL(d)$.  We associate $\mfT$ with a
transformation $\rho(\mfT):H_m\to H_n$, by defining
$\bra{p'}\rho(\mfT)\ket{p}$ for every $n$-steps path
$\ket{p}=\ket{p_1, \ldots, p_{n+1}}$ and $m$-steps path
$\ket{p'}=\ket{p'_1, \ldots, p'_{m+1}}$.  The term
$\bra{p'}\rho(\mfT)\ket{p}$ will be zero, unless the two paths are
compatible in the following sense:

\begin{deff}[compatible paths]
  Let $\mfT$ be a diagram of a basis element, or a basis element with
  closed loops,  with $n$
  in-pegs and $m$ out-pegs. Then the $n$-steps path $p$ and the
  $m$-steps path $p'$ are called \emph{compatible with respect to
  $\mfT$}, if, when labeling $\mfT$'s lower gaps with the vertices
  of the path $p$, and the upper gaps with vertices of the path
  $p'$, the result is a diagram where all connected surfaces have
  the same label. The labeling is done from the left to right with
  respect to the order of the vertices in each path. 
\end{deff}  
It is clear that the definition depends only on the isotopy class of
the diagram.

The upper path and lower path in \Fig{fig:coloring} are thus
compatible.  We define the value of $\bra{p'}\rho(\mfT)\ket{p}$ to
be zero if the two paths are not compatible. What about compatible
paths?  This is defined in the next section. \inote{I am not sure
that we need such a drastic divide of this definition into
subsections. Wouldn't it be clearer had we defined
$\bra{p'}\rho(\mcT)\ket{p}$ both for compatible and non-compatible
paths in one subsection?}

\subsection{The value $\bra{p'}\rho(\mfT)\ket{p}$ for compatible paths} 

\begin{figure}
  \center 
  \includegraphics[scale=0.6]{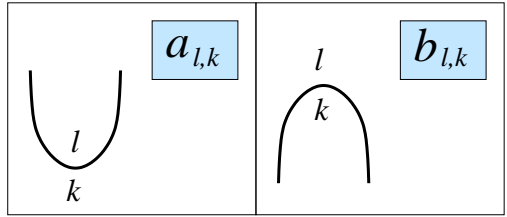} 
  \caption{Local minimum and maximum, associated with $a_{k,\ell}$
    and $b_{k,\ell}$ respectively.}
  \label{fig:cupcap}
\end{figure}

Consider a diagram $\mfT$ of a basis element and two
compatible paths $p, p'$  with respect to it. As $\mfT$ does not
contain any loops, it is easy to see that $p,p'$ determine the
labeling of all areas in $\mfT$ uniquely. To define
$\bra{p'}\rho(\mfT)\ket{p}$, we consider all the local minimums and
maximums of the diagram that represents $\mfT$. The area near each
such point is labeled by two labels - adjacent vertices from $F$. A
local minimum labeled by $\ell$ from above and $k$ from below is
associated with a constant $a_{\ell,k}$, and a maximum with the same
labeling is associated with $b_{\ell,k}$ (see Figure
\ref{fig:cupcap} for illustration). 

We will soon assign values to $a_{\ell,k}$ and $b_{\ell,k}$, and
discuss their properties. As an example, the minima and maxima of
the diagram in \Fig{fig:coloring} are associated (from top to
bottom) with the variables: $a_{4,2}, a_{2,1}, b_{1,2}, a_{1,2}$ and
$b_{2,3}$. We define $\bra{p'}\rho(\mfT)\ket{p}$ to be the product
of all the constants that appear in the diagram. The definition of
$\rho$ can thus be summarized as follows:
\begin{deff}[The path-model representation]
\label{eq:rep_def} Let $\mcT$ be a basis element with a diagram
  $\mfT$, and $p,p'$
  compatible paths with respect to $\mfT$. Then $p,p'$ determine a
  unique labeling of $\mfT$ and $\bra{p}\rho(\mfT)\ket{p'}$ is the
  product over all local maximums and minimums of $\mfT$ of the
  corresponding $a_{\ell,k}$ and $b_{\ell,k}$ coefficients, which are
  determined by the labeling.
  
  For non-compatible $p,p'$, $\bra{p}\rho(\mfT)\ket{p'}\EqDef 0$.
\end{deff}

\subsection{Extending the definition beyond basis elements of $GTL(d)$}

Thus far we have only defined $\rho(\cdot)$ for diagrams of basis
elements. We would like to extend the definition to any diagram that
represents a tangle. This is easily achieved by linearity. Recall
that the tangle diagrams have a well-defined number of in-pegs and
out-pegs, which is equivalent to say that they are a linear
combination of basis elements diagrams with the \emph{same} number
of in-pegs and the \emph{same} number of out-pegs. Consequently, the
operators that represent these diagrams have the same domain and
range spaces, and therefore we can define the representation of the
tangle's diagram by linearity. 

This defines $\rho(\mfT)$ for all tangle diagrams, and satisfies the
additivity requirement in Definition~\ref{def:GTLrep}. For
$\rho(\cdot)$ to be a $GTL(d)$ representation according to
Definition~\ref{def:GTLrep}, we still need to verify that it only
depends on the isotopy class of the diagrams and that it preserves
the multiplicative structure of the algebra.

\subsection{Conditions on $a_{k,\ell}, b_{k,\ell}$ for $\rho$ to be a
representation} 

To show that $\rho(\cdot)$ is a representation of the $GTL(d)$
algebra we must assert the following: firstly, that isotopic
elements define the same transformation - after all, our definition
of $\rho(\mfT)$ relied on a particular geometrical representation of
$\mfT$. Secondly, we must show that the representation preserves the
product rule, i.e., $\rho(\mfT_2\mfT_1) = \rho(\mfT_2)\rho(\mfT_1)$.
The following claim provides sufficient conditions for this to hold.

\begin{claim} 
  The following two conditions guarantee that $\rho(\cdot)$ is a
  representation of $GTL(d)$:
  \begin{eqnarray}
  \label{eq:cons1}
    \forall\ \ell,k: \qquad b_{\ell,k}\cdot a_{\ell,k} &=& 1 \ , \\
  \label{eq:cons2} 
    \forall\ k: \quad \sum_{\ell:\ell\sim k} 
      a_{\ell,k}b_{k,\ell} &=& d \  .
  \end{eqnarray}    
\end{claim}

\begin{proof}

  To show that two isotopic elements are given the same image by
  $\rho$, we use the fact that two isotopic elements are linked by a
  series of moves that eliminate or create a pair: a local maximum
  and a local minimum.  Accordingly it will suffice to prove
  invariance for a move that creates or eliminates a single pair.

  Let $\mfT'$ be a diagram that results from $\mfT$ by removing a
  single pair of minima and maxima. A pair of terms: $b_{\ell,k}$,
  $a_{\ell,k}$ will be missing in the expression for 
  $\bra{p'}\rho(\mfT')\ket{p}$. Now, by using \Eq{eq:cons1} we have
  $b_{\ell,k}\cdot a_{\ell,k} = 1$ and thus
  $\rho(\mfT')=\rho(\mfT)$.

  Next, we prove that $\rho(\cdot)$ is a homomorphism.  Given two
  diagrams $\mfT_1,\mfT_2$ that represent basis elements, we would
  like to prove that
  \begin{equation}
    \rho(\mfT_2\cdot\mfT_1) = \rho(\mfT_2)\cdot\rho(\mfT_1) \ .
  \end{equation}
  The diagram $\mfT_2\cdot\mfT_1$ can represent either a basis element, or
  contain some loops. Assume the first case, and let the paths
  $p,p'$ be compatible with it. Then there is only one path $p^*$ on
  the border between the two diagrams, such that $p,p^*$ are
  compatible with $\mfT_1$ and $p^*,p'$ are compatible with
  $\mfT_2$. Therefore if $\sum_{p''}$ denotes the summation over all
  possible paths on the border between the diagrams then
  \begin{equation}
    \sum_{p''} \bra{p'}\rho(\mfT_2)\ket{p''}
        \bra{p''}\rho(\mfT_1)\ket{p} 
    = \bra{p'}\rho(\mfT_2)\ket{p^*}
        \bra{p^*}\rho(\mfT_1)\ket{p}
    = \bra{p'}\rho(\mfT_2\mfT_1)\ket{p} \ .
  \end{equation}
  The first equality is due to the fact that $p^*$ is the only path
  that is compatible with both $p'$ and $p$. The second equality
  follows from the definition of $\bra{p'}\rho(\mfT_2)\ket{p^*}$ and
  $\bra{p^*}\rho(\mfT_1)\ket{p}$; they are products of the
  $a_{k,\ell}, b_{k,\ell}$ coefficients that correspond to the
  labeling induced by $p, p'$. Multiplying them gives the product of
  all the $a_{k,\ell}, b_{k,\ell}$ coefficients in the composite
  diagram of $\mfT_2\cdot\mfT_1$.
  
  Assume now that $\mfT_2\cdot\mfT_1$ contains one loop, which by
  isotopy has only one maximum and one minimum, and let $\mfT_3$ be
  equal to $\mfT_2\cdot\mfT_1$ without the loop (hence
  $d\mcT_3=\mcT_2\cdot\mcT_1$ for the corresponding $GTL(d)$
  tangles). Then there is more than one labeling that is compatible
  with $p,p'$. Consequently, the sum
  \begin{equation}
  \label{eq:loop-sum}
    \sum_{p''}\bra{p'}\rho(\mfT_2)\ket{p''} 
     \bra{p''}\rho(\mfT_1)\ket{p} \ ,
  \end{equation}
  is equal to the summation over all possible labeling of the
  diagram $\mfT_2\cdot\mfT_1$ which are compatible with $p,p'$. All are
  identical except for the label of the internal region of the loop.
  Therefore if $k$ is the label of the external region of the loop,
  and $\ell$ is the label of the internal region, then we may write
  \begin{equation}
    \sum_{p''}\bra{p'}\rho(\mfT_2)\ket{p''}
       \bra{p''}\rho(\mfT_1)\ket{p} 
    = \bra{p'}\rho(\mfT_3)\ket{p}\sum_{\ell:\ell\sim k}
      a_{\ell,k}b_{k,\ell} \ .
  \end{equation}
  By \Eq{eq:cons2}, the RHS of the equation is equal to
  $d\bra{p'}\rho(\mfT_3)\ket{p}$, and as $\mcT_1\mcT_2 = d\mcT_3$
  for the corresponding $GTL(d)$ tangles, we get 
  \begin{equation}
    \sum_{p''}\bra{p'}\rho(\mfT_2)\ket{p''}
       \bra{p''}\rho(\mfT_1)\ket{p} 
    = \bra{p'}\rho(\mfT_2\cdot\mfT_1)\ket{p} \ ,
  \end{equation}
  as required. The case of more than one loops follows easily by
  similar arguments.
\end{proof}

\subsection{Setting $a_{k,\ell}, b_{k,\ell}$: Path representations for all $d$}
\label{sec:path:all-d}

In the previous section we proved that given an auxiliary $F$ and a
corresponding set of coefficients $a_{k,\ell}, b_{k,\ell}$ that
satisfy the conditions in Definition~\ref{eq:rep_def} with respect
to $d$, we can construct a representation of the $GTL(d)$ algebra.
We will now show how to find such coefficients for every complex
$d$.

\begin{figure}
  \center
  \includegraphics[scale=0.4]{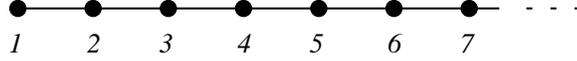}
    \caption{The auxiliary graph $F_\infty$ is one-sided
    infinite line graph.}
 \label{fig:F-infi}
\end{figure}

Consider the graph in \Fig{fig:F-infi} which is the one-sided
infinite line. Denote this graph by $F_\infty$, and let $M_\infty$
be its adjacency matrix. Given any complex $d$, we define the
infinite dimensional vector $\bar\pi=(\pi_1, \pi_2, \pi_3,\,
\ldots)$, as follows: 
\begin{eqnarray}
\label{eq:pi-1}
  \pi_1 &=& 1 \ , \\
  \pi_2 &=& d \ , \\
  \pi_3 &=& d^2-1 \ , \\
        &\vdots& \\
  \pi_n &=& d\pi_{n-1} - \pi_{n-2} \ , 
  \label{eq:pi-n}\\
  &\vdots& \nonumber
\end{eqnarray}
It is easy to see that $\bar\pi$ is an eigenvector of $M_\infty$
with an eigenvalue $d$: when we apply the adjacency matrix $M_\infty$
as an operator on $\bar\pi$ we get $(M_\infty\bar\pi)_1 = \pi_2 = d
= d \pi_1$, and for all $n \ge 2$: $(M_\infty\bar\pi)_n = \pi_{n-1}
+ \pi_{n+1} = \pi_{n-1} + ( d\pi_n - \pi_{n-1}) = d\pi_n$.

We use $\bar{\pi}$ to define the path representation for almost all
values of $d$: 
\begin{deff}
\label{def:rep} Let $d$ be such that in the above definition all
  coordinates of $\bar\pi$ are non vanishing.  For such $d$,
  the Path Representation of $GTL(d)$ using the graph $F_\infty$ is
  defined by: 
  \begin{eqnarray}
  \label{def:a}
    a_{\ell,k} &\EqDef& \sqrt{\frac{\pi_\ell}{\pi_k}} \ , \\ 
    b_{\ell,k} &\EqDef& \sqrt{\frac{\pi_k}{\pi_\ell}} \ .
  \label{def:b}
  \end{eqnarray}
\end{deff}

\begin{lem} 
\label{lem-con} 
  The map given by definition \ref{def:rep} satisfies 
  the constraints (\ref{eq:cons1}, \ref{eq:cons2}), and thus, it is a
  representation of $GTL(d)$. 
\end{lem}
\begin{proof}
  Equation~(\ref{eq:cons1}) trivially holds. For \Eq{eq:cons2}, 
  \begin{equation}
    \forall k: \qquad \sum_{\ell:\ell \sim k} 
          a_{\ell,k} b_{k,\ell} 
    = \sum_{\ell: \ell \sim k} \frac{\pi_\ell}{\pi_k} 
    = \frac{1}{\pi_k} \sum_\ell[M_\infty]_{k,\ell} {\pi_\ell} 
    = \frac{1}{\pi_k} [M_\infty\bar{\pi}]_k = d \ .
  \end{equation} 
  In the first equality we used the definition of $a_{\ell,k},
  b_{k,\ell}$. In the second equality we replaced the summation over
  $\ell$'s which are adjacent to $k$ by a summation over all $\ell$,
  using the fact that the adjacency matrix $[M_\infty]_{k,\ell}$ is
  equal to one when $k\sim\ell$ and vanishes otherwise. Finally, in
  the last equality we used the fact that $\bar{\pi}$ is an
  eigenvector of $M_\infty$ with an eigenvalue $d$. 
\end{proof}

We now need to deal with the cases in which the vector $\bar\pi$
vanishes somewhere. To do this, we simply cut the graph $F_\infty$
before the first location where the vector vanishes. Say that $m$
sites remains in the graph; call this graph $F_m$. It is easy to
see that the vector given by $\bar\pi$ cut to that point, namely the
first $m$ coordinates of $\bar\pi$, is an eigenvector of eigenvalue
$d$ of the adjacency matrix of $F_m$, which is simply the left upper
most $m\times m$ block of the matrix $M_\infty$. 

The same construction of Definition \ref{def:rep} will now work for
these values of $d$, except that now the graph being used is the
finite graph $F_m$.

\subsection{Hermitian representations} 
\label{sec:hermitian}

We conclude this section with a definition of an Hermitian
representation of the $GTL(d)$ algebra. These types of
representations have few nice properties, which are particularly 
important for the $\BQP$-hardness result.

\begin{deff}[An Hermitian representation of the $GTL(d)$ algebra]
\label{def:hermitian} A representation $\rho$ of the $GTL(d)$
  algebra is call Hermitian if $\rho(\mfA_i) = \rho(\mfB_i)^\dagger$
  for every $i$. Here $\mfA_i$ and $\mfB_i$ are diagrams which
  represent the elementary cup and cap tangles $\mcA_i$, $\mcB_i$,
  which were defined in \Sec{sec:GTL}. For the above condition to
  make sense, we require that the number of out-pegs of $\mfA_i$ be
  equal to the number of in-pegs of $\mfB_i$.
\end{deff}

For the family of path representations we have the following
corollary:
\begin{corol}
  A necessary and sufficient condition for a path representation to
  be Hermitian is that the coordinates of the eigenvector
  $\bar{\pi}$ are all positive.
\end{corol}
\begin{proof}
  It is a simple exercise to verify that a path representation is
  Hermitian if and only if $a_{\ell,k} = b^*_{k,\ell}$ for every
  $\ell\sim k$. Therefore by Eqs.~(\ref{def:a},\ref{def:b}), we get
  \begin{equation}
    \sqrt{\frac{\pi_\ell}{\pi_k}} =
    \left(\sqrt{\frac{\pi_\ell}{\pi_k}}\right)^* \ ,
  \end{equation}
  which is equivalent to the condition that $\pi_\ell/\pi_k$
  is positive for every $\ell\sim k$. Assuming that the auxiliary
  graph is connected and that we have normalized $\bar{\pi}$ such
  that $\pi_1=1$, we conclude that the representation is Hermitian
  if and only if $\bar{\pi}>0$. 
\end{proof}

We finally remark, without proof, that for the general
path-representation, which uses $F_\infty$, a \emph{sufficient}
condition for Hermiticity is that $d=2cos(\pi/k)$ for integer $k$,
or that $d>2$. \inote{give ref} These are the representations
that are mostly used in the literature. However, as we already
noted, there is no need for the representation to be Hermitian for
our quantum algorithm to work, but see \Sec{sec:universality}
for the implication of non-Hermiticity on the question of
universality. 

This conclude the section on the Path-model representation.
Throughout this and the previous section, we paid a special
attention to the distinction between $GTL(d)$ elements and the
diagrams that represent them, since the path-model representation is
only defined for the diagrams - not for the elements themselves.
From here onwards, however, we will not make that distinction when
it is clear from the context that we already picked a particular
diagram to represent a $GTL(d)$ element. We will therefore
occasionally talk about $\rho(\mcT)$ and about the in-pegs and
out-pegs of $\mcT$, even though these concepts are not-well defined.

\section{The quantum algorithm} 
\label{sec:algorithm}

\subsection{Statement of Result} 
To state the result, we assume that the graph $G$ is given to us
not as an adjacency matrix, but rather, it is embedded in the plane in some 
generic way.  

\begin{deff} {\bf Nicely Embedded} 
  A medial graph $L_G$ of a planar graph $G$ is said to be nicely
  embedded if it is given as a diagram in $\mathbbm{R}^2$, in such a
  way that if we sweep a horizontal line from the bottom of the
  tangle to the top, the horizontal line meets only one elementary
  tangle at a time - The minimum of a cup, the maximum of a cap, or a crossing.
\end{deff}

\begin{figure}
  \center
  \includegraphics[scale=0.4]{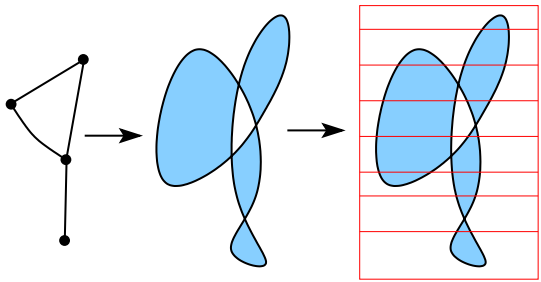}
  \caption{A nicely embedded planar graph $G$ is a graph which is
  attached with a particular nicely embedded medial graph.}
\label{fig:metamorpho}
\end{figure}

Like all isotopic facts, it is easy to see but probably much harder
to prove, that any medial graph can be given such an embedding.
Moreover, any graph can be given as a diagram in $\mathbbm{R}^2$
such that the resulting medial graph will be nicely embedded.
Therefore we abuse language and say in this case that a graph $G$ is
nicely embedded when we implicitly refer to a particular nice
embedding of $L_G$ in $\mathbbm{R}^2$. From now on we assume that
the graph $G$ is given to us in such a nicely embedded way.
Obviously, such a nice embedding induces an order on the elementary
tangles of the medial graph.  A simple example is given in
\Fig{fig:metamorpho}. 

The algorithm works by assigning to each of the elementary tangles
(a crossing, a cap or a cup) a linear operator, according to the
representation $\rho$, and applying them in the above mentioned
order. 

\begin{deff}[From a medial graph to elementary tangles]
\label{def:lin} 

  Given a nicely-embedded medial graph $L_G$, translate it to a
  product of elementary tangles by  decomposing it into
  a product of cups, caps and crossings according to the order that
  is determined by its (nice) embedding in $\mathbbm{R}^2$.

\end{deff}
\inote{I really don't know if this definition is necessary. What
exactly do we define here? - we just describe the initial step of
the algorithm no?}

The scale in which the algorithm will work is defined by

\begin{deff}[Algorithmic Scale]
\label{def:window} 

  Let $G=(V,E)$ be a nicely embedded planar graph with weights
  $\Bv=\{v_e\}$ and $q$ a complex number. In addition, let
  $\rho(\cdot)$ be the path representation of $GTL(d)$ with $d^2=q$.
  Then its medial graph, $L_G$ can be translated into a product of
  basic tangles $\mcT_1\cdot\ldots\cdot \mcT_N$. The scale of the
  algorithmic problem of approximating $Z_G(q,\Bv)$ $\Delta_{alg}$,
  is then
  \begin{equation}
    \Delta_{alg}\EqDef q^{-|V|/2}
      ||\rho(\mcT_1)||\cdot\ldots\cdot||\rho(\mcT_N)|| \ .
  \end{equation}
\end{deff} 
In other words, $\Delta_{alg}$ is the product of the norms of the
operators that correspond to the elementary tangles that make up
$L_G$, times an overall factor $q^{-|V|/2}$, which connects $\langle
L_G\rangle(d,\Bu)$ to $Z_G(q,\Bv)$ (see Claim~\ref{cl:connection}).

In accordance with the last paragraph of \Sec{sec:pathmodel}, when
we talk about $\rho(\mcT_i)$ we actually talk about the path-model
representation of a \emph{diagram} that represents the tangle
$\mcT_i$. That diagram, however, is implicitly defined by the medial
graph $L_G$, hence $\rho(\mcT_i)$ is well-defined.

We note that $\Delta_{alg}$, depends not only on the graph but also on its 
embedding. A different way to orient the graph, for example, might
lead to significant changes in the scale.  We do not
know of any algorithm to optimize the way to embed the graph so as
to get the best (smallest) possible scale.  We can now give the
precise version of Theorem~\ref{thm:alg:rough}

\begin{thm} \textbf{Quantum algorithm}:
\label{thm:alg}

  \noindent Given a finite graph $G=(V,E)$, with a nice embedding in
  $\mathbbm{R}^2$, with weights on the edges, ${\Bv}$, and a complex
  number $q$, there is an efficient quantum algorithm that
  approximates the Tutte polynomial of the graph at those weights
  and $q$, to within an additive approximation
  $\Delta_{alg}/poly(|E|)$.
\end{thm} 

After we prove this theorem, it will become clear that we have been 
quite wasteful in our quality of approximation, and that the size of
the approximation scale can be made significantly smaller if one is
dealing with non-unitary parameters. We discuss this matter in
Subsection \ref{sec:windowimproved}; We rigorously address this 
issue only towards the end of the paper, in Section
\ref{sec:complete}. 

\subsection{Proof of the Algorithm}
\label{sec:alg-proof}

To prove Theorem \ref{thm:alg}, we essentially apply the sequence of
linear operators defined before, by a quantum computer. It will be
convenient to use the following term: 

\begin{deff}[Operator circuit](see Ref.~\cite{ref:bubble})
  An operator circuit is defined just like a quantum circuit, except
  the gates are only restricted to be linear operators from $k$ to
  $\ell$ quantum registers, with no unitarity restriction. 
\end{deff}

Consider then a medial graph with a nice embedding.  There is an
operator circuit associated with $L_G$, which is simply the circuit
one gets by replacing each of the elementary tangles (cap, cup or
crossing) with the corresponding linear operator by
Definition~\ref{def:lin}.  We denote the resulting operator circuit
by $Q$. Note that this operator circuit acts on the Hilbert space
$H_0$ and takes it to $H_0$.  We claim: 

\begin{claim}
\label{cl:Q}
  $\la L_G \ra (d,{\Bu})=\la 1|Q|1\ra$.   
\end{claim} 

\begin{proof}
  By definition, $Q=\rho\Big(\Psi(L_G)\Big)$.  But by Proposition
  \ref{prop:id}, we have that $\Psi(L_G)=\la L_G \ra (d,{\Bu})\mcI$,
  and since $\rho$ is a representation, $\rho(\mcI) = \Id$. 
  Therefore, $\la 1|Q|1\ra = \la\L_G\ra(d,{\Bu})\bra{1}\Id\ket{1} =
  \la\L_G\ra(d,{\Bu})$.
\end{proof} 

In order to prove Theorem \ref{thm:alg}, we need to show that we can
approximate the value of $\la 1|Q|1\ra$ to within and additive
approximation of the scale $\Delta_{alg}$. To do this, we simply
create the state $\ket{1}\in H_0$ and apply the operators in $Q$ on
this state one by one. To approximate the inner product of the
resulting vector with $|1\ra$, we use the Hadamard test
\cite{ref:Nei00}.

{~}

\subsubsection{Moving to qubits} 

We first note that we need not consider sites in $F$ with indices
that are bigger than the number of edges in $G=(V,E)$, since we
start from $1$ and each crossing in $L_G$ can only increases the
path by one more site.  This means that we can encode the name of a
vertex in $F$ with logarithmically many qubits (in $|E|$). We view
the Hilbert space now as composed of registers; each register
contains logarithmically many qubits, and can hold a label of one
vertex in in the range $1\to |E|$. 

Notice that each elementary crossing acts locally only on three
labels of a path. Hence its operator under the path representation
is actually the tensor product of the identity with an operator that
acts on three local registers. The other elementary tangles, namely
the cup and the cap, also yield local 3-registers operators, as we
will see in the end of the next subsection. Therefore all the
operators that we shall apply act non-trivially only on a logarithmic
number of local qubits.


\subsubsection{Simulating a linear operator.} 

\begin{claim}
\label{apply_herm} Given a linear operator $M:H_k\to H_k$ that acts
  on a constant number of local registers and a quantum computer, it
  is possible to efficiently transform any normalized vector
  $\ket{\alpha}\in H_k$ to a normalized vector $\ket{\beta}\in
  H_k\otimes \mathbbm{B}$, with $\mathbbm{B}$ being the space of an
  auxiliary qubit, such that the following holds: when the auxiliary
  qubit is projected to $\ket{0}$, the resulting state on $H_k$
  becomes $\frac{1}{||M||}M\ket{\alpha}$, where $||M||$ is the
  operator norm of $M$. In other words, $\ket{\beta} =
  \frac{1}{||M||}M\ket{\alpha}\otimes\ket{0} +
  c\ket{\gamma}\otimes\ket{1}$, with $\ket{\gamma}$ being some
  residual state and $c$ a constant such that the overall state has
  unit norm.
\end{claim}

From this claim it follows that we can apply $Q$ efficiently, up to
some normalization factors. We will first prove the claim, and then
worry about the normalization factors. 
   
\begin{proofof}{Claim ~\ref{apply_herm}}

  According to the discussion in the previous section, if $M$ acts
  on on a constant number of local registers, then it can be viewed
  as an operator that acts on $\mathcal{O}(\log(|E|))$ local qubits.
  By the polar decomposition lemma, we can write $M$ as a product of
  two matrices, one unitary and the other is positive definite,
  $M=PU$. Both matrices will still act non-trivially only on
  $\mathcal{O}(\log(|E|))$ local qubits. 

  The simulation of unitary operators that act on logarithmically
  many qubits is a standard procedure in quantum computation, and
  can be done in polynomial time. Hence we can simulate $U$
  efficiently.
  
  We turn to the simulation of the positive definite matrix $P$. 
  $P$ can be diagonalized by an orthonormal basis, and so we
  assume without loss of generality that $P$ is diagonal in the
  computational basis, otherwise we can always change basis back and
  forth efficiently. Let $r_1\ge r_2\ge\ldots\ge r_m\ge 0$ be the
  eigenvalues of $P$, with the corresponding eigenvectors $\ket{1},
  \ket{2}, \ldots, \ket{m}$. We wish to apply $P/r_1$.  To do this
  we set the auxiliary qubit to $\ket{0}$, and apply a
  \emph{unitary} transformation $P_u$ that satisfies
  \begin{equation}
    P_u\big(\ket{i}\otimes\ket{0}\big)
     = \frac{r_i}{r_1}\ket{i}\otimes\ket{0}
     + \sqrt{1-(r_i/r_1)^2}\ket{i}\otimes\ket{1} \ . 
  \end{equation}

  We think of the auxiliary qubit as an indicator to the validity of
  the original transformation; up to the overall factor $r_1$, we
  get the original transformation only if the auxiliary qubit is
  $\ket{0}$. This is exactly achieved by projecting the auxiliary
  qubit on $\ket{0}$.  Therefore, the application of $P$ is
  recovered by applying $P_u$, projecting the auxiliary qubit on
  $\ket{0}$, and multiplying the final result by $r_1$. Finally,
  note that as $r_1$ is the maximal eigenvalue of $P$, it follows
  that $||M||=r_1$.
\end{proofof} 

We conclude that we can apply any desired operator, provided we
book-keep its norm: we actually apply a matrix with a unit norm. 
Moreover, for each operator we get an extra qubit. This qubit must
be projected to $\ket{0}$ for the rest of the state to hold the
desired result. 

To complete the algorithm we need to deal with the operators that
correspond to the cup and cap tangles, which are not square
operators.  Their action, however, is local. They can be viewed as
operators that transform one register to 3 registers (the cup
operator), or vice versa (the cap operator), while leaving the rest
of the registers intact. 

The cup operator is simply the concatenation of the operator
$\ket{k} \mapsto \ket{k,1,1}$, with another $3\times 3$ linear
operator: $\ket{k,1,1}\mapsto \sum_{\ell:\ell\sim k}
a_{k,l}\ket{k,\ell,k}$, and the rest of the vectors go to (the
scalar) $0$.  We can apply the first operator easily by allocating
two more registers. The second operator can be applied by the
previous lemma.  Likewise, the cap operator is defined by
$\ket{k,\ell,m}\mapsto \delta_{k,m}b_{k,\ell}\ket{k,1,1}$, which can
be applied by the lemma, after which we simply discard the last two
registers. 

\subsubsection{Completing the proof of the Algorithm with the Correct 
  Approximation window size} 

Once we have applied all operators, we need to compute the inner
product of the final state, with the state $\ket{1}$ tensor with
$\ket{0}$ on all the ancillary qubits. The final state is the
following normalized state:
\begin{equation}
  \frac{1}{\Delta'_{alg}}\Big(Q\ket{1}\Big)\otimes\ket{0}\otimes\ket{0}\ldots\ket{0}
    + c\ket{\gamma} \ ,
\end{equation}
where $\ket{\gamma}$ is some residual state in which one or more of
the ancillary qubits are different from $\ket{0}$, and
$\Delta'_{alg} = d^{|V|/2}\Delta_{alg}$, i.e., it is simply the
product of all the norms of the elementary tangles that make up $Q$.

It is a well known fact that a quantum computer can efficiently 
estimate the inner product of such a state with
$\bra{1}\otimes\bra{0}\ldots\otimes\bra{0}$ to get an approximation
of $\frac{1}{\Delta'_{alg}}\bra{1}Q\ket{1}$, using the Hadamard test.
Repeating this test $poly(|E|)$ times would result in an
approximation of $\frac{1}{\Delta'_{alg}}\bra{1}Q\ket{1}$ to within
an additive window of size $1/poly(|E|)$ with exponentially good
confidence. Finally, we use the recall from \Eq{eq:connection} and
Claim~\ref{cl:Q} that
\begin{equation}
  \frac{1}{\Delta'_{alg}}\bra{1}Q\ket{1} 
    = \frac{1}{\Delta'_{alg}}\la L_G\ra(d,\Bu)
    = \frac{1}{\Delta_{alg}}Z_G(q,\Bv) \ .
\end{equation}
This completes the proof of Theorem \ref{thm:alg}. 
\subsection{Improving the approximation window}\label{sec:windowimproved} 

When the crossings operators are all unitary, the scale
$\Delta_{alg}$ does not increase by applying them; the norm of a
product of unitary operators is equal to the product of their norms
- which is equal to $1$. However, the quality of the approximation
might worsen severely when we apply non-unitary operations.  It is
easy to be convinced that the approximation scale we define, and in
fact, our algorithm itself, are quite wasteful. For example, if one
combines several operators together, the product of their norms is
very likely to be larger than the norm of their products, and we
will lose a lot in our approximation scale.  As long as the product
of the operators is still an operator on polylogarithmically many
qubits, there is no problem in calculating the product of them all
as one matrix (this calculation is done classically on the side),
and then perform it as a whole by the quantum computer, in the same
method as in the previous subsection.  This changes the contribution
in the approximation window from the product of the norms to the
norm of the product. 

The only problem is that we do not know of a way to find the optimal
grouping that will do the job. \emph{Any} grouping would improve the 
approximation window, however. And, it is often the case, that a
classical algorithm that is given the graph can easily find an 
excellent grouping. 

In Section \ref{sec:complete} 
we will redefine the problem such that the input in fact includes 
the description of a suggested grouping, in which case the approximation 
window will be much better, namely the product of the grouped operators.

%
%

\section{BQP Hardness} 
\label{sec:universality} 

Having the algorithm at hand, we would now like to show the hardness
of the problem it solves.  In this section we will prove a precise
version of Theorem \ref{thm:hard:rough}. 

Without loss of generality, $\BQP$-hardness can be described as the
following problem. We are given a description of a quantum circuit
over $n$ qubits as a product of $N=poly(n)$ 2-qubits gates 
\begin{equation}
  \label{eq:circuit}
  U=U_N \cdot \ldots \cdot U_1 \ ,
\end{equation}
which are local operators (i.e., they act on two neighboring
qubits), taken from a well-defined basis. Then $\ket{0^{\otimes n}}$
is the tensor product of $n$ $\ket{0}$ qubits, and we are promised
that either $|\bra{0^{\otimes n}} U \ket{0^{\otimes n}}|^2 \le 1/3$
or $|\bra{0^{\otimes n}} U \ket{0^{\otimes n}}|^2 \ge 2/3$. We are
asked to decide, within $poly(n)$ time, which of these possibilities
holds.  We will show that under certain conditions, there exists an
efficient algorithm that takes the description of the quantum
circuit and outputs a weighted graph $G$ such that $Z_G(q,\Bv)$
(times a known constant) approximates $|\bra{0^{\otimes n}} U
\ket{0^{\otimes n}}|^2$. It will follow that certain additive
approximations of $Z_G(q,\Bv)$, for various choices of the 
variables, are $\BQP$ hard. Before we show that, however, we need
some preliminary definitions.

\subsection{Statement of the Result}
\label{sec:problem}

\begin{figure}
  \center
  \includegraphics[scale=0.4]{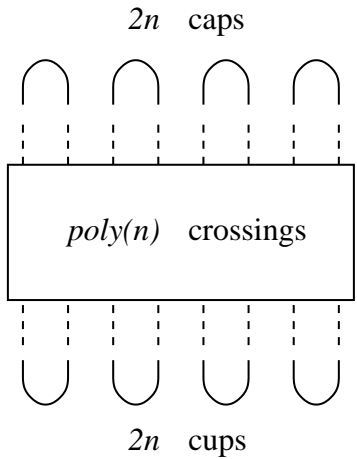}
  
  \caption{The type of medial graph that is used in the definition
     of the BQP-hard problem. Note that the lower $2n$ cups need not
     be exactly on the same horizontal line so that $L_G$ is nicely
     embedded. The same goes for the $2n$ caps.} \label{fig:plat}
\end{figure}

We wish to show that approximating the Tutte polynomial of a given
planar graph, with a given set of weights, to within some additive
approximation scale, is $\BQP$ hard.  We will be able to show this
only for weights taken from some particular sets, which will be
defined later. On the other hand, we also restrict the type of
graphs we take as input to the $\BQP$-hard problem (restricting the
type of graphs only makes the hardness result stronger): 

\begin{deff}[Input graphs] 
\label{def:ig} 
  A planar graph $G=(E,V)$ with a nice embedding in $\mathbbm{R}^2$.
  Its medial graph $L_G$ is required to have the following
  structure: it is a concatenation of $2n$ cups which are placed
  in a line one next to the other, followed by $poly(n)$ crossings
  between the resulting $4n$ strands, which are then closed by a
  line of $2n$ caps. An illustration of this kind of medial graph is
  given in \Fig{fig:plat}.\inote{Actually this is not a nicely
  embedded graph. Is it worth changing the deff a bit ?}
\end{deff} 

The weights of such graphs need to satisfy certain requirements too.  

\begin{deff}[Odd and Even Weights]
\label{def:w} The set of weights from which the edge weights of the
  input graph are taken from is denoted by $W$. It is required to be
  the union of two types of weights: odd and even weights (which
  need not be non-intersecting): $W_{odd}$ and $W_{even}$.  We
  demand that an edge that corresponds in the medial graph to a
  crossing $i\leftrightarrow (i+1)$ with odd (even) $i$, is given a
  weight from $W_{odd}$ ($W_{even}$). The subset of these edges is 
  denoted by $E_{odd}$ ($E_{even}$). Obviously $E_{odd}\cap
  E_{even}=\emptyset$ and $E_{odd}\cup E_{even}=E$.
\end{deff} 

We will restrict $W$ to be one of three sets: The unitary set, the
complex non-unitary set, and the real non-unitary set.  We will wait
with the exact definitions of these sets of parameters and provide
them when the definition is motivated. Meanwhile, we will use the
names of these sets abstractly.  

The computational problem is
defined by:
\begin{description}
  \item [Input:] An input graph $G=(V,E)$ as in Definition \ref{def:ig} 
    with edges weights from a set $W$ as in Definition \ref{def:w},
    where $W$ is either the unitary, the complex non-unitary or the
    real non-unitary set of parameters.  

  \item [Scale for BQP hardness:]
      \begin{equation}
      \label{def:Delta}
          \Delta_{hard} \EqDef q^{|V|-|E_{odd}|}
          \left(\prod_{e\in E_{odd}}v_e\right) \ .
      \end{equation}

  \item [Output:] Approximate $Z_G(q,\Bv)$ to within additive approximation 
      window of size $\frac{1}{poly(|V|)}\Delta_{hard}$.
\end{description}

Note that the scale here does not seem to be the same scale as that
the algorithm works with.  Indeed, whereas the algorithmic scale
corresponds to the product of the norms of the operators, the
hardness scale corresponds to, roughly, the product of their
determinants, as we will see later.  We will address this mismatch in the
scales only in Section \ref{sec:complete}.  We will show in
Theorem~\ref{thm:uni-complete} that in fact, in the unitary case
the two scales are equal, which follows from the fact that for a
unitary matrix, both the norm and the determinant are equal to $1$.
In the non-unitary case we will somewhat modify the problem to make
the algorithmic scale smaller, so as to match the hardness scale in
(\ref{def:Delta}).  

The following theorem is the precise version of Theorem
\ref{thm:hard:rough}, and is the second main result of the paper: 

\begin{thm}
\label{thm:hardness} 
  Solving the above problem, for $W$ which is either the unitary
  set, the complex non-unitary set or the real non-unitary set, to
  within an approximation window of size
  $\frac{1}{poly(|V|)}\Delta_{hard}$, is $\BQP$ hard. 
\end{thm}

The proof of Theorem~\ref{thm:hardness} is essentially given in the
rest of this section, and in
Sections~\ref{sec:density-proof-unitary},
\ref{sec:density-proof-nu}. For sake of simplicity we first present
its outline.

\noindent\textbf{Outline of the proof}: \\ 
To prove the above theorem we will encode the $n$-qubits strings and
the 2-qubits operators from \Eq{eq:circuit} inside the Hilbert space
of the path representation. We will then find $GTL(d)$ elements
whose image under the path-representation approximates the 2-qubits
gates. This is done by proving that using the weights in $W$ we can
construct a set of operators that generate dense group in the space
of the 2-qubits gates. Consequently we will be able to use the
famous Solovey-Kitaev algorithm to efficiently find elements
$\mcT\in GTL(d)$ whose $\rho(\mcT)$ approximates the different
gates. From the concatenation of all these $GTL(d)$ elements we will
be able to find the desired graph $G=(V,E)$.

The $GTL(d)$ operators that we consider can be either unitary or
non-unitary. Although the general structure of the proof is the same
in both cases, the details vary. Hence, we will give a single
unified proof whenever is possible, and split the discussion only at
the level of the finest details. In particular, we will show that in
the unitary case, the operators form a dense subgroup of $SU(14)$,
whereas in the non-unitary cases, it is a subgroup of
$SL(14,\mathbbm{R})$ or $SL(14,\mathbbm{C})$. In addition, the
famous Solovey-Kitaev theorem, which is used in the unitary case,
will be generalized and applied to the non-unitary cases.

\subsection{Encoding a quantum gate by Crossing operators} 

\subsubsection{The 4-steps encoding} 

Our goal is to encode the $n$-qubit Hilbert space within the Hilbert
space $H_k$ of $k$ steps (\Sec{sec:pathmodel}). We use the
linear graph $F_\infty$ that was used in \Sec{sec:path:all-d},
and choose $q$ (and hence $d$) such that non of the first 3
coordinates of $\bar{\pi}$ will vanish. This ensures us that we can
use paths over at least three different vertices in $F_\infty$. It
is easy to deduce from Eqs.~(\ref{eq:pi-1}-\ref{eq:pi-n}) that any
$q\ne 0,1$ satisfies this requirement.

The encoding is done using the so-called \emph{4-steps encoding}, 
which was independently discovered by Kitaev \cite{ref:Kit05}, and
Wocjan and Yard \cite{ref:Woc06}. In this encoding, every qubit is
represented by a 4 steps path:
\begin{eqnarray}
  \ket{\Un{0}} &\EqDef& \ket{12121} \in H_4\ , \\
  \ket{\Un{1}} &\EqDef& \ket{12321} \in H_4 \ .
\end{eqnarray}
A tensor product of qubits is naturally translated to a
concatenation of paths, hence a string of $n$ qubits is encoded in a
$4n$ steps path, and the space of $n$-qubits is encoded in $H_{4n}$.
Obviously $\dim H_{4n} > 2^n$. The $2^n$ subspace of $H_{4n}$ that
corresponds to encoded $n$-qubits is called the \emph{legitimate
subspace $L_{4n}$}, and therefore we may write $H_{4n} = L_{4n}
\oplus L^\bot_{4n}$.

We denote by $||\cdot||_{L_{4n}}$ the norm over $L_{4n}$. This norm
can be defined for \emph{every vector in $H_{4n}$} by first
projecting it onto $L_{4n}$ and calculating its norm there. This
naturally defines an operator norm in $L_{4n}$, which will be used
later. Notice that the $L_{4n}$ norm is a semi-norm, since non-zero
vectors with no projection on $L_{4n}$ will have a vanishing norm.

With the $4$-steps encoding we can write every unitary operator $U$
over $n$ qubits as an unitary operator over $H_{4n}$.  Indeed, if
$U$ is given by
\begin{equation}
  U = \sum_{i,j} U_{ij}\ket{i}\bra{j} \ ,
\end{equation} 
(here $\ket{i}$ is a short-hand for a $n$-qubits tensor)
then its encoded version is 
\begin{equation}
  \Un{U} = \sum_{i,j} U_{ij}\ket{\Un{i}}\bra{\Un{j}} +
  \mathbbm{1}_{L_{4n}^\bot} \ .
\end{equation} 
Here $\Id_{L_{4n}^\bot}$ denotes an operator that is identity on
$L_{4n}^\bot$ and zero on $L_{4n}$, and therefore $\Un{U}$ is
non-trivial only on the legitimate space $L_{4n}$. It is easy to
verify that the encoding preserves the product structure of the
quantum circuit, i.e., $\Un{U} = \Un{U}_N\cdot\ldots\cdot\Un{U}_1$.

\subsubsection{The crossing operators}

Roughly speaking, our goal is to find an element $\mcT_i\in GTL(d)$ for
every encoded gate $\Un{U}_i$, such that $\rho(\mcT_i)\simeq \Un{U}_i$
over $L_{4n}$, where $\rho(\mcT_i)$ is the image of $\mcT_i$ under the path
representation. Notice that $\mcT_i$ must have $k$ in-pegs and $k$
out-pegs for some $k\le 4n$, in order for $\rho(\mcT_i)$ to be an
operator from $H_{4n}$ into itself.

\begin{figure}
  \center
  \includegraphics[scale=0.4]{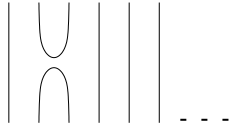}
  \caption{An illustration of the $GTL(d)$ element $\mathcal{E}_2$.}
 \label{fig:E2}
\end{figure}

The following $GTL(d)$ elements and their images under the path
representation are the building blocks of $\mcT_i$:
\begin{deff}[The $\mathcal{E}_i\in GTL(d)$ elements and the $\Phi_i$ operators]
\label{def:phi} 

  The $\mathcal{E}_i\in GTL(d)$ is the product $\mathcal{E}_i \EqDef
  \mcA_i\mcB_i$ (see Sec~\ref{sec:GTL}). In other words, it is a
  basis element with $i+1$ in-pegs and $i+1$ out pegs, that is
  defined by $i-1$ straight strands, followed by a cap-cup couple in
  the $i$,$i+1$ strands. An example of $\mathcal{E}_2$ is shown in
  \Fig{fig:E2}.
  
  Under the path representation it is be mapped to an operator
  $\Phi_i\EqDef \rho(\mathcal{E}_i)$, from $H_k$ to itself, for
  every $k\ge i+1$. 
\end{deff}

The $\mathcal{E}_i$ elements are often considered as the standard
generators of the Temperley-Lieb algebra $TL_n(d)$ (see
Ref~\cite{ref:Tem71} and \S 12.4 in \cite{ref:Bax82}). It is easy to
see that $\mathcal{E}_i^2 = d\mathcal{E}_i$ and consequently
$\Phi_i^2 = d\Phi_i$. Moreover, for Hermitian representation
$\rho(\mcB_i) = \rho(\mcA_i)^\dagger$ (see \Sec{sec:hermitian}) and
consequently $\Phi_i =
\rho(\mcA_i\mcB_i)=\rho(\mcA_i)\rho(\mcA_i)^\dagger$ is Hermitian.

Using the operators $\Phi_i$ we can now define the \emph{crossing
operators} $\sigma_i(u)$:
\begin{deff}[The crossing operators $\sigma_i(u)$]
\label{def1:crossing} 

  The crossing operators are the image, under the path
  representation, of a $GTL(d)$ elementary tangle $\mcC_i(u)$, which
  contains a crossing between the $i\leftrightarrow (i+1)$ strands
  while being trivial on the rest (see \Sec{sec:GTL}). They are denoted by
  $\sigma_i(u):H_k\to H_k$, with $u$ being the weight that is
  associated with the crossing: $\sigma_i(u)\EqDef
  \rho\Big(\mcC_i(u)\Big)$. There are two types of crossing
  operators: crossing operators with odd $i$ and crossing operators
  with even $i$. Using the black-and-white coloring convention, it
  is easy to write them in terms of $\Phi_i$:
  \begin{equation}
  \label{eq:crossing}
    \sigma_i(u) = \left\{\begin{array}{lcl}
        u\Id + \Phi_i &,& \mbox{for odd $i$'s} \ , \\
      \Id + u\Phi_i &,& \mbox{for even $i$'s} 
    \end{array}\right. \ .
  \end{equation}

\end{deff}

\subsubsection{Encoding a 2-qubit Gate: how to start?}

The main idea in the proof is to show that we can approximate every
encoded two qubit gate $\Un{U}_i$ over the space $L_8$ of legitimate
$8$ steps paths, using the 7 operators $\sigma_i(u)$, $i=1, \ldots,
7$ with a prescribed set of weights $W$, provided that $W$ satisfies
the required restrictions, which we are still to give. 

\begin{figure}
  \center \includegraphics[scale=0.3]{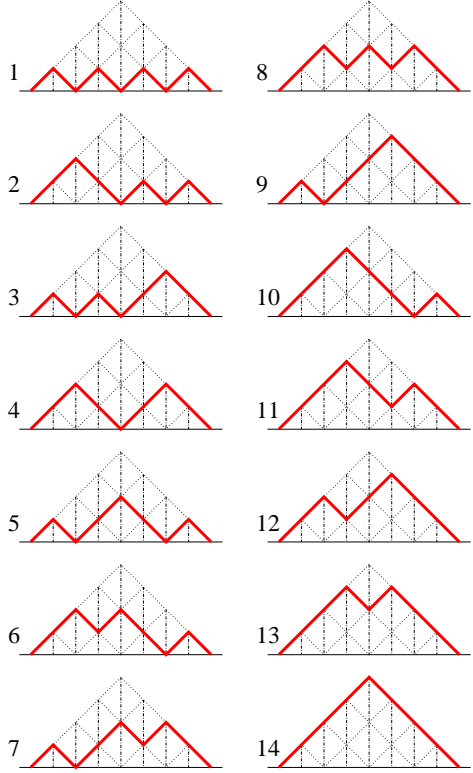} 
  \caption{The diagrams
  of 14 possible vectors $\ket{p_i}$ that span the space $K=H_{8,1\to 1}$} 
  \label{fig:8p}
\end{figure}

Note that the space $L_8$ is not invariant under these seven
operators.  Hence, we will consider a larger space, the subspace
$K=H_{8,1\to 1}$, spanned by the 8-steps paths space $H_8$ that
start and end at the first site of the auxiliary graph.  Notice that
$L_8 \subset K \subset H_8$.  In \Fig{fig:8p} we have
listed the $14$ relevant paths that span $K$ schematically.

The idea is therefore to approximate any unitary matrix on $L_8$, by
approximating its extension to $K$, which applies the identity on
the remaining $10$ paths. Clearly, if $||T-V||_{H_{8,1\to 1}} \leq
\delta$ then also $||T-V||_{L_8} \leq \delta$, as required. In order
to do this, we show that the subgroup generated by the seven
operators is dense in $SU(K)$ - the group of unitary transformations
over $K$.  For that matter we will have to restrict the parameters we
will deal with in various ways.  The treatment will be completely
different depending on the type of operators that the parameters
correspond to; we discuss the different types of parameters next.

\subsection{Parameters}
\subsubsection{Sets of parameters Closed to Inverses}

The reason for restricting the parameters set $W$ is so that we are
able to prove the density, in some relevant group, of the subgroup 
generated by the crossing operators.  More specifically, every $U_i$
in the quantum circuit (\ref{eq:circuit}) is a 2-qubits gate, and
its encoded gate $\Un{U}_i$ works on the space $L_8$, of legitimate
8-steps paths. We wish to approximate, using the crossing operators,
either the unitary or the orthogonal group over $L_8$ using the 7
operators $\sigma_i(u)$, $i=1, \ldots, 7$ with a prescribed set of
weights. 

Our first restriction on the set guarantees that we can treat the 
set generated by these operators as a group, by requiring that the set is 
closed to inverses:

\begin{deff}[A set of weights closed to inverses with respect to $q$]
\label{def:weights} 

   For a given $q$, a set of weights $W=W_{odd} \cup W_{even}$ is
   said to be closed to inverses with respect to $q$, if $W_{odd}$ and
   $W_{even}$ are non-empty and satisfy the following requirements:
  \begin{itemize}
    \item For every odd weight $v\in W_{odd}$ there must be another odd
      weight $w\in W_{odd}$ such that $w+v+q=0$.  $w$ is then called
      the odd inverse weight of $v$, and vice versa.
    \item For every even weight $v\in W_{even}$ there must be another even
      weight $w\in W_{even}$ such that $w+v+wv=0$.  $w$ is then
      called the even inverse weight of $v$, and vice versa. 
  \end{itemize}
\end{deff}

The following claim justifies why we call these weights inverses of 
each other. 

\begin{claim} 
  Consider two odd weights $v,w\in W_{odd}$ such that $w+v+q=0$.
  Then the (odd) crossing operators $\sigma_i(v/d)$ and
  $\sigma_i(w/d)$ satisfy 
  \begin{equation}
  \label{eq:oddinv}
    \sigma_i(v/d)\sigma_i(w/d) = \frac{vw}{q}\Id \ . 
  \end{equation}
  Similarly, for two even weights $v,w\in W_{even}$ such that 
  $v+w+wv=0$, we have 
  \begin{equation}
  \label{eq:eveninv}
    \sigma_i(v/d)\sigma_i(w/d) = \Id \ .
  \end{equation}
\end{claim} 

\begin{proof} 
  We use \Eq{eq:crossing} and the identity $\Phi_i^2=d\Phi_i$. 
\end{proof}

We note that there is a slight abuse of language here; in the odd
case, the operators are not really inverses but there is an overall
factor $wv/q$ in front of the identity, which means that
$\sigma_i(w/d)$ is only proportional to the inverse of
$\sigma_i(v/d)$. As we shall see, this is not a significant problem.

%
%

\subsubsection{Types of parameters} \label{sec:uni-non-uni}

Essentially, the parameters we can handle are divided to three
different types, according to the three different types of operators
they correspond to.

\begin{deff}[Parameters of unitary type]
\label{def:uni} 

  A set of parameters, $(q,W)$ is said to be of \emph{unitary type}
  if for $d=\sqrt{q}$, the path-model representation is Hermitian
  (i.e., $d=2\cos\pi/k$ for an integer $k>0$ or
  $d>2$, according to Definition~\ref{def:hermitian})
  
  we have  and additionally,
  \begin{eqnarray}
  \label{eq:odd-uni-con}
    |v+q| &=& |v|\ , \quad \mbox{for all  $v\in W_{odd}$} \ , \\
    |1+v| &=& 1 \ , \quad \mbox{for all $v\in W_{even}$} \ .
    \label{eq:even-uni-con}
  \end{eqnarray}
\end{deff} 

\begin{claim} 
  Fix a set of parameters which is of unitary type.  Then when
  restricting attention to the subspace $K$, for any even $i$ and
  $v\in W_{even}$ we have that $\sigma_i(v/d)$ is unitary, and for
  every odd $i$ and every $v\in W_{odd}$ we have that
  $\sigma_i(v/d)$ is equal to a unitary times an overall factor
  $u=v/d$. 
\end{claim} 

\begin{proof} 
  For the values of $d$ for which the path representation is
  Hermitian (see \Sec{sec:hermitian}), the unitarity is easy
  to prove. In this case, the $\Phi_i$'s operators are Hermitian
  (see discussion below Definition~\ref{def:phi}), and can
  diagonalized by an orthonormal basis. Therefore, the crossing
  operators, being a linear combination of $\Phi_i$ and $\Id$ are
  diagonalized by the same basis. Moreover, as $\Phi_i^2 = d\Phi_i$,
  its eigenvalues are either $d$ or $0$. Therefore the eigenvalues
  of the odd crossings operator are $\{v/d, v/d+d\}$, which by
  \Eq{eq:odd-uni-con} are of the same magnitude - $|v/d|$. It
  follows that in the odd case $\sigma_i(v/d)$ equals a unitary
  operators time an overall factor $v/d$.
  
  Similarly, in the even case the eigenvalues of the crossing
  operators are $\{1,1+v\}$, and the restrictions in
  \Eq{eq:even-uni-con} guarantee that both have a unit magnitude.

  \inote{Removed this part:
  The reason that unitarity holds also under the weaker condition 
  on $d$ is that we restrict ourselves to the subspace $K=H_{8,1\to
  1}$. Indeed recall from definition~\ref{def:rep} that the path
  representation for $d=\sqrt{q}$ is determined by the coordinates
  of the infinite eigenvector $\bar{\pi}$ of the auxiliary graph
  $F_\infty$ for eigenvalue $d$. The first five coordinates are
  given by $\bar{\pi}=(1,d,d^2-1,d^3-2d, d^4-3d^2+1, \ldots)$, and
  it is easy to see that for $d>2\cos\pi/5$ they are all positive.
  These are the coordinates that define the $\Phi_i$ operators over
  $K$. Their positivity implies that the matrices of $\Phi_i$ in $K$
  are real. Since they are also symmetric, it follows that the
  operators $\Phi_i$ are Hermitian inside the subspace $K$.}
\end{proof} 

\begin{deff}[A set of parameters of complex non-unitary type]  

  A set of parameters $(q,W)$ is said to be of complex non-unitary
  type if none of its weights satisfies the conditions of Definition
  \ref{def:uni}, \emph{and} at least one of the following happens:
  there is at least one weight which is not real, or $q$ is not
  real.  
\end{deff} 

\begin{deff}[A set of parameters of real non-unitary type]
  is defined similarly to the above, except all weights, and $q$,
  must be real. 
\end{deff} 

In the following section we further restrict these three types of
parameters to get three families of parameters, one of each type.

\subsubsection{Three families of parameters} 

The core of the universality proof is to show that the generators
defined above, for three families of parameters, generate a dense
subgroup of the unitary group on the $14$ dimensional subspace $K$
(or the orthogonal group in the real non-unitary set of parameters.)
To show density, we have to restrict our parameters even further.
The exact definition only requires that $W$ would contain four
weights: $v_1\in W_{odd}$, $v_2\in W_{even}$ and their inverse
weights. The definition can be elegantly casted by using the
parameters $\alpha$ and $\beta$:
\begin{equation}
\label{eq:alphabeta}
  \alpha = \sqrt{1+qv_1^{-1}} \ , \quad \beta=\sqrt{1+v_2} \ .
\end{equation}
Indeed, we first notice that in this notation, $|\alpha|=|\beta|=1$
is equivalent to the conditions of unitarity in
Definition~\ref{def:uni}.

\begin{deff}{\bf [The unitary set of parameters  (case I)]}
\label{def:case-II}

  We define the sets of parameters $(q,W)$ which are contained in 
  our unitary family. First, $W$ is closed to inverses with respect
  to $q$, and $(q,W)$ satisfy the requirements of Definition
  \ref{def:uni}. In addition we require that $q\ne 0,1, 2,
  \frac{3\pm\sqrt{5}}{2}$, and that there exist $v_1\in
  W_{odd}$, $v_2\in W_{even}$ such that the following holds:
  \begin{itemize}
    \item $\alpha^2, \beta^2 \ne 1$, and
    \item At least one of the two real numbers $s_1, s_2 \in [0,1)$ which
      are defined by
      \begin{equation}
        e^{2\pi i s_1} = \alpha^2 = 1+qv_1^{-1}\ , \quad 
        e^{2\pi i s_2} = \beta^2 = 1+v_2
      \end{equation}
      is \emph{not} a rational number $p/r$ with $\gcd(p,r)=1$ and
      $r=1,2,3,4,5$.
  \end{itemize}
\end{deff} 

\begin{deff}[The complex non-unitary case (case II)]
\label{def:case-III} 
  $|\alpha|,|\beta|\ne 1$ and at least one of the following
  inequalities holds:
        \begin{eqnarray}
        \label{eq:jorg1}
          \left| \alpha-\frac{1}{\alpha}\right|^2 +
          \left|\frac{q-1}{q^2}\right|\cdot
          \left| \alpha-\frac{1}{\alpha}\right|^2\cdot
          \left| \beta-\frac{1}{\beta}\right|^2 &<& 1 \ , \mbox{or} \\
          \left| \beta-\frac{1}{\beta}\right|^2 +
          \left|\frac{q-1}{q^2}\right|\cdot
          \left| \alpha-\frac{1}{\alpha}\right|^2\cdot
          \left| \beta-\frac{1}{\beta}\right|^2 &<& 1 \ .
        \label{eq:jorg2}          
        \end{eqnarray}
        
    In addition to that, at least one of the numbers $q,v_1,v_2$ is
    not a pure real. 
\end{deff} 

\begin{deff}[The real non-unitary case (case (3))] 
    Same conditions on $\alpha,\beta$ as in the previous case, but
    now $v_1, v_2, q$ are real numbers with $q>4\cos^2\pi/5$.
\end{deff}  
\inote{I  dont like the way it is now. It is very very
confusing. Too many definitions - I mean the unitary case and the
unitary family etc.}

\subsection{Density and Efficiency on the subspace $K$}
\label{sec:density-proof-unitary} 

In this section we state and prove the central theorem on which 
universality result builds. This is the density and efficiency
theorem on the subspace $K$, Theorem~\ref{thm:8-strands}.
Essentially it assures us that any unitary gate on $K$ can be
efficiently approximated by the crossing operators to within any
accuracy. To prove it, we first prove density, and then use the
Solovey-Kitaev theorem for efficiency. The density theorems,
however, are given in terms of the \emph{normalized} versions of the
crossing operators that we define below.
\begin{deff}[The normalized crossing operators] 
\label{def:normalized} 
  We define the normalized operators $\hat{\sigma}_i$ to be equal to
  $\sigma_i$, up to a constant, determined by the requirement that
  $det(\hat{\sigma}_i)=1$, where the determinant is calculated only
  in the subspace $K$. In other words,
  $\hat{\sigma}_i\EqDef\sigma_i/\left[\det(\sigma_i)\right]^{1/14}$.
\end{deff} 

\subsubsection{Density Theorems} 

Using the normalized generators for the above three families of
parameters, we can prove density in the relevant groups: 

\dnote{throughout the paper, there is a big balagan with respect to 
when do we use d and when do we use q. This should be fixed...}
\inote{I change almost everywhere to $q$. Only left $d$ where it is
really needed, or where using $q$ makes things too cumbersome. }

\begin{thm}[Density on $K$, the unitary case]
\label{thm:uni-dense} 

  For $(q,W)$ a unitary set of parameters (case I), the normalized
  operators $\{\hat{\sigma}_1(v/d),\cdots,
  \hat{\sigma}_7(v/d)\}_{v\in W}$ generate a dense subgroup of
  $SU(K)$. 
\end{thm}

\begin{thm}[Density on $K$, the complex non-unitary case]
\label{thm:com-nu-dense} 
  For $(q,W)$ a complex non-unitary set of parameters (case II), the
  normalized operators $\{\hat{\sigma}_1(v/d),\cdots,
  \hat{\sigma}_7(v/d)\}_{v\in W}$ generate a dense subgroup of
  $SL(K,\mathbbm{C})$. 
\end{thm}

\begin{thm}[Density on $K$, the real non-unitary case]
\label{thm:real-nu-dense} 

  For $(q,W)$ a real non-unitary set of parameters (case III), the
  normalized operators $\{\hat{\sigma}_1(v/d),\cdots,
  \hat{\sigma}_7(v/d)\}_{v\in W}$ generate a dense subgroup of
  $SL(K,\mathbbm{R})$.  
\end{thm}

The proof of these density theorems is non-trivial, and is deferred
to the next two sections of the paper.

\subsubsection{From Density to Efficiency} 

We proceed assuming these density results.  The idea is that if we
have density, then any quantum gate can be written as a product of a
sequence of crossing operators.  Thus, the quantum circuit can be
approximated by a sequence of crossing operators.  For this, we
should of course claim that the approximation is efficient, and can
be found efficiently.  The proof is essentially a simple application
of the famous Solovay-Kitaev theorem, where we also have to account
for the normalization factors correctly, which requires using a
simple trick with commutators. In addition, since the Solovay-Kitaev
theorem applies only for $SU(K)$, we need to generalize it to our
purposes. 

Let us first recall the Solovay-Kitaev theorem in the version which
will be most useful to us: 

\begin{thm}[The Solovay-Kitaev theorem]
\label{thm:SK} 
  Let $M$ be a Hilbert space with $\dim M$ $\ge 2$ over
  $\mathbbm{C}$. Then there exists an $\epsilon_0>0$ which only
  depends on $\dim M$ such that if $\mathcal{G}$ is an
  $\epsilon_0$-net over the special unitary group $SU(M)$, then for
  every transformation $V\in SU(M)$ and every $\delta>0$ there
  exists a sequence $S$ of $poly(\log \delta^{-1})$ elements from
  $\mathcal{G}$, which can be found in $poly(\log \delta^{-1})$
  time, such that $||V-S||<\delta$.
\end{thm}
To deduce efficiency from density in the non-unitary case, we can no
longer apply the Solovay kitaev theorem, which holds for $SU(K)$. 
To this end, we will prove a generalization of the Solovey-Kitaev
theorem that holds for density in $SL(K,\mathbbm{C})$, in which the
generators of the $\epsilon_0$-net may be close - but not
necessarily inside $SU(K)$. For the $SL(K,\mathbbm{R})$ case we will
prove yet another generalization of the algorithm which works on the
special-orthogonal groups $SO(K)$ instead of $SU(K)$. The
modifications of the SK theorem are not difficult technically, and
so their proof is deferred to the Appendix. 

\begin{thm}[The non-unitary Solovay-Kitaev theorem]
\label{thm:non-SK} 
  Let $M$ be a Hilbert space with $\dim M$ $\ge 2$ over the field
  $F$, which can either be $\mathbbm{C}$ or $\mathbbm{R}$, 
  let $R>0$ be some finite radius,
  and let $\mathcal{G}=\{g_1, \ldots, g_N\}$ be a finite set of
  generators in $SL(M,F)$.
  Define $B_R(M)$ to be the set of all transformation in
  $SL(M,F)$  whose distance
  from the unity transformation is smaller than $R$. Then there
  exists an $\epsilon_0>0$ which only depends on $\dim M$ and on
  $R$, such that if $\mathcal{G}$ forms an $\epsilon_0$-net over
  $B_R(M)$, then for every transformation $V\in B_R(M)$ and every
  $\delta>0$ there exists a finite sequence $S$ of $poly(\log
  \delta^{-1})$ generators from $\mathcal{G}$, which can be found in
  $poly(\log \delta^{-1})$ time, such that $||V-S||<\delta$.
\end{thm}

\begin{proof} 
  The proof is given in Appendix~\ref{sec:non-SK-proof}. 
\end{proof} 

We now use the Solovay-Kitaev Theorem and its non-unitary variants
to prove that indeed, the generators can be used to approximate any
given gate efficiently, with the appropriate overall factor.  In
addition to the Solovay Kitaev theorems and the density theorems, 
we also need to take a correct account of the overall factors. 

\begin{thm}[The density and efficiency theorem]
\label{thm:8-strands} 
  Let $q=d^2$ and $W$ belong to one of the allowed families.  Then
  if $W$ is unitary or complex non-unitary, then there exists a
  classical algorithm that takes a 2-qubits operator $V\in SU(4)$
  and a number $\epsilon>0$, and outputs the description of a tangle
  $\mcT \in GTL(d)$ with 8 in-pegs and 8 out-pegs such that
  \begin{equation}
    \left\| \Un{V} - \rho(\mcT)/\Delta_\mcT \right\|_{L_{8}} \le \epsilon \ ,
  \end{equation}
  in $poly(\log\epsilon^{-1})$ time. $\mcT$ is equal to a product of
  $poly(\log\epsilon^{-1})$ elementary crossings with weights taken
  from $W/d$, and $\Delta_\mcT$ is an overall factor, given by the
  product of all the odd weights $v/d$ that appear in $\mcT$.
  
  If $W$ is real non-unitary then there is a similar algorithm for
  any operator $V\in SO(4)$. 
 \end{thm}

\begin{proof}
  Let us start
  by proving the density and efficiency theorem for the unitary
  case. We would like to use the density theorem to create an
  $\epsilon_0$ net that would later be used by the Solovey-Kitaev
  theorem. The problem is that the density theorem uses the
  normalized crossing operators, whereas we want the density and
  efficiency theorem to be written in terms of the original crossing
  operators. The solution is to approximate the $\epsilon_0$-net
  with products in which for every normalized crossing operator
  there appears its inverse. This would either cancel the
  normalizing factors, or would make it easy to book-keep them. We
  achieve that using the following trick: let $\epsilon_0$ be as in
  Theorem \ref{thm:SK}, and consider an arbitrary $\epsilon_0/2$-net
  in $SU(K)$. Each element in the net can be written as a finite
  product of commutators in $SU(K)$, by the fact that the commutator
  group of $SU(K)$ is equal to $SU(K)$\footnote{This follows from
  the following argument: the commutator group is a normal closed
  subgroup of $SU(K)$, which contains an infinite number of
  elements. Now $SU(K)$ is an ``almost simple Lie group'', that is,
  the quotient group of $SU(K)$ divided by its center is a simple
  abstract group. In addition, it is also a connected group. It
  follows that any normal closed subgroup of it must either be in
  its center (hence, be finite) or be the whole $SU(K)$, and
  therefore the commutator group of $SU(K)$ is equal to $SU(K)$.}.
  
  Consider then an element $X$ in the $\epsilon_0/2$-net which is
  written as a product of $m$ commutators. All together $X$ is
  written as a product of $4m$ \dnote{corrected to 4 from 2}
 transformations from $SU(K)$. By
  Theorem~\ref{thm:uni-dense}, the set of normalized crossing
  operators generates a dense subgroup in $SU(K)$. Therefore by 
  approximating every element in the product to within
  $\epsilon_0/(8m)$, we can approximate $X$ to within $\epsilon_0/2$
  by a finite product of normalized crossing operators. If we now
  replace all the elements in the $\epsilon_0/2$-net with their
  $\epsilon_0/2$ approximations we get an $\epsilon_0$-net that
  consists of finite products of crossing operators. As we shall see
  shortly, the important thing is that because we use commutators,
  then each element in the new net contains, for each normalized
  generator, its inverse. 

  We can now apply the Solovay-Kitaev Theorem for this
  $\epsilon_0$-net.  This implies that we can approximate $\Un{V}$ 
  by $poly(\log\epsilon^{-1})$ normalized operators with weights
  taken from $W/d$, where each normalized operator appears in the 
  approximating sequence the same number of times as its inverse. 

  The operators we actually have to use are the unnormalized ones. 
  Since $\sigma_i=\det(\sigma_i)^\frac{1}{14} \hat{\sigma_i}$, we
  get that we have to calculate the product of the determinants of
  all the crossing operators involved, and then take the power
  $1/14$.  To prove the theorem, we need to show that this factor is
  indeed equal to $\Delta_\mcT$, the product of all the odd weights
  $v/d$ that appears in the approximation sequence.  This is where
  we use the fact that for each generator we also have its inverse. 

  The determinant of a product of matrices is equal to the product
  of the determinants; to calculate the determinant, we group
  together the \emph{unnormalized} versions of $\hat{\sigma}_i$ and
  $\hat{\sigma}_i^{-1}$.  For $i$ even, the product of those
  unnormalized operators is $1$, by \Eq{eq:eveninv}. For
  $i$ odd, the product of $\sigma_i(v/d)$ and $\sigma_i(w/d)$ for
  $v,w$ being inverses with respect to $q$, is equal to
  $\frac{vw}{q}\Id$, by \Eq{eq:oddinv}.  The determinant
  of this operator, calculated on $K$, is $(\frac{vw}{q})^{14}$.
  Taking the power $1/14$, we get that the contribution of this pair
  to the overall factor is $\frac{vw}{q}= \frac{v}{d}\frac{w}{d}$
  which is indeed the correct contribution to $\Delta_\mcT$.

  We now address the complex non-unitary case. The proof follows exactly
 the same lines as for the unitary case, and is based on 
  the non-unitary version of the Solovay-Kitaev theorem, Theorem
  \ref{thm:non-SK} (The only differences are minor technicalities, as 
we shall see).  
  We fix $R$ in the non unitary Solovay Kitaev theorem to be $3$, so that
  $B_R(K)$ contains $SU(K,\mathbbm{C})$ (this follows from the fact
  that any unitary matrix is within distance $2$ from the identity).
  We consider an $\epsilon_0/2$ net in $B_R(K)$, where each element $X$ in the 
net is a product of $m$ commutators, as before (we  use the fact
  that the commutator group of $SL(K,\mathbbm{C})$ is equal to 
   $SL(K,\mathbbm{C})$, just like in the unitary group). 
   Hence, each element is a product of $4m$ elements in  $SL(K,\mathbbm{C})$.
 By Theorem~\ref{thm:com-nu-dense}, the normalized generators
  generate a dense subgroup in $SL(K,\mathbbm{C})$, and so we can
  approximate each of the $4m$ elements to within
  $\epsilon_0/m3^m$, using a sequence (of bounded length) of the
  normalized operators. The improved accuracy of $\epsilon_0/m3^m$ is
  required because of the non-unitarity, which makes the error non-additive. 
  It is  never the less easy to bound the error using a telescopic sum, and 
  get that the  product of the approximating sequences is within
  $(\epsilon_0/m3^m)m3^{m-1}\le \epsilon_0/2$ of the
  net element.  The rest of the argument is the same.
  \inote{I don't understand this part, and why it is different from
  the unitary thing. I therefore didn't fix the error in the 
  commutator reasoning.}

  For the real non-unitary case, the argument is the same, except
  that unitary matrices are replaced by orthogonal matrices. 
\end{proof} 

It is quite easy to finish the proof of Theorem \ref{thm:hardness} 
using the density and efficiency theorem, and this is what we will
do next. This will complete the proof of hardness, except that it
remains to prove the density theorems; as mentioned before, this
will be done in Sections \ref{sec:density-proof-unitary} and
\ref{sec:density-proof-nu}. 

\subsection{Proof of Theorem~\ref{thm:hardness}}
\label{sec:hardness-proof}

Suppose we are given the quantum circuit~(\ref{eq:circuit}) over $n$
qubits, which applies the unitary matrix $U$. We will describe an
efficient reduction from approximating the Tutte polynomial to
evaluating the outcome of this circuit. Specifically, the reduction
constructs a weighted graph $G=(V,E)$ of the form that is given in
the theorem, such that an additive approximation of $Z_G(q,\Bv)$,
within the window size of the theorem, will enable to decide whether
$\big|\bra{0^{\otimes n}}U\ket{0^{\otimes n}}\big|^2\le 1/3$ or
$\big|\bra{0^{\otimes n}}U\ket{0^{\otimes n}}\big|^2 \ge 2/3$.

\subsubsection{The reduction: The unitary and the complex non-unitary case} 
For now, we restrict attention to the case of unitary or complex
non-unitary sets of parameters.  The real non-unitary case requires
some additional step, and we will deal with it at the end of this
subsection.

The reduction constructs the graph $G$ as follows. 
\begin{enumerate}
  \item For every one of the $N$ gates $U_i$ in
    \Eq{eq:circuit}, use the density and efficiency theorem
    (Theorem~\ref{thm:8-strands}) with
    $q=d^2$ and weights $W$ to find tangles $\mcT_i\in GTL(d)$ of
    polylogarithmic size such that
    $||\Delta^{-1}_{\mcT_i}\rho(\mcT_i) - \Un{U}_i||_{L_8} \le
    1/1000N$. As $N=poly(n)$ this step is done in $poly(n)$
    time.\inote{changed $1/1000N \to 1/100N$, here and in the next}
 \dnote{it is a silly issue... the error is not additive, in the non unitary case, so you have to take it into acount. returned the 1000 therefore...} 
    
  \item Every tangle $\mcT_i\in GTL(d)$ is then padded with an
    appropriate number of straight strands from its left to match
    the original two local qubits on which $U_i$ operates.
    Specifically, if $U_i$ operates on the qubits $j,j+1$ then pad
    $\mcT_i$ with $4(j-1)$ strands. It is easy to verify that now
    $||\Delta^{-1}_{\mcT_i}\rho(\mcT_i) - \Un{U}_i||_{L_{4n}} \le
    1/100N$. In other words, we now use the $L_{4n}$ norm instead
    of the $L_8$ norm.
    
  \item The tangle $\mcT=\mcT_N\cdot\ldots\cdot \mcT_1$ is
    calculated. 
    
    From the assertion
    $||\Delta^{-1}_{\mcT_i}\rho(\mcT_i)-\Un{U}_i||_{L_{4n}}\le
    1/1000N$, and using a standard telescopic argument, we have
    $||\Delta^{-1}_\mcT\rho(\mcT)- \Un{U}||_{L_{4n}}\le 1/100$,
    where $\Delta_\mcT \EqDef \prod \Delta_{\mcT_i}$ is the product
    of all odd weights $v/d$ that appear in $\mcT$. Therefore, as
    $\ket{\Un{0}^{\otimes n}}\in L_{4n}$, we get
    \begin{equation}
    \label{eq:distance2}
      \Big|\Delta^{-1}_{\mcT}\bra{\Un{0}^{\otimes n}}
         \rho(\mcT)\ket{\Un{0}^{\otimes n}} 
         - \bra{0^{\otimes n}}U\ket{0^{\otimes n}}\Big|\le 1/100 \ .
    \end{equation}
    
  \item To calculate $\bra{\Un{0}^{\otimes n}}\rho(\mcT)
    \ket{\Un{0}^{\otimes n}}$ we define a new $GTL(d)$ element $\mcT_p
    = \mcB\cdot \mcT\cdot \mcA$ by multiplying a $2n$
    cups element $\mcA$ by $\mcT$ and then by a $2n$ caps element
    $\mcB$.  $\mcT_p$ has the form of the medial graph that
    was defined in the hardness problem (see, for example,
    \Fig{fig:plat}). 
    We can therefore find a weighted graph
    $G=(E,V)$ whose medial graph $L_G$ corresponds to $\mcT_p$, and its
    weights are taken from $W$. Notice that in accordance with the
    conditions of the computational problem, every $e\in E_{odd}$ in
    $G$ has a weight taken from $W_{odd}$ and similarly for the even
    case.
    
\end{enumerate} 

We now prove 

\begin{claim} 
\label{claim:approx-U}
  The graph generated by the above reduction satisfies 
  \begin{equation}
  \label{eq:R}
     \left| \frac{1}{\Delta_{hard}} Z_G(q,\Bv) 
   - \bra{0^{\otimes n}}U\ket{0^{\otimes n}}
        \right| \le 1/100 \ .
  \end{equation}
\end{claim} 

It follows from this claim 
that it suffices to provide an additive approximation 
of  $Z_G(q,\Bv)$ to within $\Delta_{hard}/10$, in order to be able to 
distinguish between the two cases: 
$|\bra{0^{\otimes n}}U\ket{0^{\otimes n}}|^2\ge 2/3, 
|\bra{0^{\otimes n}}U\ket{0^{\otimes n}}|^2\le 1/3$. 
We now prove the claim. 

\begin{proof} 
    According to Proposition~\ref{prop:id}, $\mcT_p = \langle L_G
    \rangle \mcI$, and so $\rho(\mcT_p) = \langle L_G \rangle \Id$. As
    $\rho(\mcT_p)$ is an operator from $H_1$ to $H_1$, it follows
    that $\bra{1}\rho(\mcT_p)\ket{1} = \langle L_G \rangle$.
    
    Let us now evaluate $\bra{\Un{0}^{\otimes n}}\rho(\mcT)
    \ket{\Un{0}^{\otimes n}}$: returning to the identity $\mcT_p =
    \mcB\cdot \mcT\cdot \mcA$, we have $\rho(\mcT_p) =
    \rho(\mcB)\rho(\mcT)\rho(\mcA)$ with
    $\rho(\mcA):H_1\to H_{4n}$, $\rho(\mcT):H_{4n}\to H_{4n}$
    and $\rho(\mcB):H_{4n}\to H_1$. From \Sec{sec:pathmodel}
    we see that:
    \begin{eqnarray}
      \rho(\mcA) &=& (a_{21})^{2n}\ket{121212\ldots}\bra{1} 
       = d^{n}\ket{\Un{0}^{\otimes n}}\bra{1} \ , \\
      \rho(\mcB) &=& (b_{12})^{2n}\ket{1}\bra{121212\ldots} 
       = d^{n}\ket{1}\bra{\Un{0}^{\otimes n}} \ .
    \end{eqnarray}

    Here $\ket{1212\ldots}$ corresponds to the ``zig-zag'' path
    $1\to 2\to 1\to 2\ldots$. It is the only path in $H_{4n}$ that
    is compatible with the path $\ket{1}\in H_1$ with respect to the
    tangles $\mcA$ and $\mcB$. 
    
    We conclude by Claim \ref{cl:connection} that 
    \begin{equation}
         \bra{\Un{0}^{\otimes n}}\rho(\mcT)\ket{\Un{0}^{\otimes n}}
          = d^{-2n}\bra{1}\rho(\mcT_p)\ket{1} = d^{-2n}\langle L_G \rangle 
          = q^{-n-|V|/2} Z_G(q,\Bv) \ , 
    \end{equation}
    and therefore by \Eq{eq:distance2} we get
    \begin{equation}
    \label{eq:balagan}
      \left| 
         \Delta^{-1}_\mcT q^{-n-|V|/2} Z_G(q,\Bv) -
           \bra{0^{\otimes n}}U\ket{0^{\otimes n}}
      \right| \le 1/100 \ .
    \end{equation}
    
    Consider now the factor $\Delta^{-1}_\mcT q^{-n-|V|/2}$ that
    multiplies $Z_G(q,\Bv)$. We claim that it is exactly
    $1/\Delta_{hard}$ for $\Delta_{hard}$ defined in \Eq{def:Delta}.
    Indeed, by the definition of $\Delta_{\mcT}$ it is equal to
    \begin{equation}
      q^{-n-|V|/2} \left(\prod_{e\in E_{odd}} \frac{v_e}{d}\right)^{-1}
        = q^{|E_{odd}|/2-n-|V|/2}\prod_{e\in E_{odd}}v^{-1}_e \ .
    \end{equation}
    In addition, it is easy to verify by going from the medial graph
    back to $G$ that $|V| = 2n + |E_{odd}|$, and therefore the
    factor is simply
    \begin{equation}
      q^{|E_{odd}| - |V|}\prod_{e\in E_{odd}}v^{-1}_e =
      \frac{1}{\Delta_{hard}} \ ,
    \end{equation}
    as required.
    
\end{proof}

\subsubsection{The real non-unitary case} 

We now have to deal with the real non-unitary case.  The problem is
that in this case, the density and efficiency theorem, only
guarantees that we can approximate any gate in the orthogonal group
$SO(4)$, whereas the gates of the quantum circuit are in $SU(4)$.
We therefore start the reduction by first translating the quantum
circuit using unitary gates in $SU(4)$ to an equivalent circuit over
$n+1$ qubits whose gates are real matrices in $SO(4)$. The reduction
from unitary gates to orthogonal gates is fairly standard: 
 
\begin{claim}
  Universal quantum computation can be performed using an array of
  qubits, on which two qubit gates taken from the orthogonal group
  are applied on nearest neighbor qubits only. Moreover, any quantum
  circuit can be translated to a circuit of this form with only
  polynomial overhead. 
\end{claim}
 
\begin{proof}
  Recall the proof in \cite{bv} that real matrices suffice to
  achieve universality in quantum computation; This is done by
  adding an extra qubit which carries the information of whether the
  state is in the imaginary part of the space or the real part. We
  now use this idea in the following way. We will start with a
  quantum circuit (not necessarily one dimensional) which uses only
  one qubit gates plus CNOT gates - which is well known to be a
  universal set.

  We translate those gates to their real versions, as is done in
  \cite{bv}\inote{Ref ??}. We note that the CNOT 
  gates have not changed: in the new
  version, a CNOT gate goes to a CNOT tensor with identity on the extra
  qubit. The one qubit gate have now become two qubit gates. Hence, we
  have expressed the circuit using two qubits gates in the
  orthogonal group (namely, real). The only problem is that they may
  not be applied to nearest neighbors. We solve this in the standard
  way by adding linearly many SWAP gates for any gate; luckily, the
  SWAP gate is in the orthogonal group and is a two qubit gate, so we
  are allowed to use it.
\end{proof}

\section{Density in The unitary case}
\label{sec:density-proof-unitary} \label{sec:uni-case}

The proof of the density in the unitary case essentially follows the
same lines of Sec.~4 of Ref~\cite{ref:Aha06}.  For completeness, we
will provide most of the proof from scratch, as we will build on
this proof in the non-unitary case. We will be using two basic tools
from Ref~\cite{ref:Aha06}, without reproving them: The bridge lemma
and the decoupling lemma. 

\subsection{The overall structure of the density proof}

We now want to show density of the group generated by the normalized
operators.  Recall that the space in which we want to prove density 
is spanned by $14$ vectors, which are denoted by $\ket{p_i}$ and are
shown schematically in \Fig{fig:8p}. 

Let us analyze the action of the normalized crossing operators
$\hat{\sigma}_i(u)$ on these vectors. From \Eq{eq:crossing} we
deduce that $\hat{\sigma}_i(u)$ has the same block structure of
$\Phi_i$. The latter has a very simple structure: denoting by
$\ket{\ldots jk\ell\ldots}$ a path whose $i$, $i+1$ steps are
$j\to k\to \ell$, we find that $\Phi_i$ mixes the vectors
$\ket{\ldots j,j+1,j\ldots}$ and $\ket{\ldots j,j-1,j\ldots}$ while
returning zero whenever $\ell \ne j$ or $k\ne j\pm 1$. When $j=1$,
the path $\ket{\ldots j,j-1,j\ldots}$ does not exists, making
$\ket{\ldots j,j+1,j\ldots}$ a non-trivial eigenvector of $\Phi_i$.

Therefore $\Phi_i$ works on $1\times 1$ and $2\times 2$ blocks of
$K$. In Table~\ref{tab:blocks} we list the blocks on
which the different $\Phi_i$'s are non-zero.

\begin{table}[htbp]
  \center
    \begin{tabular}[htbp]{llllll} 
        $\Phi_1:$ & $(1)$   & $(3)$   & $(5)$    & $(7)$    & $(9)$ \\
        $\Phi_2:$ & $(1,2)$ & $(3,4)$ & $(5,6)$  & $(7,8)$  & $(9,12)$ \\
        $\Phi_3:$ & $(1)$   & $(3)$   & $(6,10)$ & $(8,11)$ & $(12,13)$ \\
        $\Phi_4:$ & $(1,5)$ & $(2,6)$ & $(3,7)$  & $(4,8)$  & $(13,14)$ \\
        $\Phi_5:$ & $(1)$   & $(2)$   & $(7,9)$  & $(8,12)$
        & $(11,13)$ \\
        $\Phi_6:$ & $(1,3)$ & $(2,4)$ & $(5,7)$  & $(6,8)$  & $(10,11)$ \\
        $\Phi_7:$ & $(1)$   & $(2)$   & $(5)$    & $(6)$    & $(10)$ 
    \end{tabular}
\caption{The block structure of the $\Phi_i$ operators in
  $K=H_{8,1\to 1}$. Vectors that do not appear in a certain row
  vanish by the action of the corresponding $\Phi_i$.}
\label{tab:blocks}
\end{table}

Consider then the group that is generated by $\hat{\sigma}_1,
\hat{\sigma}_2$. Table~\ref{tab:blocks} shows that there are five
$2\times 2$ blocks on which these operators act
non-trivially: 
\begin{itemize}
  \item $\{ \ket{p_1}, \ket{p_2}\}$  
  \item $\{ \ket{p_3}, \ket{p_4}\}$  
  \item $\{ \ket{p_5}, \ket{p_6}\}$  
  \item $\{ \ket{p_7}, \ket{p_8}\}$  
  \item $\{ \ket{p_9}, \ket{p_{12}}\}$  
\end{itemize}
The following subsection shows that under the conditions of
Theorem~\ref{thm:8-strands}, $\hat{\sigma}_1,\hat{\sigma}_2$
generate a dense subgroup of $SU(2)$ inside each block. We call this
the {\it seed}; once we have a seed of dense unitary group on some
small subspace, climbing up in the dimensionality can be done using
fairly general tools. 

\subsection{Constructing the Seed: Density in $SU(2)$}  
\label{sec:uni-seeding-proof}

We start by generating the unitary group on a two dimensional subspace. 


\begin{lem}[The unitary seeding lemma]
\label{lem:uni-seeding} 
  For $q$ and $W$ as in the unitary set of parameters, the
  normalized operators $\hat{\sigma}_1, \hat{\sigma}_2$ generate a
  dense subgroup of $SU(2)$ in each of the two-dimensional blocks
  that are listed above.
\end{lem}

\begin{proof} 
Fix a particular $SU(2)$.    
Observe that  $\hat{\sigma}_1, \hat{\sigma}_2$ are not in the $SU(2)$ 
group even though they are in $SU(K)$. This
  is because their determinant in each block need not be equal to 1
  in order for the determinant in $K$ is equal to 1. 
Let $\tau_1,\tau_2$ be the operators normalized so that their determinant 
in the two dimensional subspace is $1$. 
We will show that $\tau_1,\tau_2$ and their inverses generate 
a dense subgroup in $SU(2)$. 
By the fact that the commutator group of $SU(2)$ is equal to $SU(2)$, 
it is sufficient to approximate commutators of $SU(2)$, and thus, 
in each approximating sequence we have for every $\tau_i$ also 
its inverse. Thus, if we replace $\tau_i$ by $\hat{\sigma_i}$ 
and likewise for its inverse, we will not change the product.

Hence, it suffices to prove density of $\tau_1,\tau_2$ and their
inverses.  According to the reasoning of Theorem~4.1 in
Ref~\cite{ref:Aha06}, which is based on Theorem~5.1, page 262 in
Ref~\cite{ref:Jon83}, it is sufficient show that $\tau_1$ and
$\tau_2$ are non-commuting and generate an infinite group. The first
part is simple; $\tau_1$ and $\tau_2$ commute if and only if
$\llbracket\tau_1, \tau_2\rrbracket=\llbracket\hat{\sigma}_1,
\hat{\sigma}_2\rrbracket= \Id$. A straightforward calculation yields
\begin{equation}
  Tr \Big(\llbracket\hat{\sigma}_1, \hat{\sigma}_2\rrbracket 
    - \Id\Big)= -\frac{q-1}{q^2}
    \left(\beta-\frac{1}{\beta}\right)^2
    \left(\alpha-\frac{1}{\alpha}\right)^2 \ .
\end{equation}
Obviously, the two matrices are non-commuting provided that $q\ne
1$ and $\alpha,\beta \ne \pm 1$ - all of which cannot happen for
the unitary family of parameters, Definition \ref{def:uni}.  

To prove the other condition we use the canonical homomorphism
$\varphi:SU(2)\to SO(3)$ for which $\ker\varphi=\{\mathbbm{1},
-\mathbbm{1}\}$, and the fact that all finite subgroups of $SO(3)$
are well known and have been classified.  Let us denote by $G$ the
subgroup which is generated by $\tau_1, \tau_2$ and
assume that it is finite. In such a case, $\varphi(G)$ is a finite
subgroup of $SO(3)$, and must be one of the following five
subgroups: \inote{add refs}
  \begin{enumerate}
    \item The cyclic group $Z_n$, for some $n>1$.
    \item The Dihedral group $D_n$, for some $n>1$.
    \item The Alternating group $A_4$.
    \item The Permutation group $S_4$.
    \item The Alternating group $A_5$.
  \end{enumerate}

Each one of these possibilities is contradictory under the
conditions of the theorem. For start, the eigenvalues of
$\tau_1$ are $\alpha$ and $1/\alpha$ which are equal to
$e^{\pm i\pi s_1}$. Similarly, the eigenvalues of $\tau_2$
are $e^{\pm i\pi s_2}$. Therefore if either $s_1$ or $s_2$ are
irrational then the period of $\tau_1$ or $\tau_2$ is
infinite and consequently $G$ and $\varphi(G)$ will be infinite.

Assume then that $s_1 = p_1/q_1$ and $s_2=p_2/q_2$ with $\gcd(p_1,
q_1) = \gcd(p_2, q_2) = 1$ and that $q_1 > 5$ (the case of $q_2>5$
is similar). Then it is easy to verify that the period of
$\varphi(\tau_1)$ is $q_1>5$. However, the maximal period in
$A_4, S_4$ and $A_5$ is $3, 4$ and $5$ respectively, hence
$\varphi(G)$ is not one of these groups.  $\varphi(G)$ is also not
equal to $Z_n$ because that would imply that there exists an element
$U\in SU(2)$ and two integers $0\le \ell,m < n$ such that
$\tau_1\ker\varphi = U^\ell\ker\varphi$ and 
$\tau_2\ker\varphi = U^m\ker\varphi$, but as
$\ker\varphi=\{\pm\mathbbm{1}\}$, this would imply that
$\tau_1, \tau_2$ commute. A similar argument can be
applied to the $D_n$ case.
\end{proof}

\subsection{Building up in the Dimensionality} 

Let us now consider what happens when we also act with
$\hat{\sigma}_3$. Table~\ref{tab:blocks} tells us that
$\hat{\sigma}_3$ mixes $\ket{p_{10}}$ with $\ket{p_6}$, while, we
already know that $\hat{\sigma}_1$ and $\hat{\sigma}_2$ generate a
dense group in the $SU(2)$ of $span\{\ket{p_5}, \ket{p_6}\}$ direct
sum with the identity on $\ket{p_{10}}$.\footnote{Note that 
$\hat{\sigma}_1, \hat{\sigma}_2$ do not leave $\ket{p_{10}}$
invariant, but instead give it a phase. This phase, however, is
eliminated when we consider approximating commutators, in which 
operators appear the same number of times as their inverses.} The
three operators therefore operate inside the space $C\EqDef
span\{\ket{p_5}, \ket{p_6}, \ket{p_{10}}\}$ which is a direct sum of
two subspaces: $A\EqDef span\{\ket{p_5}, \ket{p_6}\}$ and $B\EqDef
span\{\ket{p_{10}}\}$. We can generate $SU(A)$ and $SU(B)$ (which is
trivial), and we have a transformation that bridges these two spaces
and leaves $C$ invariant, where we consider again the normalized
version of $\hat{\sigma}_3$, that has determinant $1$ on $C$.  The
following lemma assures us that in such case we can also generate a
dense group in $SU(C)$. The lemma is proved in Appendix~A.2 of
Ref~\cite{ref:Aha06}.

\begin{lem}[The unitary Bridge lemma \cite{ref:Aha06}] 
\label{thm:uni-bridge} Consider a linear space $C$ which is a direct
  sum of two subspaces $A$ and $B$, and assume that $\dim B > \dim A
  \ge 1$. Let $W\in SU(C)$ be a transformation that mixes the two
  subspaces, i.e., $W(B)\ne B$. Then any $U\in SU(C)$ can be
  approximated to an arbitrary precision using a finite sequence of
  transformations from $SU(A)$, $SU(B)$ and $W$. Consequently, the
  group generated by $SU(A)$, $SU(B)$ and $W$ is dense in $SU(C)$.
\end{lem}

Using once again similar commutator reasoning, we can get rid of the 
normalizing factor and deduce that the three generators 
$\hat{\sigma}_1,\hat{\sigma}_2,\hat{\sigma}_3$ span 
a dense subgroup of $SU(C)$. 

Returning to the space $K$, identical reasoning shows that we can
also generate $SU(3)$ over the subspaces $span\{\ket{p_7},
\ket{p_8}, \ket{p_{11}}\}$, and $span\{\ket{p_9}, \ket{p_{12}},
\ket{p_{13}}\}$. 

Let us now consider the $\hat{\sigma}_4$ transformation.
Table~\ref{tab:blocks} reveals that it bridges the subspaces
$span\{\ket{p_1},$ $\ket{p_2}\}$, and $span\{\ket{p_5}, \ket{p_6},
\ket{p_{10}}\}$, where we already have $SU(2)$ and $SU(3)$ densities
respectively. We would therefore like to use the Bridge lemma to
show that we have a $SU(5)$ density on the direct sum of these
spaces. However, the Bridge lemma cannot be directly applied here
because it assumes that we can generate $SU(2)$ and $SU(3)$
\emph{independently} of each other, whereas the transformations in
these two subspaces are generated simultaneously by $\hat{\sigma}_1,
\hat{\sigma}_2, \hat{\sigma}_3$, and hence might be coupled to each
other. The following lemma comes to our rescue; it guarantees that
we can always approximate any transformation on $span\{\ket{p_1},
\ket{p_2}\}$ while leaving $span\{\ket{p_5}, \ket{p_6},
\ket{p_{10}}\}$ invariant, and vice versa. The proof can be found in
the A.2 appendix of Ref~\cite{ref:Aha06}.

\begin{lem}[The unitary decoupling lemma \cite{ref:Aha06}] 
 \label{lem:uni-decouple} 
 Let $G$ be an infinite countable group, and let $A$, $B$ be two
 finite Linear spaces with different dimensionality. Let $\tau_a$
 and $\tau_b$ be two homomorphisms of $G$ into $SU(A)$ and $SU(B)$
 respectively and assume that $\tau_a(G)$ is dense in $SU(A)$ and
 $\tau_b(G)$ is dense in $SU(B)$. Then for any $U\in SU(A)$ there
 exist a series $\{\sigma_n\}$ in $G$ such that
  \begin{eqnarray}
    \tau_a(\sigma_n) &\to& U \, \\
    \tau_b(\sigma_n) &\to& \Id \ ,
  \end{eqnarray}
  and vice versa.
\end{lem}

From here onward, the proof is straightforward. Together with
$\hat{\sigma}_4$, we generate a dense $SU(5)$ on $span\{\ket{p_1},
\ket{p_2}, \ket{p_5}, \ket{p_6}, \ket{p_{10}}\}$ and on
$span\{\ket{p_3}, \ket{p_4}, \ket{p_7}, \ket{p_8}, \ket{p_{11}}\}$
and $SU(4)$ on $span\{\ket{p_9}$, $\ket{p_{12}}, \ket{p_{13}},
\ket{p_{14}}\}$. Finally, $\hat{\sigma}_5$ and $\hat{\sigma}_6$
bridge these three subspaces, enabling us to generate a group which
is dense in $SU(K)\sim SU(14)$. This completes the proof of
Theorem~\ref{thm:uni-dense} for the unitary case.

\section{Density in  the Non-Unitary Cases}
\label{sec:density-proof-nu}

The proof follows the same general structure of the unitary case,
but with fundamental differences.  It uses a modified versions of 
Lemmas~\ref{lem:uni-seeding}-\ref{lem:uni-decouple} to the
non-unitary case. The main difference is in the seeding lemma. 
Whereas in the unitary case, the generic behavior is that the
crossing operators generate a dense subgroup, in the non-unitary case
this is no longer true.  Density is much more difficult to achieve,
and requires restricting the parameters severely.  Naturally, the
techniques for proving density are very different than in the 
unitary case.  We use results by Jorgensen \cite{ref:Jor76} and
Sullivan \cite{ref:Sul85}, which enable us to show density in a
restricted set of parameters.  We note that providing a full
characterization of all cases which are dense remains open. 

Once density is established in the seeding lemma, the methods that
enable us to climb in dimensionality to the entire subspace $K$ are
fairly similar to the unitary case, though technical modifications
are required in the proofs of the bridge lemma and the decoupling
lemma. 

\subsection{Constructing the Seed: density in $SL(2, \mathbbm{C})$
  and in $SL(2, \mathbbm{R})$}
\label{sec:non-seeding-proof}

It is the seeding lemma that truly distinguishes the unitary from 
the non-unitary cases. The behavior of the two cases is entirely
different. As a result, the proof uses different techniques, and
applies to a much more restricted set of parameters. 
The main result of this section is: 

\begin{lem}[The non-unitary seeding lemma]
\label{lem:non-seeding} 
  For $q$ and $W$ as in the complex non-unitary case, the 
  normalized crossing operators $\hat{\sigma}_1, \hat{\sigma}_2$
  generate a group which is dense in $SL(2,\mathbbm{C})$ in each of
  the five $2\times 2$ blocks in $K$ on which they act non
  trivially:
  \begin{enumerate}
    \item $\{\ket{p_1}, \ket{p_2}\}$
    \item $\{\ket{p_3}, \ket{p_4}\}$
    \item $\{\ket{p_5}, \ket{p_6}\}$
    \item $\{\ket{p_7}, \ket{p_8}\}$
    \item $\{\ket{p_9}, \ket{p_{12}}\}$
  \end{enumerate}
  
  For $q$ and $W$ as in the real non-unitary case, the
  group is dense in $SL(2,\mathbbm{R})$. 
  
\end{lem}

Lemma~\ref{lem:non-seeding} is essentially an application of two
results by J{\o}rgensen \cite{ref:Jor76} and Sullivan
\cite{ref:Sul85}. To state them let us first define the concept of
\emph{elementary subgroup of $SL(2,\mathbbm{C})$}. The full
definition of this object is intimately connected to the theory of
complex M\"obius transformations, for which a standard introduction
can be found in Ref~\cite{ref:Bea83}. Instead of dwelling into this
rich and beautiful theory, we shall quote a theorem by Baribeau and
Ransford from Ref~\cite{ref:Bar00} that gives a necessary and
sufficient condition for a subgroup of $SL(2,\mathbbm{C})$ that
is generated by two elements $X,Y\in SL(2,\mathbbm{C})$ to be an 
elementary subgroup.
\begin{thm}[Proposition 2.1 from Ref~\cite{ref:Bar00}]
\label{thm:Baribeau} 
  Let $G=\langle X,Y\rangle$ be a subgroup of $SL(2,\mathbbm{C})$,
  and define the following 3 complex numbers:
  \begin{equation}
    \tau \EqDef Tr^2(X)-4 \ , \quad
    \tau' \EqDef Tr^2(Y)-4 \ , \quad
    \gamma \EqDef Tr^2(\llbracket X,Y\rrbracket)-2 \ .
  \end{equation}
  
  Then $G$ is an elementary group if and only if one of the three
  conditions hold
  \begin{itemize}
    \item $\tau,\tau'\in [-4,0]$ and $\gamma \in [-\tau\tau'/4,0]$ ,
    \item $\gamma=0$ , 
    \item $\tau=\gamma$ and $\tau'=-4$, or
      $\tau=-4$ and $\tau'=\gamma$,  or
      $\tau=-4$ and $\tau'=-4$ .
  \end{itemize}
\end{thm}
With this ``definition'' of an elementary group, we can now state
the J{\o}rgensen inequality:
\begin{thm}[J{\o}rgensen, \cite{ref:Jor76}]
  \label{thn:trace} If two matrices $X,Y \in SL(2,\mathbbm{C})$
  generate a non-elementary and discrete group, then
  \begin{equation}
    |\tr^2(X)-4| + |\tr(XYX^{-1}Y^{-1})-2| \ge 1 \ .
  \end{equation}
\end{thm}

Finally, we use a well-known result of Sullivan, which can be
restated in the following form
\begin{thm}[Proposition in Section 1 of Sullivan, \cite{ref:Sul85}]
  \label{thn:sl2den}
  Let $G$ be a non-elementary and non-discrete subgroup in
  $SL(2,\mathbbm{C})$. Then one of the following holds:
  \begin{itemize}
    \item $G$ is dense in $SL(2,\mathbbm{C})$.
    \item The connected component of the identity of the topological
    closure of $G$ is conjugate to $SL(2,\mathbbm{R})$, and
    consequently $G$ has a subgroup which is conjugate to a dense
    subgroup of $SL(2,\mathbbm{R})$.
  \end{itemize}
\end{thm}

A direct corollary of these three results is then
\begin{corol}
  If $X,Y\in SL(2,\mathbbm{C})$ are two non-commuting,
  diagonalizable matrices with eigenvalues $\{\alpha, 1/\alpha\}$
  and $\{\beta, 1/\beta\}$ respectively, such that
  $|\alpha|,|\beta|\ne 1$, and if either
  \begin{equation}
    |\tr^2(X)-4| + |\tr(XYX^{-1}Y^{-1})-2| < 1 \ , \mbox{or}
  \end{equation}
  \begin{equation}
    |\tr^2(Y)-4| + |\tr(XYX^{-1}Y^{-1})-2| < 1 \ , 
  \end{equation}
  then $\langle X,Y\rangle$ is dense in $SL(2,\mathbbm{C})$, or
  $\langle X,Y\rangle$ contains a subgroup which is conjugate to a dense
  subgroup of $SL(2,\mathbbm{R})$.
\end{corol}
Indeed it is easy to verify that the condition on $X,Y$ guarantees,
according to Theorem~\ref{thm:Baribeau}, that $\langle X,Y\rangle$
is a non-elementary group, hence using the J{\o}rgensen inequality
in conjunction with Sullivan's result gives the corollary. 

Returning to the proof of the seeding lemma, let us now consider, the action
of $\hat{\sigma}_1$ and $\hat{\sigma}_2$ on the $2\times 2$ block of
$\{\ket{p_1}, \ket{p_2}\}$ (the structure of these two transformations
in the other blocks is identical). There we have 
\begin{equation}
  \Phi_1 = \mat{d}{0}{0}{0} \ , \quad 
  \Phi_2 = \frac{1}{d}\mat{1}{\sqrt{d^2-1}}{\sqrt{d^2-1}}{d^2-1} \ ,
\end{equation}
and consequently
\begin{eqnarray}
  \hat{\sigma}_1 &=& \mat{\alpha}{0}{0}{1/\alpha} \ , \\
  \hat{\sigma}_2 &=& \frac{1}{(1+v_2)^{1/2}}\left(\Id + \frac{v_2}{d}\Phi_2\right) 
   \sim \mat{\beta}{0}{0}{1/\beta} \ ,
\end{eqnarray}
with $\alpha,\beta$ that are defined in Theorem~\ref{thm:hardness}.
A straight forward calculation then gives
\begin{eqnarray}
  |Tr^2\hat{\sigma}_1 - 4| 
     &=& \left|\alpha-\frac{1}{\alpha}\right|^2 \ , \\
  |Tr^2\hat{\sigma}_2 - 4| 
     &=& \left|\beta-\frac{1}{\beta}\right|^2 \ , \\
  |Tr\llbracket\hat{\sigma}_1,\hat{\sigma}_2\rrbracket - 2| 
     &=& \left|\frac{q-1}{q^2}\right|\cdot
     \left|\alpha-\frac{1}{\alpha}\right|^2\cdot
     \left|\alpha-\frac{1}{\alpha}\right|^2  \ ,
\end{eqnarray}
and plugging this into the previous Corollary proves most of the
seeding Lemma. 

What remains to show is that for $W$ in Case~(II) we have a
$SL(2,\mathbbm{C})$ density while for Case~(III) we have a
$SL(2,\mathbbm{R})$ density. To prove the first claim we use the
following lemma:
\begin{lem}
  If $H=gSL(2,\mathbbm{R})g^{-1}$ is a normal subgroup of $G$ then
  the trace of every $f\in G$ is either real or is purely imaginary.
\end{lem}
\begin{proof}
  Define $h=g^{-1}fg$. Then $Tr(h) = Tr(f)$ and 
  $h\,SL(2,\mathbbm{R})h^{-1} = SL(2,\mathbbm{R})$. Therefore for
  every $x\in SL(2,\mathbbm{R}$, $hxh^{-1}$ is a real matrix, hence
  $h^*x(h^{-1})^* = hxh^{-1}$. Therefore if we define $q=h^{-1}h^*$
  then $qxq^*=x$. This equality should hold for every $x\in
  SL(2,\mathbbm{R})$, and in particular to the family of 
  \begin{equation}
    x=\mat{0}{\lambda}{-1/\lambda}{0} \ , \quad
      \lambda\in \mathbbm{R} \ .
  \end{equation}
  Then it is a simple exercise to verify that $q$ must be equal to
  $\Id$ or $-\Id$. Consequently, $h=\pm h^*$ and therefore $Tr(h)$ is
  either real or purely imaginary.
\end{proof}

Now assume that $W$ is in case (II), yet the $G=\langle
\hat{\sigma}_1, \hat{\sigma}_2\rangle$ is not dense in
$SL(2,\mathbbm{C})$.  Let $\bar{G}$ denote the topological closure
of $\langle \hat{\sigma}_1, \hat{\sigma}_2\rangle$. Then the
connected component of the identity of $\bar{G}$ is conjugate to
$SL(2,\mathbbm{R})$, and is also a normal subgroup of
$\bar{G}$. Therefore the trace of every element of $G$ must be
either real or purely imaginary.  However, if $v_1$, $v_2$ or $q$
are not strictly real then it is easy to verify that
$Tr\hat{\sigma}_1$, $Tr\hat{\sigma}_1$, or
$Tr(\hat{\sigma}_1\hat{\sigma}_2)$, are not real or purely
imaginary, which is a contradiction.

Lastly, if $W$ is in case~(III), yet $G=\langle \hat{\sigma}_1,
\hat{\sigma}_2\rangle$ is dense in $SL(2,\mathbbm{C})$, then we can
continue the proof of Theorem~\ref{thm:hardness} in exactly the same
manner as in case~(II). We would deduce that for every unitary
circuit $U$ which is given by \Eq{eq:circuit} we could find a graph
$G=(V,E)$ with weights in $W$ such that $q^{-2n}\left(\prod_{e\in
E_{odd}}v^{-1}_e\right)Z_G(q,\Bv)$ approximates $\bra{0^{\otimes
n}}U\ket{0^{\otimes n}}$ up to $1/100$. However, under the
conditions of case~(III), the former expression is purely real
whereas the latter can be any complex number in the unit disc.
Therefore $G$ has a subgroup which is conjugate to a dense subgroup
of $SL(2,\mathbbm{R})$. Finally, note that the conjugation matrix
must be trivial since for $W$ in case~(III) the crossing operators
in $K$ are real matrices. This concludes the proof of the
Lemma~\ref{lem:non-seeding}.

\subsection{Building up in the Dimensionality: the non-Unitary case} 

The methods for building up in the dimensionality, once we have the
seed, are much more similar to those in the unitary case, even
though, there too, there are many complications; however, those seem
more technical and less fundamental than in the case of the seeding
lemmas. We delay the proof of the non-unitary bridge and decoupling 
lemmas to 
Appendices~\ref{sec:non-bridge-proof}-\ref{sec:non-decouple-proof}.
Given those lemmas, the proof of density follows the almost same
lines as in the unitary case.

\begin{lem}[The non-unitary bridge Lemma]
\label{lem:non-bridge} 

  Let $C$ be an Hilbert space over the field $F=\mathbbm{R}$ or
  $F=\mathbbm{C}$, and Let $A,B$ be subspaces of C such that
  $C=A\oplus B$ with $\dim B > \dim A$.
  
  Let $W\in SL(C, F)$ be a bridge transformation between $A$ and
  $B$, i.e., $WA\ne A$ and $WB\ne B$. Furthermore, let $SL(A,F)$ denote all the
  $\det=1$ transformations that act on $A$ while leaving $B$
  invariant, and similarly $SL(B,F)$ act only on $B$ while leaving
  $A$ invariant. Then $SL(C,F)$ is generated by $W$, $SL(A,F)$,
  and $SL(B,F)$.  More over, every element in $SL(C,F)$ can be
  represented as a \emph{finite} product of elements from $SL(A,F)$,
  $SL(B,F)$ and $W$.
\end{lem}

\begin{lem}[The non-unitary decoupling Lemma]
\label{lem:non-decouple} Let $A,B$ be linear spaces (over
  $F=\mathbbm{C}$ or $F=\mathbbm{R}$), and let $G$ be an abstract
  group.  In addition, let $\rho_A:G \to SL(A,F)$ and $\rho_B:G\to
  SL(B,F)$ be two homomorphisms, and assume that $\rho_A(G)$ is
  dense in $SL(A,F)$ while $\rho_B(G)$ is dense in $SL(B,F)$.
  Finally, assume that if $\rho_A(g_n)$ is a Cauchy series in
  $SL(A,F)$ then there exists a subseries $n_k$ for which
  $\rho_B(g_{n_k})$ is Cauchy in $SL(B,F)$ and vice versa.

  Then the following holds:
  \begin{itemize}
    \item  If $\dim A \ne \dim B$ then we have a full decoupling: 
      for every $V_A\in SL(A,F)$ and $V_B\in SL(B,F)$ we can find a
      series $\{g_n\}\in G$ such that $\rho_A(g_n) \to V_A$ while
      $\rho_B(g_n) \to V_B$.
    \item If $\dim A = \dim B$ then either we have a full
    decoupling, or, for every series $\{g_n\}$ in $G$ for which
    $\rho_A(g_n) \to \mathbbm{1}$ we can find a subseries $\{g_{n_k}
    \}$ for which $\rho_B(g_{n_k})$ converges to some element in the
    \emph{center} of $SL(B,F)$, and vice versa.

  \end{itemize}
\end{lem}

Notice that in order to use the bridge Lemma in case (3), we must be
sure that the bridge operators are in $SL(K,\mathbbm{R})$. This is
where the condition $q>4\cos^2\pi/5$ is needed: recall from
definition~\ref{def:rep} that the path representation for
$d=\sqrt{q}$, is determined by the coordinates of the infinite
eigenvector $\bar{\pi}$ of the auxiliary graph $F_\infty$ for
eigenvalue $d$. The first five coordinates are given by
$\bar{\pi}=(1,d,d^2-1,d^3-2d, d^4-3d^2+1, \ldots)$, and it is easy
to see that for $d>2\cos\pi/5$ they are all positive. These are the
coordinates that define the $\Phi_i$ operators over $K$. Their
positivity implies that the matrices of $\Phi_i$ in $K$ are real. 

If in addition to that, $v_1, v_2$ (and consequently $w_1,w_2$) are
real then it follows that crossing operators have real matrices in
$K$. The normalized crossing operators in that case will have
matrices which are either real (when the determinant of the
un-normalized operator is positive) or purely imaginary (when the
determinant is negative). In the first case, they belong to
$SL(K,\mathbbm{R})$, while in the second case we can always use the
square of matrix, which will be in $SL(K,\mathbbm{R})$, as a bridge
instead of matrix itself. 

Another important difference between the unitary and the non-unitary
cases is found in the extra condition of the non-unitary decoupling
lemma: to use that lemma, we must assume that for any element in
$SL(A)$ we can find a series that converges to it while also
converging in $SL(B)$, and vice-versa.

This always happen for the following reason: we start the proof by
looking at the subgroups that are generated by $\hat{\sigma}_1$ and
$\hat{\sigma}_2$. These are either $2\times 2$ subgroups that are
dense in $SL(2)$, or they are trivial unit transformation on the one
dimensional subspaces. But since every $2\times 2$ subspaces are
isomorphic to each other by conjugation (this follows from the fact
that $\hat{\sigma}_1$ and $\hat{\sigma}_2$ look the same at every
subspace up to a change of basis), it follows that at least at that
initial step, our subgroups fulfill the requirements of the
non-unitary decoupling lemma. 

But what about the next steps, after we use the bridge lemma? Also
in those steps, the subgroups in question always fulfill that extra
condition. This follows from the fact that the non-unitary bridge
lemma promises us that it is possible to generate every
transformation in $SL(A\oplus B)$ using a \emph{finite} number of
transformations from $SL(A)$, $SL(B)$ and the bridge $W$. Therefore,
if the extra condition holds for $SL(A)$ and $SL(B)$ it must also
hold for $SL(A\oplus B)$. \inote{its a bit vauge, I know, but its
true}.

This completes the proof of the density theorem in the non-unitary
cases.


\section{$\BQP$-Completeness: matching the different sizes of the 
  approximation window}
\label{sec:complete}  

In this section we will match the algorithmic scale of
\Sec{sec:algorithm} with the hardness scale of
\Sec{sec:universality}, thereby proving completeness. As stated
previously, the two scales do not match in all cases; a
straightforward matching can be done only in the unitary case. In
the non-unitary case, a slight modification of the computational
problem has to be done first; the so-called ``grouping'' of the
crossings. We will show that in this form, the approximation of the
Tutte polynomial for $q=3$ in the non-unitary cases and weights for
which we have hardness, is also $\BQP$-complete \dnote{which
parameters? and we should put a sentence about this in the
introduction, perhaps shorten the sentence here once this is done.}
\inote{is that more clear now?} We believe that our techniques can
be further generalized to many other values of $q$. Such a
generalization, however, is outside the scope of this paper.

\subsection{Completeness in the unitary case}

In the unitary case, the computational problem of
Theorem~\ref{thm:alg} is $\BQP$-complete. Formally, we have the
following theorem:
\begin{thm}[$\BQP$-completeness in the unitary case]
\label{thm:uni-complete}
  Consider a unitary set of parameters $(q,W)$, as in
  Definition~\ref{def:uni}. Then the following computational problem
  is $\BQP$-complete:
  \begin{description}
    \item [Input:] A graph $G=(E,V)$ which is nicely embedded in
      $\mathbbm{R}^2$, with edge weights from $W$.
      
    \item [Output:] Approximate the multivariate Tutte-polynomial
      $Z_G(q,\Bv)$ to within an additive approximation
      $\Delta_{alg}(G)/poly(|V|)$, where $\Delta_{alg}$ is given in
      Definition~\ref{def:window}.  
  \end{description}
\end{thm}

\begin{proof}
  By Theorem~\ref{thm:alg} the above problem is in $\BQP$. It
  remains to show that it is also $\BQP$-hard. Consider then the
  conditions of Theorem~\ref{thm:hardness}. We will show that
  $\Delta_{alg}=\Delta_{hard}$ for the special type of graphs that
  are used in the $\BQP$-hard problem.
    
  Assume then that $G=(V,E)$ a graph that is used in the $\BQP$-hard
  problem, and consider its medial graph $L_G$.  It is a
  concatenation of $2n$ cups, followed by $poly(n)$ crossings (with
  weights taken from $W$), which are then closed by $2n$ caps (see
  \Fig{fig:plat}). Let us calculate the norm of each one of these
  elements. We begin with the cups/caps:
  \begin{lem}
    \label{lem:hermitian-norm}
      For an Hermitian representation, the norm of the cup (cap)
      operator is exactly $\sqrt{d}=q^{1/4}$.
  \end{lem}
  \begin{proof}
      For every path $p$ and a cup element $\mcA_i$ we have
      \begin{equation}
        || \rho(\mcA_i)\ket{p}||^2 = 
        \bra{p}\rho(\mcA_i)^\dagger \rho(\mcA_i)\ket{p} 
        = \bra{p}\rho(\mcB_i)\rho(\mcA_i)\ket{p} = 
        \bra{p}\rho(\mcB_i \mcA_i)\ket{p} = \bra{p}d\Id\ket{p} = d \ .
      \end{equation}
      The second equality follows from the Hermiticity of the
      representation (Definition~\ref{def:hermitian}), whereas the
      forth equality follows from the fact that $\mcB_i\mcA_i$ is
      simply a loop, hence is equal to $d\mcI$. Therefore
      $||\rho(\mcA_i)||=\sqrt{d}$. 
      
      Similarly, for $\rho(\mcB_i)$ we have
      \begin{equation}
        || \rho(\mcB_i)\ket{p}||^2 =
        \bra{p}\rho(\mcA_i \mcB_i)\ket{p} \ ,
      \end{equation}
      But $\mcA_i\mcB_i=\mathcal{E}_i$, hence $\rho(\mcA_i\mcB_i) =
      \Phi_i$, which, for Hermitian representation, is an Hermitian
      operator with eigenvalues $d,0$ (see \Sec{sec:uni-non-uni}).
      Consequently, $||\rho(\mcB_i)||=\sqrt{d}$ as claimed.
  \end{proof}
    
  Returning to the main proof, we are left with $poly(n)$ crossings
  of two types, odds and evens. Odd crossings are crossings between
  $i\leftrightarrow (i+1)$ strands with odd $i$. Recall from
  Definition~\ref{def1:crossing}, that the operators that correspond
  to these crossings are $\sigma(u)=u\Id + \Phi_i$ with eigenvalues
  $\{u+d, u\}$ for $u\in W/d$. In addition, the unitarity condition
  implies that $|u+d|=|u|$, and therefore the norm of the crossing
  operators is 
  \begin{equation}
      || \sigma(u) || = |u| = |q|^{-1/2}|v| \ , \quad 
        \mbox{odd case} \ .
  \end{equation}
  Similarly, in the even case $\sigma(u) = \Id + u\Phi$ with
  eigenvalues $\{1, 1+ud\}$ and a unitary condition $|1+ud|=1$ that
  implies
  \begin{equation}
      || \sigma(u) || = 1  \ , \quad 
        \mbox{even case} \ .
  \end{equation}
    
  Gathering all terms, we find that the product of all norms is
  \begin{equation}
      \left|q^{n-|E_{odd}|/2} \prod_{e\in E_{odd}} v_e\right| \ ,
  \end{equation}
  hence
  \begin{equation}
      \Delta_{alg} = \left|q^{|V|/2 + n-|E_{odd}|/2} \prod_{e\in E_{odd}} v_e\right| \ .
  \end{equation}
  Finally, as already noted in the proof of
  Theorem~\ref{thm:hardness}, the particular structure of $G$
  yields $|V| = 2n + |E_{odd}|$, hence
  \begin{equation}
      \Delta_{alg} = \left|q^{|V|-|E_{odd}|} \prod_{e\in E_{odd}}
      v_e\right| = \Delta_{hard} \ .
  \end{equation}
\end{proof}

\subsection{Completeness in the non-unitary case}

Before giving an exact definition of the completeness result in the
non-unitary case, let us try to explain the problem in a somewhat heuristic
manner.

\subsubsection{Why the scales do not match in the non-unitary case}
\dnote{i am not pleased with this subsection... will change it later} 
As mentioned in the beginning of this section, in the non-unitary
case, it is no longer true that $\Delta_{alg}=\Delta_{hard}$. 
 To understand this, let us try a naive approach to prove the
equality.

Re-examining the argument that lead to \Eq{eq:distance2} in
\Sec{sec:universality}, reveals that when we construct the $GTL(d)$
element that corresponds to a quantum circuit $U$, we find
elements $\mcT_i$ such that
\begin{equation}
\label{eq:hard-res}
  || \rho(\mcT_1\cdot \mcT_2\cdot\ldots) - \Delta_\mcT \Un{U} ||_{L_{4n}} 
    \le 1/100 \ .
\end{equation}
Note, however, that unlike the notation in \Eq{eq:distance2}, here
we consider $\mcT_i$ to be the elementary crossings rather than
the product of crossings that results from the density and
efficiency theorem (Theorem \ref{thm:8-strands}).

In the unitary case, we examined every crossing $\mcT_i$, and
concluded that
\begin{equation}
\label{eq:break-norm}
  ||\rho(\mcT_1)||\cdot||\rho(\mcT_2)||\cdot\ldots = |\Delta_\mcT| \ .
\end{equation}
Then after taking into account the norms of the cup and caps
operator, it easily follows that concluded that
$\Delta_{alg}=\Delta_{hard}$. Can we do the same in the non-unitary
case? Equation~(\ref{eq:hard-res}) gives us
\begin{equation}
\label{eq:L4-equality}
  || \rho(\mcT_1\cdot \mcT_2\cdot\ldots)||_{L_{4n}} \simeq
    || \Delta_\mcT \Un{U} ||_{L_{4n}} = |\Delta_\mcT| \ ,
\end{equation}
where $\simeq$ means equality up to an order of unity. We would
like to show that the LHS is equal to $||\rho(\mcT_1)||\cdot
||\rho(\mcT_2)||$. This is, however, not true in the unitary case,
and the reason is twofold: firstly, in the non-unitary case, the
norm of a product of operators is always smaller than or equal to
the norm of the product. Therefore $||\rho(\mcT_1\cdot
\mcT_2\cdot\ldots)||_{L_{4n}} \le ||\rho(\mcT_1)||_{L_{4n}} \cdot
||\rho(\mcT_2)||_{L_{4n}}\cdot\ldots$. Secondly, the norm over $L_{4n}$
is always smaller than or equal to the overall norm:
$||\rho(\mcT_1)||_{L_{4n}} \cdot ||\rho(\mcT_2)||_{L_{4n}}\cdot\ldots \le
||\rho(\mcT_1)||\cdot ||\rho(\mcT_2)||\cdot\ldots$.

\subsubsection{Motivating the Solution} 
Our way to overcome these problems is to take the product of
elementary tangles that a nicely embedded medial graph defines, and
group neighboring elements: 
\begin{equation}
\label{eq:grouping}
  L_G \mapsto
  \overbrace{\mcT_1\cdot\ldots}^{\mcS_1}\cdot
  \overbrace{\mcT_{i_2}\cdot\ldots}^{\mcS_2}\cdot
  \overbrace{\mcT_{i_3}\cdot\ldots}^{\mcS_3} 
  = \mcS_1\cdot\mcS_2\cdot\mcS_3\cdot\ldots \ ,
\end{equation}
where
\begin{equation}
\label{def:group}
  \mcS_j \EqDef \mcT_{i_j}\cdot\ldots\cdot 
         \mcT_{i_{j+1}-1} \ .
\end{equation}
In other words, the quantum algorithm applies the matrices that
correspond to $\mcS_j$, rather then applying the matrices that
correspond to each $\mcT_i$ individually.  The new algorithmic scale
is thus the product of $||\rho(\mcS_j)||$ instead of the product of
the norms of the individual operators $\rho(\mcT_i)$.
$||\rho(\mcS_j)||$ is smaller than or equal to the norms of the
operators that make up this product. This forces us to \emph{change
the definition of the computational problem} by adding the grouping
information to the input to the problem. The grouping, however, must
be done in such a way that the resulting operators $\rho(\mcS_i)$
still work on logarithmically many strands, so that we can first
compute the exact matrix form of the operator efficiently, and then
simulate the relevant operator efficiently by the quantum algorithm.

The grouping of the operators gives us a lot of additional power.
Even if the crossing operators are non unitary, we may still group
them into an almost unitary (up to an overall constant) operators,
thereby bounding their norm. This is precisely the idea of the
non-unitary completeness. After all, the density and
efficiency theorem~\ref{thm:8-strands} provides us with a natural
grouping of the crossing operators. It gives us a product of a
polylogarithmic number of crossing operators that approximate a
unitary gate $U_i$. It seems that if we use this grouping as an
input to the quantum algorithm, the grouped operators $\rho(\mcS_i)$
will become (approximately) unitary (up to an overall constant), and
we will be able to prove completeness. There is a small caveat here,
however. The product of the crossing operators approximate the
(encoded) unitary gate $\Un{U}_i$ only over the subspace
$K=H_{8,1\to 1}$ of the 8-strands space $H_8$. In order for the
product to be (approximately) unitary, we must ``fix'' its behavior
on the other subspaces that make up $H_8$. We must make sure that
the eigenvalues of $\rho(S_i)$ have (approximately) the same
magnitude in all the subspaces of $H_8$. Changing the definition of
the computational problem is therefore not enough; we must also
improve the result of the density and efficiency theorem.
The exact definition of the new computational problem, and the
necessary modifications to the universality theorem are now given
below.

\subsubsection{Statement of Results} 

\begin{deff}[A nicely embedded graph with grouping and its
  scale] Consider a graph $G=(V,E)$ whose medial $L_G$ is nicely
  embedded in $\mathbbm{R}^2$. The embedding maps $L_G$ to a product
  of elementary tangles (cups, caps and crossings): $L_G\mapsto
  \mcT_1\cdot\mcT_2\cdot \ldots\cdot\mcT_\ell$. A grouping of these
  elements is a partition of the product into sub-products
  $\mcS_j=\mcT_{i_j}\cdot \ldots\cdot\mcT_{i_{j+1}-1}$, as shown in
  Eqs.~(\ref{eq:grouping}, \ref{def:group}). The partition must be
  such that any $\rho(S_j)$ operates on the tensor product of at
  most polylogarithmically many local registers.
  
  The scale of such a grouped graph is then
  \begin{equation}
    \Delta_{grp} \EqDef q^{|V|/2}\cdot||\rho(\mcS_1)|| \cdot
       ||\rho(\mcS_1)|| \cdot\ldots
       \cdot||\rho(\mcS_N)|| \ .
  \end{equation}
\end{deff}
The universality result for the non-unitary case can now be
stated exactly,
\begin{thm}[$\BQP$-completeness in the non-unitary case]
\label{thm:non-uni-complete} 

  For $q=3$ and a corresponding non-unitary set of weights $W$
  (either complex or real, as given in
  Definitions~\ref{def:case-II},\ref{def:case-III}), the following
  problem is $\BQP$-complete:
  \begin{description}
    \item[Input:] A nicely embedded graph $G=(E,V)$ with grouping.
    \item[Output:] An additive approximation of $Z_G(q,\Bv)$ to
    within an error of $\Delta_{grp}/poly(|G|)$
  \end{description}
\end{thm}

To prove the theorem, we will need a stronger version of the density
and efficiency theorem. As hinted above, we must use an alternative
version of the density and efficiency theorem in which we
approximate any two-qubits gate on the subspace $K=H_{8,1\to 1}$,
\emph{while controlling the behavior of the other subspaces of $H_8$, so
that the operator there has the same, or smaller norm as in
$K$.}

\begin{thm}[The improved density and efficiency theorem]
\label{thm:improved}

  Let $q=3$ and $W$ be a corresponding set of non-unitary weights
  (either real or complex, as given in
  Definitions~\ref{def:case-II},\ref{def:case-III}). Then there
  exists a classical algorithm that takes a two-qubits gate $U$
  ($U\in SU(4)$ in the complex case and $U\in SO(4)$ in the real
  case), and a number $\epsilon>0$ and outputs the description of a
  $\mcT \in GTL(d)$ with 8 in-pegs and 8 out-pegs such that
  \begin{equation}
     \left\|\Un{U} - \Delta_\mcT^{-1}\rho(\mcT)\right\|_{L_{8}}
      \le \epsilon \ ,
  \end{equation}
  and in addition
  \begin{equation}
    \left\|\Delta_\mcT^{-1}\rho(\mcT)\right\|_{H^*_8} 
      \le 1+\epsilon \ .
  \end{equation}
  $H^*_8$ is the space of \emph{all} 8-steps paths over the
  auxiliary graph of $d=\sqrt{3}$ - not just those that start at
  $1$. $\mcT$ is equal to a product of $poly(\log\epsilon^{-1})$
  crossings with weights taken from $W/d$. The factor
  $\Delta_\mcT$ is the absolute value of the product of all the odd
  weights $v/d$ that appear in $\mcT$. Finally, the running time of
  the algorithm is $poly(\log\epsilon^{-1})$.
  
\end{thm}

\begin{proof} 
The proof is given in Appendix~\ref{sec:improved}.
\end{proof} 

With this result, the proof of Theorem~\ref{thm:non-uni-complete} is
fairly straightforwards:

\begin{proofof}{Theorem~\ref{thm:non-uni-complete}}

  It is easy to see that the problem is in $\BQP$. Indeed, we can
  use the proof of the quantum algorithm in \Sec{sec:alg-proof}
  with the only difference that now we use the quantum computer
  to implement the operators $\rho(\mcS_j)$ of grouped elements
  $\mcS_j$ instead of the individual cups/caps and crossings. This
  is possible because we are guaranteed that each $\rho(\mcS_j)$
  works only on the tensor product of polylogarithmically many
  registers. Therefore all the classical computations that we need
  to do on $\rho(\mcS_j)$, such as computing its norm, or its polar
  decomposition, can be done efficiently.
  
  Let us now prove that the problem is $\BQP$-hard. The proof
  follows closely the proof of the universality
  theorem~\ref{thm:hardness} (see \Sec{sec:hardness-proof}). We
  are given a description of a quantum circuit over $n$ qubits as a
  product of $N$ two-qubits gate $U=U_N\cdot\ldots\cdot U_1$, and
  are asked to decide whether $|\bra{0^{\otimes n}} U\ket{0^{\otimes
  n}}|^2 \le 1/3$ or $|\bra{0^{\otimes n}} U\ket{0^{\otimes n}}|^2
  \ge 2/3$. The only difference is that now we use the improved
  density and efficiency theorem~\ref{thm:improved} to approximate
  every two-qubits gate $U_i$. The result is a graph $G$ which,
  according to claim~\ref{claim:approx-U} satisfies
  \begin{equation}
       \left| \frac{1}{\Delta_{hard}} Z_G(q,\Bv) 
       - \bra{0^{\otimes n}}U\ket{0^{\otimes n}}
      \right| \le 1/100 \ .
  \end{equation}
  As previously mentioned, $G$ is endowed with a natural grouping
  that is provided by the improved density and efficiency theorem:
  the crossings that approximate a single two-qubits gate are
  grouped together, while every cup/cap element has a group of its
  own. This is a legitimate grouping of $G$ since every product of
  crossing acts on the tensor product of 9 local registers (8
  strands). We will prove that the problem is $\BQP$-hard by showing
  that for this particular grouping, $\Delta_{hard}\simeq
  \Delta_{grp}$ up to an overall factor of the order one.
  
  Let $\mcS_j$ be the product of elementary crossings that
  corresponds to the gate $U_i$. Following the universality proof
  from \Sec{sec:hardness-proof}, it is created by using the improved
  density and efficiency theorem with $\epsilon=1/100N$.
  Consequently, we are promised that $||\rho(\mcS_j)||_{H^*_8}
  \simeq \Delta_{\mcS_j}$ up to a factor of order $1\pm 1/100N$. The
  crucial point here is that the above condition is casted in terms
  of the $H_8^*$ norm rather than the, weaker, $L_8$ norm.  $H_8^*$ is
  the space of \emph{all} 8-steps paths on the auxiliary graph of
  $d=\sqrt{3}$, hence $||\rho(\mcS_j)||_{H^*_8}=||\rho(\mcS_j)||$.
  Consequently,
  \begin{equation}
    \prod_{j=1}^N ||\rho(\mcS_j)||
      \simeq \prod_{j=1}^N \Delta_{\mcS_j} 
      = \prod_{e\in E_{odd}} |v_e/d| \ ,
  \end{equation}
  with $\simeq$ indicating equality up to a factor of order one, and
  $E_{odd}\subset E$ is the set of edges with odd weights in $G$.
  
  Lastly, we note that for $q=3$, we have $d=2\cos\pi/6$, therefore
  the path representation is Hermitian (see \Sec{sec:hermitian}). In
  such case, according to Lemma~\ref{lem:hermitian-norm}, the norm
  of every cup/cap operator is exactly $d^{1/2}=q^{1/4}$.
  Multiplying all these norms together, we get
  \begin{equation}
    \Delta_{grp} = q^{|V|/2}\cdot \left(q^{1/4}\right)^{4n}
          \cdot\prod_{j=1}^N ||\rho(\mcS_j)||
      \simeq q^{n+|V|/2}\prod_{e\in E_{odd}} |v_e/d| \ ,
  \end{equation}
  and as $|V|=2n+|E_{odd}|$, we conclude that
  \begin{equation}
    \Delta_{grp} \simeq
      q^{|V|-|E_{odd}}\left(\prod_{e\in E_{odd}}|v_e|\right) =
      \Delta_{hard} \ .
  \end{equation}
\end{proofof}

\section{Acknowledgments} 

We are grateful to Vaughan Jones for inspiring discussions; 
D.A. is grateful also to Umesh Vazirani for important seemingly 
unrelated discussions.




\appendix

\section{Tools for Building up the Dimensionality}

\subsection{Proof of the non-unitary Bridge lemma
  (Lemma~\ref{lem:non-bridge})}
\label{sec:non-bridge-proof}

The non-unitary Bridge lemma can be proved with the aid of 4 smaller
lemmas. For brevity, we do not write which of the cases
$F=\mathbbm{C}$ or $F=\mathbbm{R}$ we treat since the proof is the
same in both cases. 

\begin{lem}[Lemma A]
  Let $C=A\oplus B$ with $\dim B> \dim A$, and let $W\in SL(C)$.
  Then there exists a vector $b\in B$ such that $W\ket{b}\in B$.
\end{lem}
\begin{proof}
  Consider the subspace $WB$. If $WB\cap B = \emptyset$ then
  $\dim(WB\oplus B)=\dim(WB) + \dim(B) = 2\dim(B) > \dim C$, which is
  a contradiction.  
\end{proof}

\begin{lem}[Lemma B]
  Under the same conditions of the bridge lemma, it is possible to
  generate a transformation $T_{\psi\phi}\in SL(C)$ that would take
  the vector $\ket{\psi}$ to the vector $\ket{\phi}$.
\end{lem}

\begin{proof}
  For each vector $\ket{\psi}\in C$ we will generate a
  transformation $T_{\psi b}$ where $\ket{b}\in B$ is some fixed
  vector. This will prove the lemma since we can define
  $T_{\psi\phi} = T^{-1}_{\phi b}T_{\psi b}$.
  
  \begin{itemize}
    \item \emph{The $\dim A=1$ case}
    
    Let $A$ be spanned by the vector $\ket{a}$. As $WB\ne B$, we
    pick $\ket{b}\in B$ such that $W\ket{b} = \alpha\ket{a} +
    \beta\ket{b'}$ with $\alpha\ne 0$. By virtue of Lemma A we can
    also assume that $\beta\ne 0$ for we can always add to
    $\ket{\beta}$ a vector from $B$ whose image under $W$ is inside
    $B$.
    
    Let $\ket{\psi}=\alpha_0\ket{a} + \beta_0\ket{b_0}$, and assume
    for a start that $\alpha_0\ne 0$ and $\beta_0 \ne 0$. Then we
    perform an $SL(B)$ transformation that takes $\ket{b_0} \to
    \frac{\alpha_0\beta}{\alpha\beta_0}\ket{b'}$. We get
    $\ket{\psi_1}=
    \frac{\alpha_0}{\alpha}(\alpha\ket{a}+\beta\ket{b})$. Acting
    with $W^{-1}$ we obtain $\ket{\psi_2} =
    \frac{\alpha_0}{\alpha}\ket{b}$ which can be then scaled to
    $\ket{b}$ using yet another $SL(B)$ transformation.

    If $\alpha_0=0$ then $\ket{\psi}\in B$ and we can simply move it
    to $\ket{b}$. If $\beta_0=0$ then $\ket{\psi}\in A$ and so
    $W\ket{\psi}$ must have some projection on $B$, and we return to
    the previous cases.
    
    \item \emph{The $\dim A>1$ case}
    
    As in the first case, we pick pick $\ket{b}\in B$ such that
    $W\ket{b} = \alpha\ket{a'} + \beta\ket{b'}$ with $\alpha\ne 0$
    and $\beta\ne 0$.
    
    Let $\ket{\psi}=\alpha_0\ket{a} + \beta_0\ket{b_0}$, and assume
    for a start that $\alpha_0\ne 0$ and $\beta_0 \ne 0$. Then we
    perform an $SL(B)$ transformation that takes $\ket{b_0} \to
    \frac{\beta}{\beta_0}\ket{b'}$, and an $SL(A)$ transformation
    that takes $\ket{a_0}$ to $\frac{\alpha}{\alpha_0}\ket{a}$. We
    get $\ket{\psi_1}=(\alpha\ket{a}+\beta\ket{b})=W\ket{b}$. Then
    Acting with $W^{-1}$ we obtain $\ket{b}$.
    
    If $\alpha_0=0$ is handled as in the
    $\dim A=1$ case. In the $\beta_0=0$ case, $\ket{\psi}\in A$ and
    hence using a $SL(A)$ transformation it can be moved to another
    vector in $A$ whose image under $W$ has some projection on $B$,
    after which we proceed as before.
    
  \end{itemize}    
 
\end{proof}

\begin{lem}[Lemma C]
  Under the same conditions of the bridge lemma, there exists a
  basis (not necessarily orthogonal) $\{\ket{a_1}, \dots,
  \ket{a_m}\}$ of $A$, a basis $\{\ket{b_1}, \dots, \ket{b_n}\}$ of
  $B$, and a transformation $V\in SL(C)$ that we can generate, which
  is defined as follows: $V\ket{a_1}=\ket{b_1}$,
  $V\ket{b_1}=-\ket{a_1}$ and $V$ is leaves the rest of the bases
  elements unchanged
\end{lem}

\begin{proof}
  
  Pick any vector $\ket{a_1}\in A$ and a vector $\ket{b'}\in B$.
  Then by Lemma B there exists a transformation $T$ that takes
  $\ket{a_1}$ to $\ket{b'}$. Use lemma A with the transformation $T$
  to find vectors $\ket{b_1},\ket{b''}\in B$ such that
  $T\ket{b_1}=\ket{b''}$. Obviously $\ket{b''}\ne \ket{b'}$ and
  therefore they span a subspace of $B$ with dimension 2. Denote by
  $S$ the orthogonal complementary of this subspace in $C$. Then
  $\dim S = \dim A + \dim B -2$. We now construct a transformation
  $U\in SL(B)$ such that $U\ket{b'}=\ket{b''}$ and
  $U\ket{b''}=-\ket{b'}$, while leaving $S$ unchanged. 
  
  Now consider the transformation $V=T^{-1}UT$. It has the following
  properties:
  \begin{itemize}
    \item $V\ket{a_1} = \ket{b_1}$
    \item $V\ket{b_1} = -\ket{a_1}$
    \item $V$ leaves the space $R\EqDef T^{-1}S$ unchanged. This is because
    for every $\ket{s}\in S$, we have 
    \begin{equation}
      V T^{-1}\ket{s} = T^{-1}UT T^{-1}\ket{s}= T^{-1}U\ket{s}
      = T^{-1}\ket{s} \ .
    \end{equation}
  \end{itemize}

  $V$ is a non-singular transformation hence $\dim R = \dim S = \dim
  A + \dim B - 2$. Moreover, as $\ket{a_1}, \ket{b_1}$ are not in
  $R$ then $C=span\{R, \ket{a_1}, \ket{b_1}\}$. It follows that
  $\dim (R\cap A) = \dim A -1$. Indeed if $\dim(R\cap A) \ge \dim A$
  then $R$ must include $A$, contradicting the fact that
  $\ket{a_1}\notin R$. On the other hand, if $\dim(R\cap A)<\dim A
  -1$ then $A$ must contains a vector other than $\ket{a_1}$ which
  is not in $R$ - contradicting the fact that $C=span\{R, \ket{a_1},
  \ket{b_1}\}$. Similarly, we may prove that $\dim(R\cap B) = \dim B
  -1$. 
  
  Finally, we choose a basis $\{\ket{a_2}, \ldots, \ket{a_m}\}$ for
  $R\cap A$ and a basis $\{\ket{b_2}, \ldots, \ket{b_n}\}$ for
  $R\cap B$. Then $\{\ket{a_1}, \ldots, \ket{a_m}\}$ is a basis of
  $A$ and $\{\ket{b_1}, \ldots, \ket{b_n}\}$ is a basis of $B$ as
  required.
\end{proof}

\begin{lem}[Lemma D]
  Let $B$ be a linear space with $\dim B>1$, and let $\ket{a}$ be a
  vector outside of $B$ and define the space $C=B\oplus
  span\{\ket{a}\}$. Let $W\in SL(C)$ be a transformation that mixes
  $B$ with $span\{\ket{a}\}$. Then using $W$ and transformations in
  $SL(B)$ (which do not affect the $\ket{a}$ vector), we can generate
  $SL(C)$.
\end{lem}

\begin{proof}
  Let $V$ be a transformation in $SL(C)$. Define
  $\ket{\psi}=V\ket{a}$. Then according to Lemma B we can generate a
  transformation $T$ that also takes $\ket{a}$ to $\ket{\psi}$. 
  
  Consider now the transformation $X=T^{-1}V$. By generating it we
  will prove the lemma. $X$ leaves $\ket{a}$ invariant. If, in
  addition, $X\in SL(B)$ then we are done. Assume then that $X\notin
  SL(B)$, and pick a basis $\ket{b_1}, \ldots, \ket{b_n}$ of $B$. 
  The action of $X$ on $B$ can be described by 
  \begin{equation}
     X\ket{b_i} = \sum_j X^{r}_{ij}\ket{b_i} + x_i\ket{\alpha}
  \end{equation} 
  Here $X^{r}$ is a $n\times n$ matrix. Since $\det X = 1$, and
  $X\ket{a}=\ket{a}$, it follows that $\det X^r = 1$. Also notice
  that as $XB\ne B$ then at least of $x_i$ must be non-zero.
  
  We now wish to generate a transformation $Y$ that similarly to $X$,
  leaves $\ket{\alpha}$ invariant while $YB\ne B$. Let $\tilde{T}$
  be a transformation that takes $\ket{a}\to \ket{b_1}$, and
  consider transformations $Y$ of the form
  $Y=\tilde{T}^{-1}U\tilde{T}$ where $U$ is any transformation in
  $SL(B)$ that leaves $\ket{b_1}$ invariant.  Then
  $Y\ket{a}=\ket{a}$. We claim that there must be a $U$ that creates
  $Y$ such that $YB\ne B$. Indeed if $YB=B$ then
  $U\tilde{T}B=\tilde{T}B$, i.e., $U$ preserves the subspace
  $\tilde{T}B$. However, this is a contradiction since
  $\ket{b_1}\notin \tilde{T}B$ and at the same time by Lemma A there
  are $\ket{b'}, \ket{b''}\in B$ such that
  $\tilde{T}\ket{b'}=\ket{b''}$. So we can construct a $U$ that
  takes $\ket{b''}\to \ket{b''}+\ket{b_1}$ (which is outside
  $\tilde{T}B$) while leaving $\ket{b_1}$ invariant.
  
  We conclude that
  \begin{equation}
     Y\ket{b_i} = \sum_j Y^r_{ij}\ket{b_i} + y_i\ket{\alpha}
  \end{equation} 
  with $Y^r$ being an $n\times n$ matrix  with $\det=1$ and at least
  one $y_i$ is non-zero. 
  
  Let $N\in SL(B)$ be the transformation that in the $\{\ket{b_1},
  \ldots, \ket{b_n}\}$ basis is given by $N^r_{ij}$ such that 
  \begin{equation}
    \sum_j N^r_{ij}y_j = x_i \ .
  \end{equation}
  Similarly let $M\in SL(B)$ be such that 
  \begin{equation}
     M^r Y^r N^r = X^r \ .
  \end{equation}
  Then it is easy to see that $MYN = X$. We have thus generated $X$.
\end{proof}

We are now in position to prove the Bridge lemma:

\begin{proof}
  Using lemma C we generate a transformation $V_1$ that mixes a
  vector $\ket{a_1}\in A$ with $B$ while leaving the rest of $A$
  unchanged. We then use lemma D to generate $SL(B\oplus
  span\{\ket{a_1}\})$. Repeating this process we add more and more
  vectors from $A$ until we generate all $SL(A\oplus B)$. Finally, a
  quick glance at the proof of Lemmas A-D reveals that in each one
  of them we used a finite number of transformations. Therefore
  every transformation in $SL(A\oplus B)$ can be represented as a
  finite product of transformations from $SL(A), SL(B)$ and the
  bridge $W$.
\end{proof}

\subsection{Proof of the non-unitary Decoupling
  lemma (Lemma~\ref{lem:non-decouple})}
\label{sec:non-decouple-proof}

  Consider the subgroups $H_A$ of $SL(A)$ and $H_B$ of $SL(B)$ which
  is defined by
  \begin{eqnarray}
    H_A &\EqDef& \left\{ X\in SL(A) | \exists \{g_n\}\in G \mbox{ s.t. }
    \rho_A(g_n)\to X \mbox{\ and\ } \rho_B(g_n)\to \Id\right\}
       \ , \\
    H_B &\EqDef& \left\{ X\in SL(B) | \exists \{g_n\}\in G \mbox{ s.t. }
    \rho_B(g_n)\to X \mbox{\ and\ } \rho_A(g_n)\to \Id\right\}
  \end{eqnarray}
  We would like to prove that if $\dim A\ne\dim B$ then $H_A=SL(A)$
  and $H_B=SL(B)$ (complete decoupling), whereas if $\dim A=\dim B$
  then either we have a complete decoupling, or $H_A$ is in the
  center of $SL(A)$ and $H_B$ is in the center of $H_B$.

  We first notice that $H_A$ and $H_B$ are normal closed subgroups
  of $SL(A)$ and $SL(B)$ respectively. The proof is straight
  forwards, and follows exactly a similar claim in the unitary
  decoupling lemma. 
  
  We now use fact that $SL(A)$ and $SL(B)$ are ``almost simple Lie
  groups'', both for $F=\mathbbm{R}$ and for $F=\mathbbm{C}$, which
  are also connected groups. Then every normal closed subgroup of
  them is either the whole group or is in their center. It is easy
  to see that if one of them is the whole master group then so must
  be the other. In such case the lemma is proved. Let us therefore
  assume that both subgroups are in the center of their
  master groups. We would like to show that in such case we must have
  $\dim A= \dim B$.
  
  We now define a homomorphism $M: SL(A)/H_A \to SL(B)/H_B$. For
  every coset $V_A H_A$ in $SL(A)/H_A$ we find a series $\{g_n\}$ in
  $G$ such that $\rho_A(g_n)$ converges to some element in $V_A H_A$
  and $\rho_B(g_n)$ converges to some $V_B\in SL(B)$. The existence
  of such series is promised by the conditions of the
  lemma\footnote{Notice that this is essential the only difference
  between the unitary and non-unitary decoupling lemmas. In the
  unitary case, the existence of such converging series is always
  promised due to the compactness of the unitary groups. We do not
  know if this is so in the non-unitary case, hence we used the
  extra condition.} Following the same argument of the unitary
  decoupling lemma we prove that $M$ is a continuous $1-1$
  homomorphism between $SL(A)/H_A$ and $SL(B)/H_B$. But these quotient
  groups are also Lie groups which can also be viewed as smooth
  manifolds. The isomorphism is thus turned into diffeomorphism
  between two manifolds, which can only happen if they have the same
  dimensions. But $\dim SL(A)/H_A = \dim SL(A)$ because $H_A$ is
  finite, and similarly $\dim SL(B)/H_B = \dim SL(B)$. Therefore
  $\dim SL(A)=\dim SL(B)$ which gives us $\dim A=\dim B$.

\section{Variants of the Solovay-Kitaev algorithm}\label{sec:non-SK-proof}

We now prove Theorem~\ref{thm:non-SK}.  We follow very closely the
derivation of Dawson \& Nielsen in Ref~\cite{ref:Daw05}.

We are given an $\epsilon_0$-net over $B_R(M)$, whose members are
transformations in $SL(M,\mathbbm{C})$ (the complex case) or $SL(M,
\mathbbm{R})$ (the real case). The size of $\epsilon_0$ depends on
$m\EqDef \dim M$ and on $R$. We will not give an explicit formula
for it, but instead assume that is small enough (but finite!). By a 
close inspection of the derivation, one can easily find the desired
size of $\epsilon_0$.

The SK algorithm is basically a recursive routine that receives a
transformation $V$ (not necessary unitary - but inside $B_R(M)$),
and an integer $n$, and returns $V_n$ - the $n$'th order
approximation for $V$. The distance between $V_n$ and $V$ is smaller
than $\epsilon_n \EqDef c^{-1}(c\epsilon_0)^{(3/2)^n}$, for some
constant $c<1/\epsilon_0$, thereby providing a super-exponential
convergence.

The main idea of the algorithm is unchanged. The goal is to find an
$\epsilon_n$-approximation for $\Delta \EqDef VV^{-1}_{n-1}$. This
is done by finding a product of commutators that approximate
$\Delta$ to a factor of $\mcO(\epsilon_{n-1}^{3/2})$, while making
sure that the matrices in the commutators are at most
$\mcO(\epsilon_{n-1}^{1/2})$ away from unity.  This approximation,
however, is \emph{not} given in terms of the generators.  Therefore
in the last step, we find a $\epsilon_{n-1}$ approximation to the
every matrix that appears in the commutators \emph{in terms of the
generators}, and use the following general property of the group
commutator: if $\tilde{V}, \tilde{W}$ are are an $\epsilon$
approximation for $V,W$, and in addition the distance of $V,W$ from
unity is $\delta$, then 
\begin{equation}
    \left\| \llbracket V,W \rrbracket - \llbracket\tilde{V},
    \tilde{W}\rrbracket\right\| = \mcO(\epsilon\delta) \ .
\end{equation}
(see Lemma~\ref{lem:commutator}). This allows us to give an
$\epsilon_{n-1}^{3/2}$ approximation of $\Delta$ in terms of the
generators.

The overall result is that we find $V_n$ such that $||V-V_n|| \le
\epsilon_0^{(3/2)^n}$. $V_n$ is a product of $13^n$ generators, and
constructing it takes a similar number of steps. Yet the
super-exponential convergence provides the promised asymptomatic.
This asymptotic is probably not an optimal one, yet it is simple to
derive and suffice for our purpose.

We begin by presenting a pseudo-code version of the algorithm,
followed by an analysis of its routines. Most of them are the same
for the complex and real cases, except for the
\texttt{GC-Unitary-Approx} routine, for which we supply a different
version for each case.

\subsection{A pseudo-code of the algorithm}

\noindent
\texttt{
function SK( Gate $V\in B_R(M)$, depth $n$)\\
if $(n==0)$: \\
.\hspace{1.5cm}Return Basic-Approximation($V$)\\
else \\
.\hspace{1.5cm}Set $V_n$ = SK($V,n-1$)\\
.\hspace{1.5cm}Set $\Delta = V V^{-1}_{n-1}$\\
.\hspace{1.5cm}Set $A,P$ = Polar-Decomp($\Delta$)\\
.\hspace{1.5cm}Set $V_A, W_A$ = GC-Unitary-Approx($A$)\\
.\hspace{1.5cm}Set $V_o, W_o, V_e, W_e$ = GC-Hermitian-Approx($A$)\\
.\hspace{1.5cm}Set $V_A', W_A', V_o', W_o', V_e', W_e'$ 
            = SK($V_A, W_A, V_o, W_o, V_e, W_e, n-1$)  \\
.\hspace{1.5cm}Set $\Delta' =  \llbracket V'_A, W'_A\rrbracket
  \cdot\llbracket V'_o, W'_o\rrbracket\cdot\llbracket V'_e, W'_e\rrbracket$ \\
.\hspace{1.5cm}Return $\Delta' V_{n-1}$
}

\subsection{The \texttt{Basic-Approximation($V$)} function}

This function returns the zero approximation to $V\in B_R(M)$. This
is a single generator whose distance from $V$ is smaller than
$\epsilon_0$. Such a generator must exist by the condition of the
$\epsilon_0$-net. As this net is finite, the search can be done in a
finite time with a naive search.

The necessity of the $\epsilon_0$-net to be finite is the reason why
we limit ourselves for transformation in $B_R(M)$ instead of the
full group $SL(M,\mathbbm{C})$ (or $SL(M, \mathbbm{R})$), whose
volume is infinite. 

\subsection{The \texttt{Polar-Decomp($\Delta$)} function}

The \texttt{Polar-Decomp($\Delta$)} routine gets a general
invertible matrix $\Delta$ with $||\Delta-\Id||\le \epsilon_{n-1}$
and returns its polar-decomposition
\begin{equation}
  \Delta = AP \ ,  
\end{equation}
with $||A-\Id||, ||P-\Id|| = \mcO(\epsilon_{n-1})$. The
polar-decomposition is a standard procedure which can be done
efficiently.  In the complex case, $A$ is a unitary matrix and $P$
is a positive-definite Hermitian matrix. In the real case, $A$ is an
orthogonal matrix while $P$ is a real positive-definite matrix.

The next lemma promises us that if $\Delta$ is close enough to
$\Id$, then also $P$ and $A$ will be close to $\Id$. 

\begin{lem}
 For small enough $\epsilon$, if $||\Delta-\Id|| \le \epsilon$ and
 $\Delta=AP$ with $\det \Delta=1$ then $||A-\Id||\le C\epsilon$ and
 $||P-\Id||\le C\epsilon$ for some constant $C$ which only depends
 on $m$.
\end{lem}
\begin{proof}
  Since $A$ is unitary (or orthogonal) then $||P||=||\Delta||$. But
  $||\Delta|| = ||\Delta -\Id+\Id|| \le ||\Delta-\Id|| + 1\le
  1+\epsilon$. Therefore $||P||\le 1+\epsilon$.
  
  $P$ is positive-definite Hermitian matrix and is therefore
  diagonalizable with eigenvalues $r_1\ge r_2\ge\ldots \ge r_m> 0$.
  In addition, as $\det P=1$, we have $\prod_i r_i =1$. It follows
  that $||P|| = r_1 \le 1+\epsilon$. But since $\prod_i r_i = 1$ it
  also follows that $r_m \ge (1+\epsilon)^{-(m-1)}\ge
  1-C_1\epsilon$, with $C'$ being a constant that depends only on
  $m$.  Combining these two results, we find that there are two
  $m$-dependent constants $C_1, C_2$ such that $||P-\Id|| \le
  C_1\epsilon$ and $||P^{-1}-\Id||\le C_2\epsilon$. Finally, as
  $A=\Delta P^{-1}$, we there is an $m$-dependent constant $C_3$
  such that $|A-\Id|| \le C_3\epsilon$. Taking $C=\max(C_1,C_2,C_3)$
  proves the lemma.
\end{proof}

\subsection{The \texttt{GC-Unitary-Approx($A$)} routine (complex case)}

This routine receives a special-unitary matrix $A$ with
$||A-\Id||\le \epsilon$, and (for epsilon small enough) returns two
special-unitary matrices $V_A,W_A$ such that $||W_A-\Id||,
||V_A-\Id|| \le C_1 \epsilon^{1/2}$ and $||A - \llbracket
V_A,W_A\rrbracket|| \le C_2\epsilon^{3/2}$. Note that as $V_A, W_A$
have unit determinant, then for small enough $\epsilon$, they are
inside $B_R(M)$. Again, this condition is guaranteed to hold
provided $\epsilon_0$ is small enough.  

The description of this function is given in \cite{ref:Daw05}.

\subsection{The \texttt{GC-Unitary-Approx($A$)} routine (real case)}

The real case is similar to the complex case, only that here $A$ is
a real orthogonal matrix, and resulting matrices $V_A, W_A$ are real
matrices (in order to have them in $B(M,\mathbbm{R})$) - though not
orthogonal.

Consider then an orthogonal matrix $A$ with $||A-\Id||\le \epsilon$.
It is well-known that $A$ admits a canonical form in which it
is block-diagonal with $2\times 2$ blocks of the form
\begin{equation}
  O(\theta)=\mat{\cos\theta}{\sin\theta}{-\sin\theta}{\cos\theta} \ ,
\end{equation}
and an optional diagonal term of $1$ (when $\dim M$ is odd).
\inote{Give ref. Look at the papers in the ref of the Wikipedia
entry for orthogonal matrices}. The
canonical basis is connected to the standard basis of $A$ via an
orthogonal transformation, hence we will assume without loss of
generality that $A$ is already given in its canonical form.
We will now work in every block separately. 

The one-dimensional block, if exists, is trivially taken care of by
also setting $V_A, W_A$ to be $1$. In the $2\times 2$ blocks, we
first work in the diagonalizing basis using the transformation
\begin{equation}
  S \EqDef \frac{1}{\sqrt{2}}\mat{1}{i}{i}{1} \ ,
\end{equation}
which gives us
\begin{equation}
  S^\dagger O(\theta) S = \mat{e^{i\theta}}{0}{0}{e^{-i\theta}} 
  = \exp \mat{i\theta}{0}{0}{-i\theta} \EqDef e^{iH}\ .
\end{equation}
Notice that the condition $||O(\theta)-\Id|| = \mcO(\epsilon)$
implies $||H||=|\theta|=\mcO(\epsilon)$. 

Consider the two traceless and Hermitian matrices
\begin{equation}
  F\EqDef \sqrt{\theta/2}\mat{0}{1}{1}{0} \ , \quad
  G\EqDef i\sqrt{\theta/2}\mat{0}{1}{-1}{0} \ .
\end{equation}
It is easy to verify that $[F,G]=iH$ and that $||F||,||G|| =
\mcO(\sqrt{||H||})=\mcO(\epsilon^{1/2})$. Here $[F,G]\EqDef FG-GF$ is the Lie
commutator, not to be confused with the group commutator. Defining
$V'_A\EqDef e^{F}$ and $W'_A\EqDef e^{G}$, it is easy to see that
$||V_A'-\Id||, ||W_A'-\Id||=\mcO(\epsilon^{1/2})$. We now use a
general exponential expansion to deduce that for every
matrices $A,B$ with $||A||, ||B|| \le \delta$, necessarily,
\begin{equation}
  \left\|e^{[A,B]} - \llbracket e^A, e^B\rrbracket\right\| 
    \le c\delta^3 \ ,
\end{equation}
for some $c\approx 4$ \inote{check this, or give a proper ref}. This implies that
\begin{equation}
  \left\| e^{iH} - \llbracket V'_A,W'_A\rrbracket \right\|
  = \mcO(\epsilon^{3/2}) \ .
\end{equation}
Finally, going back to the original basis, we define
\begin{equation}
    V_A \EqDef SV'_A S^\dagger = e^{SFS^\dagger} \ , \mbox{ and} \quad
    W_A \EqDef SW'_A S^\dagger = e^{SGS^\dagger} \ , 
\end{equation}
which gives
\begin{equation}
  \Big\| O(\theta) - \llbracket V_A,W_A\rrbracket \Big\|
  = \mcO(\epsilon^{3/2}) \ .
\end{equation}
It is easy to verify that $SFS^\dagger=F$, while
$SGS^\dagger=\mbox{diag}(\sqrt{\theta/2}, -\sqrt{\theta/2})$.
Therefore $V_A,W_A$ are in $SL(M, \mathbbm{R})$ as required.

\subsection{The \texttt{GC-Hermitian-Approx($P$)} routine}

This routine gets an Hermitian matrix $P$ with $||P-\Id||\le
\epsilon$ and returns four matrices $V_o, W_o, V_e, W_e \in B_R(M)$,
whose distance from $\Id$ is $\mcO(\epsilon^{1/2})$, and
\begin{equation}
  \label{eq:hermitian-approx}
  \Big\| \llbracket V_o, W_o\rrbracket
    \cdot\llbracket V_e, W_e\rrbracket - P\Big\| = \mcO(\epsilon^{3/2})
    \ .
\end{equation}

Its structure is similar to the structure of the
\texttt{GC-Unitary-Approx} routine \emph{in the real case}, only
that now we work with a positive-definite, Hermitian matrix $P$ (in
the real case $P$ is symmetric). 

We assume without loss of generality that $P$ is given in its
diagonal form $P=\mbox{diag}(r_1, \ldots, r_m)$ with $r_1\ge
r_2\ge\ldots \ge r_m >0$ and $r_1\cdot r_2\cdot\ldots\cdot r_m=1$.
This is because the diagonalizing matrix is either unitary (in the
complex case) or is orthogonal (in the real case).

As in the previous routine, we would like to work with pairs of
eigenvalues with opposite signs. We do that by decomposing $P$ into
a product of two diagonal matrices $P=P_o\cdot P_e$ where
$P_{o}=\mbox{diag}(\lambda_1, 1/\lambda_1, \lambda_3, 1/\lambda_3,
\ldots)$ (odd indices) and $P_e = \mbox{diag}(1, \lambda_2,
1/\lambda_2, \lambda_4, 1/\lambda_4, \ldots)$ (even indices). It is
easy to verify from the condition $||P-\Id||\le \epsilon$ that
$||P_o-\Id||, ||P_e-\Id|| = \mcO(\epsilon)$.

Consider now a $(\lambda, 1/\lambda)$ block in either matrices. It
can be written as $e^H$ with $H=\mbox{diag}(\theta, -\theta)$, and
$e^{\theta}=\lambda$. Using the same trick that we used in the
previous routine, we may write $H=[F,G]$ with 
\begin{equation}
  F\EqDef \sqrt{\theta/2}\mat{0}{1}{1}{0} \ , \quad
  G\EqDef \sqrt{\theta/2}\mat{0}{1}{-1}{0} \ .
\end{equation}
By the same arguments that were given in the previous routine,
$V=e^F$ and $W=e^G$ are both $\det=1$, \emph{real matrices}, with
$||V-\Id||, ||W-\Id||=\mcO(\epsilon^{1/2})$, and
\begin{equation}
  \left\| \mat{\lambda}{0}{0}{1/\lambda} - \llbracket V, W
  \rrbracket\right\| = \mcO(\epsilon^{3/2}) \ .
\end{equation}
The fact that $V,W$ are real matrices is crucial to the real case
where we must work with matrices from $SL(M,\mathbbm{R})$.

Performing the above decomposition for every $2\times 2$ block in
$P_o, P_e$, it follows that we can find matrices $V_o, W_o, V_e, V_e$
in $B_R(M)$ whose distance from the unity is $\mcO(\epsilon^{1/2})$
and which satisfy \Eq{eq:hermitian-approx}.

\subsection{Approximating the commutators in the main routine}

The distance between the output of the main routine and $V$ is
$||\Delta' V_{n-1} - V|| = ||\Delta' V_{n-1} - \Delta V_{n-1}|| \le
(R+1)||\Delta' -\Delta||$. The last equality follows from the fact
that $V_{n-1}\in B_R(M)$, hence $||V_{n-1}||\le R+1$.

Let us now estimate $||\Delta'-\Delta||$. We use the following
result that generalizes Lemma~1 in page~9 of Ref~\cite{ref:Daw05}:
\begin{lem}
\label{lem:commutator}
  Let $V,W, \tilde{V}, \tilde{W}$ be four matrices such that
  $||V-\tilde{V}||, ||W-\tilde{W}||\le \epsilon$ and $||V-\Id||,
  ||W-\Id||\le \delta$. Then (for $\epsilon, \delta$ small enough)
  there exists a constant $C$ such that
  \begin{equation}
    \left\| \llbracket V,W \rrbracket - \llbracket\tilde{V},
    \tilde{W}\rrbracket\right\| \le C\epsilon\delta \ .
  \end{equation}
\end{lem}

\begin{proof}

We expand $\tilde{V}, \tilde{W}$ in $\epsilon$ by
\begin{eqnarray}
  \tilde{V} &=& V + \epsilon A + O(\epsilon^2) \ , \\
  \tilde{W} &=& W + \epsilon B + O(\epsilon^2) \ .
\end{eqnarray}
Then it is easy to verify that
\begin{eqnarray}
  \tilde{V}^{-1} &=& V^{-1} - \epsilon V^{-1}AV^{-1} + O(\epsilon^2) \ , \\
  \tilde{W}^{-1} &=& W^{-1} - \epsilon W^{-1}BW^{-1} + O(\epsilon^2) \ .
\end{eqnarray}
Plugging this into $\tilde{V}\tilde{W}\tilde{V}^{-1}\tilde{W}^{-1}$
we obtain
\begin{eqnarray}
  \tilde{V}\tilde{W}\tilde{V}^{-1}\tilde{W}^{-1} &=&
     VWV^{-1}W^{-1} + \epsilon A WV^{-1}W^{-1}
     + \epsilon V BV^{-1}W^{-1} \\
    &-& \epsilon V WV^{-1}AV^{-1}W^{-1}
   - \epsilon V WV^{-1}W^{-1}BW^{-1} + O(\epsilon^2)
\end{eqnarray}
We wish to bound the norm of the terms linear in $\epsilon$. There
are four such terms.  Consider the terms that involve $A$ (other two
terms are treated similarly). As $||W||, ||W^{-1}||$ are bounded
from above, we get: 
\begin{equation}
  ||A WV^{-1}W^{-1} - WV^{-1}AV^{-1}W^{-1}|| 
    \le C ||A WV^{-1} - WV^{-1}AV^{-1}|| \ .
\end{equation}
Now expand $W=\Id+\delta D$ and plug it in the RHS of the above
equation. We get
\begin{eqnarray}
  && ||A (\Id+\delta D) V^{-1} - (\Id+\delta D)V^{-1}AV^{-1}||
  = ||AV^{-1} - V^{-1}AV^{-1} + O(\delta)|| \\
  && \le ||\Id-V^{-1}||\cdot ||A||\cdot||V^{-1}|| + O(\delta)\ .
\end{eqnarray}
However, by virtue of our $\epsilon$-expansion, $||A||=O(1)$ and
$||V^{-1}||$ is also bounded since $||V-\Id||\le \delta$. Therefore,
we are left with $||V^{-1}-\Id||$ which is $O(\delta)$. We have thus
shown that lowest non-trivial order is $\epsilon\delta$.

\end{proof}

With the last Lemma, we are able to show that $||\Delta'-\Delta||\le
C\epsilon_{n-1}^{3/2}$. Indeed by the previous reasoning we have
$\Delta=\llbracket V_A,W_A\rrbracket \cdot \llbracket
V_o,W_o\rrbracket\cdot \llbracket V_e,W_e\rrbracket$, and
$\Delta'=\llbracket V'_A,W'_A\rrbracket \cdot \llbracket
V'_o,W'_o\rrbracket\cdot \llbracket V'_e,W'_e\rrbracket$, with all
the unprimed matrices within distance $\mcO(\epsilon^{1/2}_{n-1})$
from $\Id$. Also, by the recursive application of the \texttt{SK}
routine, the distance of the primed matrices from the unprimed
matrices is $\mcO(\epsilon_{n-1})$. 
Using the last lemma, we get
\begin{equation}
  \Big\| \llbracket V_A,W_A\rrbracket 
     - \llbracket V'_A,W'_A\rrbracket\Big\|
     = \mcO(\epsilon^{3/2}) \ ,
\end{equation}
and similarly
\begin{equation}
  \Big\| \llbracket V_o,W_o\rrbracket 
     - \llbracket V'_o,W'_o\rrbracket\Big\|
     = \mcO(\epsilon^{3/2}) \ , \quad
  \Big\| \llbracket V_e,W_e\rrbracket 
     - \llbracket V'_e,W'_e\rrbracket\Big\|
     = \mcO(\epsilon^{3/2}) \ .
\end{equation}
Therefore $||\Delta'-\Delta|| =\mcO(\epsilon_{n-1}^{3/2})$. A
careful examination of the previous steps (which we omitted for sake
of clarity) reveals that the statement can be written as 
\begin{equation}
   ||\Delta'-\Delta|| \le C\epsilon_{n-1}^{3/2} \ ,
\end{equation}
with $C$ being some constant that depends on $m$ and $R$.

\subsection{Verifying the asymptotic}

We will now prove by induction that $||V_n-V|| \le \epsilon_n\EqDef
c^{-1}(c\epsilon_0)^{(3/2)^n}$, for some constant $c$, and that
$V_n$ is a product of $11^n$ generators.

For $n=0$ the statement is trivially true. Assume that it is true
for $n-1$ and consider the $n$'th case. Using the result of the last
section, we have
\begin{equation}
  ||V_n-V|| \le C\epsilon^{3/2}_{n-1}
    = C \left[ c^{-1}(c\epsilon_0)^{(3/2)^{n-1}} \right]^{3/2}
    = C c^{-1/2} \epsilon_n \ .
\end{equation}
Therefore choosing $c=C^2$ and making sure that $c\epsilon_0 < 1$
gives the desired result.

Finally, if $V_{n-1}, V'_A, W'_A, V'_o, W'_o, V'_e, W'_e$ are all
products of $13^{n-1}$ generators then it is clear from the formula
of $V_n$ that it contains $13^n$ generators.

This completes the proof of the non-unitary Solovey-Kitaev theorem.

\section{Proof of improved 8-strands density and efficiency
  theorem (Theorem~\ref{thm:improved})}
\label{sec:improved}

We first list the different subspaces that make up $H_8^*$ when
$q=3$. In such case $d=\sqrt{3}=2\cos\pi/6$, and we have an
Hermitian representation that is defined using a \emph{finite}
auxiliary graph with 5 sites. Indeed, the first 6 entries of the
general eigenvalue that is described in
Eqs.~(\ref{eq:pi-1}-\ref{eq:pi-n}) are $\pi=(1,\sqrt{3}, 2,
\sqrt{3}, 1, 0, \ldots)$. Therefore $H^*_8$ has a finite dimension,
and can write down its decomposition to the subspaces
$H_{8,k\to\ell}$ with $k,\ell \in\{1,2,3,4,5\}$ and even $|k-\ell|$.
An easy calculation shows that there are exactly 13 such subspaces,
with a total dimensionality of 378. They are fully listed in
Table~\ref{tab:spaces}.

\begin{table}
\center
  \begin{tabular}{l|l}
    dim & Subspaces \\
    \hline
    $14$ & $H_{8,1\to 1}$, $H_{8,5\to 5}$ \\
    $13$ & $H_{8,1\to 5}$, $H_{8,5\to 1}$ \\
    $27$ & $H_{8,1\to 3}$, $H_{8,3\to 1}$, 
           $H_{8,3\to 5}$, $H_{8,5\to 3}$ \\
    $40$ & $H_{8,2\to 4}$, $H_{8,4\to 2}$ \\
    $41$ & $H_{8,2\to 2}$, $H_{8,4\to 4}$ \\                
    $54$ & $H_{8,3\to 3}$
  \end{tabular}
  \caption{The list of subspaces that make up the full 8-steps space
  $H^*_8$ for $q=3$. Subspaces with the same dimensionality are
  isomorphic to each other with respect to the path representation.
      \label{tab:spaces}}
\end{table}

Because of the fact that $\pi_i = \pi_{6-i}$, it is easy to see that
the spaces $H_{8,i\to j}$ are isomorphic to $H_{8,6-i\to 6-j}$ with
respect to the $\Phi_i$ operators, and consequently the crossings
operators. Moreover, under the transformation $\ket{p_1,p_2,\ldots,
p_9} \mapsto \ket{6-p_1, 6-p_2, \ldots, 6-p_9}$, these operators
retain the same matrix form. In addition we can prove that the
spaces $H_{8,i\to j}$ are isomorphic to $H_{8,j\to i}$. 

It follows that all subspaces with the same dimensionality are
isomorphic to each other. They are therefore grouped in same line in
Table~\ref{tab:spaces}.

The first step of the proof is to show that in every subspace
$H_{8,k\to\ell}$, the normalized crossing operators $\hat{\sigma}_1,
\ldots, \hat{\sigma}_7$ create a dense subgroup, either in
$SL(H_{8,k\to\ell}, \mathbbm{C})$ or in $SL(H_{8,k\to\ell},
\mathbbm{R})$. Notice that according to Lemma~XXX\inote{write down
the commutator lemma}, we need not care about the actual
normalization of these operators, because in the end we always use
the commutator group, which cancels them out. In addition, due to
the isomorphism between subspaces of equal dimensionality, it is
enough to prove density over six subspaces of different
dimensionality. 

To prove density in a general $H_{8,k\to\ell}$, we follow the same
steps we took in the density proof of $H_{8,1\to 1}$. We first
establish an $SL(2,\mathbbm{C})$ or $SL(2,\mathbbm{R})$ density on a
two-dimensional subspaces (the seeding), and then build up density
on larger and larger subspaces using the non-unitary Bridge and
Decoupling lemmas. The trickiest part of the proof is the first
part, the seeding. We need to find a two dimensional subspace and a
pair of operators $\Phi_i, \Phi_{i+1}$ which operate inside that
subspace. Then the corresponding crossing operators will also
operate inside that subspace. In the $H_{8,1\to 1}$ case this was
done by finding a path ($\ket{p_1}$ in \Fig{fig:8p}) which is a
non-trivial eigenvector of some $\Phi_i$ ($\Phi_1$ in the $H_{8,1\to
1}$ case). This vector is mixed with another vector ($\ket{p_2}$ in
\Fig{fig:8p}) by either $\Phi_{i+1}$ or $\Phi_{i-1}$, while at the
same time, the other vector is sent to zero under the action of
$\Phi_i$.

Such non-trivial eigenvector can always be found if $H_{8,k\to\ell}$
contains ``zigzag'' paths that ``hit the ceiling or the floor''.
More precisely, we are looking for paths that contain a passage like
$1\to 2\to 1$ or $5\to 4\to 5$. When this happens at the $i$, $i+1$
steps, then such a path is an eigenvalue of $\Phi_i$ with eigenvalue
$d$. The reason is that when $\Phi_i$ acts on a path
$\ket{\ldots\ell,\ell+1,\ell\ldots}$, or
$\ket{\ldots\ell,\ell-1,\ell\ldots}$, it will transform it to a linear
combinations of $\ket{\ldots\ell,\ell\pm 1,\ell\ldots}$. But when
$\ell$ is the first or last site ($\ell=1$ or $\ell=5$
respectively), one of the eigenvectors does not exists, hence the
other becomes an eigenvector.

It is easy to see that such ``zigzag'' paths must exist in all of
the subspaces of Table~\ref{tab:spaces} because we are dealing with
8-steps paths over a graph with only 5 vertices; no matter where we
start, we can always reach the one end and bounce back. Finally, it
is also easy to check that the two-dimensional matrices always have
the same form - independently of the subspace that we work in - and
therefore the density criteria that is given in
Definition~XXX\inote{add label} is valid also here.

The next step is the use density on the two-dimensional space and
prove density over larger and larger spaces using the Bridge lemma
and the decoupling lemma. This step can probably be done generally
for a large class of subspaces, using some kind of an induction
argument. However, as we are interested solely in $q=3$, we opted 
for a brute-force solution. We designed a simple computer code that
found a series of moves consisting of an altering application of the
Bridge and Decoupling lemmas that establish density in each
subspace. Not surprisingly, the details of the derivation in each
subspace is far too long and tedious to be given here.\inote{I don't
know if this is the right hand-waving we should use here.}

After we established density on every subspace, we can use the
non-unitary Decoupling Lemma~\ref{lem:non-decouple} to prove that
spaces with different dimensionality decouple. In other words, we
can approximate any operator on $V$ on $H_8^*$ using the normalized
crossing operators provided that: 1) $V$ is block-diagonal on the
subspaces of Table~\ref{tab:spaces}, and 2) blocks with the same
dimensionality are coupled to each other by the similarity
transformation that defines the isomorphism of these subspaces. 

We now wish to use the Solovey-Kitaev theorem to generate such
approximations efficiently. Looking at the proof of the non-unitary
Solovey-Kitaev theorem that is given in
Appendix~\ref{sec:non-SK-proof}, shows that it is easy to generalize
it to a direct sum of spaces. More precisely, the generalization
goes as follows: we define $M_1, \ldots M_k$ to be finite subspaces,
and look at $M=M_1\oplus\ldots\oplus M_k$. We wish to approximate
any operator in $B_R(M)$, which is the set of transformations $V$
such that: 1) $V$ is block diagonal on $M$ according to the $M_i$
blocks, 2) at every block it belongs to $SL(M_i, \mathbbm{C})$ (or
$SL(M_i, \mathbbm{R})$), and 3) $||V-\Id||<R$. This is done using a
set of generators that create an $\epsilon_0$-net over $B_R(M)$ for
some $\epsilon_0$ that depends on the dimensionality of the various
$M_i$ spaces, as well as on $R$.

The fact that we establish density, independently, over all the
subspaces of $H_8^*$ with different dimensionality, means that we
can use the generalized Solovey-Kitaev for a direct sum of spaces
with different dimensionality. In other words, we take $M_1, \ldots,
M_6$ to be representatives from each dimensionality class in
Table~\ref{tab:spaces}, with $M_1=H_{8,1\to 1}$. We will use the
generalized Solovey-Kitaev on $M=M_1\oplus\ldots\oplus M_6$.

Consider a two-qubits gate $U$ and its encoded version $\Un{U}$
which operates on $H_{8,1\to 1}$. We can extend $\Un{U}$ to be a
block operator on $M$ by setting to $\Id$ over all the other five
blocks. Then we use the extended Solovey-Kitaev theorem to find an
$\epsilon$-approximation $V$ to $Un{U}$ for some $\epsilon>0$, using
the normalized crossing operators. How does $V$ behaves on the
subspaces which are included in $M$? By virtue of the isomorphism
between these spaces to their representative in $M$, we conclude
that $V$ is either conjugate to a $\epsilon$-approximation of the
unity matrix or to a $\epsilon$-approximation of $\Un{U}$. In both
cases, since the conjugation is done by a unitary matrix, the norm
of $V$ is bounded by $1+\epsilon$. To summarize, we found an
approximation $V$ in terms of the normalized crossing operators,
such that
\begin{equation}
  || V-\Un{U}||_{H_{8,1\to 1}} \le \epsilon \ , 
\end{equation}
and 
\begin{equation}
  || V||_{H_8^*} \le 1+\epsilon \ .
\end{equation}

Lastly, we write the approximation $V$ in terms of the un-normalized
crossing operators, and we recover the $GTL(d)$ element $\mcT$ such
that $V=\Delta^{-1}_\mcT \rho(\mcT)$. Substituting this in the two
equations above, and using the fact that $||\cdot||_{L_8}
\le||\cdot||_{H_{8,1\to 1}}$ proves the theorem.

\end{document}